\documentclass[a4paper,11pt,headings=big,DIV=12]{scrartcl}
\pdfoutput=1

\usepackage[a4paper, top=1.2in, bottom=1.2in, left=1.1in, right=1.1in]{geometry}
\usepackage[utf8]{inputenc}
\usepackage{amsmath, amssymb}
\usepackage{color}
\usepackage[dvipsnames, table]{xcolor}
\definecolor{blue}{rgb}{0,0,0.5}
\usepackage[colorlinks,linkcolor=blue,urlcolor=blue,citecolor=blue]{hyperref}
\usepackage{graphicx}
\graphicspath{{./Figs/}}
\usepackage{slashed}
\usepackage{cite}
\usepackage{enumitem}
\usepackage{bm}

\newcommand{\be}{\begin{equation}}
\newcommand{\ee}{\end{equation}}
\newcommand{\bea}{\begin{eqnarray}}
\newcommand{\eea}{\end{eqnarray}}
\newcommand{\<}{\langle}
\renewcommand{\>}{\rangle}
\newcommand{\mc}{\mathcal}

\newcommand{\nn}{\nonumber}

\DeclareOldFontCommand{\rm}{\normalfont\rmfamily}{\mathrm}
\DeclareOldFontCommand{\sf}{\normalfont\sffamily}{\mathsf}
\DeclareOldFontCommand{\tt}{\normalfont\ttfamily}{\mathtt}
\DeclareOldFontCommand{\bf}{\normalfont\bfseries}{\mathbf}
\DeclareOldFontCommand{\it}{\normalfont\itshape}{\mathit}
\DeclareOldFontCommand{\sl}{\normalfont\slshape}{\@nomath\sl}
\DeclareOldFontCommand{\sc}{\normalfont\scshape}{\@nomath\sc}

\begin{document}

%%%% TITLE PAGE
\begin{flushright}
\small
LAPTH-024/20
\end{flushright}
\vskip0.5cm

\begin{center}
%%%% TITLE
{\sffamily \bfseries \LARGE \boldmath
The Dark Side of 4321
}\\[0.8 cm]
%%%% AUTHORS
{\normalsize \sffamily \bfseries Diego Guadagnoli, Méril Reboud, Peter Stangl} \\[0.5 cm]
\small
{\em LAPTh, Universit\'{e} Savoie Mont-Blanc et CNRS, Annecy, France}
%\\[0.1cm]
\end{center}

\medskip

\begin{abstract}

\noindent The evidence of Dark Matter (DM) is one of the strongest observational arguments in favour of physics beyond the Standard Model.
Despite expectations, a similar evidence has been lacking so far in collider searches, with the possible exception of $B$-physics discrepancies, a coherent set of persistent deviations in a homogeneous dataset consisting of $b \to c$ and $b \to s$ semi-leptonic transitions.
We explore the question whether DM and the $B$~discrepancies may have a common origin. We do so in the context of the so-called 4321 gauge model, a UV-complete and calculable setup that yields a $U_1$ leptoquark, the by far most successful single mediator able to explain the $B$~anomalies, along with other new gauge bosons, including a $Z^\prime$. Adding to this setup a `minimal' DM fermionic multiplet, consisting of a ${\bf 4}$ under the 4321's $SU(4)$, we find the resulting model in natural agreement with the relic-density observation and with the most severe direct-detection bounds, in the sense that the parameter space selected by $B$~physics is also the one favoured by DM phenomenology. The DM candidate is a particle with a mass in the WIMP range, freeze-out dynamics includes a co-annihilator (the `rest' of the {\bf 4} multiplet), and the most important gauge mediator in the DM sector is the $Z^\prime$.

\end{abstract}

%%%% BEGIN OF TEXTBODY %%%%

\section{Introduction} \label{sec:intro}

After the end of LHC Run 2, no sign of new physics (NP) has been observed in direct searches.
There are, however, several indirect hints for NP.
Flavour physics experiments have reported a large set of deviations from Standard Model (SM) predictions in $B$-meson decays, which are also known as $B$-meson anomalies.
They amount to discrepancies both in neutral current $b\to s\ell\ell$ decays~\cite{Aaij:2013qta,Aaij:2014pli,Aaij:2015esa,Aaij:2014ora,Aaij:2015oid,ATLAS:2017dlm,CMS:2017ivg,Khachatryan:2015isa,Aaij:2017vbb,Aaij:2019wad,Abdesselam:2019wac} and in charged current $b\to c\ell\nu$ decays~\cite{Lees:2012xj,Lees:2013uzd,Huschle:2015rga,Aaij:2015yra,Sato:2016svk,Hirose:2016wfn,Aaij:2017tyk,Aaij:2017uff,Abdesselam:2019dgh,Belle:2019rba
}.
It was realized that the presence of a $U_1$ leptoquark (LQ) with SM quantum numbers $({\bf 3},{\bf 1},2/3)$ could simultaneously explain both of these sets of discrepancies~\cite{Alonso:2015sja,Calibbi:2015kma,Barbieri:2015yvd,Hiller:2016kry,Bhattacharya:2016mcc,Buttazzo:2017ixm,Calibbi:2017qbu,Angelescu:2018tyl,Kumar:2018kmr}.
While other simultaneous solutions are possible (see e.g.~\cite{Das:2016vkr,Crivellin:2017zlb,Marzocca:2018wcf,Becirevic:2018afm,Bigaran:2019bqv,Datta:2019bzu,Altmannshofer:2020axr}), an explanation in terms of a $U_1$ LQ has become even more consistent with recent data~\cite{Aebischer:2019mlg,Alguero:2019ptt,Crivellin:2018yvo}.
Being a massive vector boson, the $U_1$ requires a UV completion from which it arises either as a gauge boson of a spontaneously broken gauge symmetry or as a composite vector boson (see e.g.~\cite{Barbieri:2016las}).
In fact, the $U_1$ is a well-known prediction of Pati-Salam models~\cite{Pati:1974yy}, which extend the SM color $SU(3)_c$ gauge group to $SU(4)$, under which quarks and leptons transform in unified multiplets.
Since traditional Pati-Salam models cannot accommodate the flavour structure needed for explaining the $B$-meson anomalies, variants of the Pati-Salam models based on the gauge group $SU(4)\times SU(3)'\times SU(2)_L\times U(1)_X$ have been constructed to this end, the so-called 4321 models~\cite{Georgi:2016xhm,Diaz:2017lit,DiLuzio:2017vat,Blanke:2018sro,DiLuzio:2018zxy,Bordone:2017bld,Greljo:2018tuh,Bordone:2018nbg,Cornella:2019hct,Fuentes-Martin:2020bnh}.
In these models, the SM arises after the group $SU(4)\times SU(3)'\times U(1)_X$ is spontaneously broken to its $SU(3)_c\times U(1)_Y$ subgroup.
The heavy vector bosons resulting from this symmetry breaking include the $U_1$ LQ but in addition also an $SU(3)_c$ octet $G'$ dubbed ``coloron'' and a SM singlet $Z'$.
Among the 4321 models, those that unify the third family of SM quarks and leptons are of special interest since they imply an approximate global $U(2)^5$ flavour symmetry \cite{Barbieri:2011ci,Blankenburg:2012nx,Barbieri:2012uh}.
Such a symmetry is particularly useful for explaining the $B$-meson anomalies without violating other flavour bounds while at the same time reproducing the SM fermion masses and CKM elements~\cite{Bordone:2017bld,Greljo:2018tuh,Bordone:2018nbg,Cornella:2019hct,Fuentes-Martin:2020bnh}.

Apart from the hints of NP provided by the $B$-meson anomalies, there are other observations that suggest an extension of the SM.
One of the most solid indications of physics beyond the SM is provided by the strong evidence for the existence of Dark Matter (DM) \cite{Bertone:2004pz,Roszkowski:2017nbc}.
An immediate question is thus whether any of the new heavy vector bosons in 4321 models could be related to the generation of a DM thermal relic.
More specifically, we would like to address the possibility that these vector bosons serve as mediators between SM fermionic currents and a DM current.
The latter current may be either bosonic or fermionic.
However, a scalar DM candidate can annihilate to SM particles via a Higgs portal such that the DM phenomenology would not rely on the new vector bosons.
Therefore, we restrict the discussion to fermionic DM.\footnote{%
Other interesting cases that are beyond the scope of the present article include composite bosonic DM that could naturally arise if the 4321 gauge symmetry is broken by a new strong interaction~\cite{Fuentes-Martin:2020bnh}.
}

To be specific, we consider a fermionic DM candidate $\chi_0$ that fulfils the following assumptions (cf.\ e.g.\cite{Cirelli:2005uq,Baker:2015qna})
\begin{enumerate}[label={\em (\roman*)}]
 \item it is a thermal relic,
 \item it is colorless and electrically neutral,
 \item it has zero hypercharge to avoid direct-detection bounds,
 \item it is the component of a massive fermion multiplet $\Psi_{\rm DM}$ that is vector-like (VL) under the 4321 gauge symmetry,
 \item (co-)annihilation proceeds via 2 $\rightarrow$ 2 processes induced at tree level through the new vector bosons $U_1$, $G'$, and $Z'$.
\end{enumerate}
These assumptions put restrictions on the possible 4321 quantum numbers of $\Psi_{\rm DM}$.
Conditions {\em (ii)} and {\em (iii)}, i.e.\ zero electric charge and zero hypercharge of $\chi_0$ require that $\Psi_{\rm DM}$ transforms under an $SU(2)_L$ representation with odd dimension.
Condition {\em (iii)} further implies that the $\chi_0$ eigenvalue of the hypercharge generator $Y$ vanishes.
The definition of $Y$ in terms of the $U(1)_X$ charge $X$ and the diagonal $SU(4)$ generator $T^{15}$, $Y=X+\sqrt{\frac{2}{3}}\,T^{15}$
then fixes $X$ for a given $SU(4)$ representation.

Guided by minimality, we restrict our discussion to singlets of $SU(3)'$ and to the smallest non-trivial representation of $SU(4)$, the fundamental $\mathbf{4}$ representation, which leads to\footnote{%
If $\Psi_{\rm DM}$ is a singlet of $SU(4)$, then its 4321 quantum numbers are fixed to $({\bf 1}, {\bf 1}, {\bm N}, 0)$, and couplings to $U_1$, $G'$, and $Z'$ are absent.
This corresponds to ``Minimal Dark Matter'', discussed in~\cite{Cirelli:2005uq}.
}
\begin{equation}
 \Psi_{\rm DM} \sim ({\bf 4}, {\bf 1}, {\bm N}, +1/2),
    \quad\quad
    {\bm N}\in\{{\bf 1}, {\bf 3}, {\bf 5},...\}
\end{equation}
under the 4321 gauge group.
After the 4321 symmetry breaking, the $\Psi_{\rm DM}$ multiplet splits into the two components $\chi$ and $\psi$, which transform under the SM gauge group as
\be
\label{eq:chi-psi-SU2L}
    \chi\sim({\bf 1}, {\bm N}, 0),
    \quad\quad
    \psi\sim({\bf 3}, {\bm N}, 2/3),
    \quad\quad
    {\bm N}\in\{{\bf 1}, {\bf 3}, {\bf 5},...\}.
\ee
The dark matter candidate $\chi_0$ is then identified with the electrically neutral component of the $SU(2)_L$ $N$-plet $\chi$.
For ${\bm N}={\bf 1}$ and ${\bm N}={\bf 3}$, renormalizable couplings between SM particles and the DM candidate exist, such that the latter is in general not stable on time scales vastly below the age of the Universe~\cite{Cirelli:2005uq}.
In such cases one has to advocate extra symmetries in order for it to be a viable relic.
Within our setup, in the ${\bm N}={\bf 1}$ case the field $\psi$ has the same quantum numbers as right-handed up-type quarks and mixing between $\psi_R$ and $u_R^i$ make $\chi$ unstable.
In the ${\bm N}={\bf 3}$ case, a coupling of $\chi$ to the Higgs and lepton doublets is allowed such that the DM candidate could decay to a Higgs and a neutrino.
The smallest $N$ for which the DM candidate is stable because of the absence of renormalizable couplings that would allow it to decay is ${\bm N}={\bf 5}$ \cite{Cirelli:2005uq}.\footnote{
Other less minimal scenarios that even for ${\bm N}={\bf 1},{\bf 3}$ do not contain renormalizable couplings that destabilize DM would require $\Psi_{\rm DM}$ to transform under larger $SU(4)$ representations or under non-trivial representations of both $SU(4)$ and $SU(3)'$.
}

In the rest of this paper, we will analyze the phenomenology of all the ${\bm N}={\bf 1}, {\bf 3}, {\bf 5}$ cases, bearing in mind that ${\bm N}={\bf 1},{\bf 3}$ require the additional assumption that renormalizable couplings that destabilize DM are absent. In Sec.~\ref{sec:model} we will describe our model setup, paying particular attention to the fermionic sector, that includes the DM multiplet.
An extended discussion about the different possibilities for implementing the SM fermions in such a setup is included in Appendix~\ref{app:fermions}. Sec.~\ref{sec:Omega_DM} discusses our analytic approach towards estimating the DM relic within our model, including in particular the impact of mass splittings between the DM and its co-annihilator, and our procedure towards estimating the thermally averaged cross section. Mass splittings are discussed within a more general approach in Appendix~\ref{app:mass_splitting}, and the cross sections relevant for the thermal average are collected in Appendix~\ref{app:Sigma_eff}. In Sec.~\ref{sec:DD} we then move on to describe our approach towards the estimate of direct-detection signals. Sec.~\ref{sec:results} collects our results, addressing the question to what extent $B$-physics discrepancies and DM phenomenology are compatible with one another within our setup. We conclude in Sec.~\ref{sec:conclusions}.

\section{Model setup} \label{sec:model}

We consider a `4321' model \cite{DiLuzio:2017vat,Diaz:2017lit} based on the gauge group
\be
    SU(4)\times SU(3)'\times SU(2)_L\times U(1)_X.
\ee
At a scale\footnote{%
If the breaking of the gauge group is due to vacuum expectation values (VEVs) of scalar fields, different scalars can contribute to the breaking at slightly different scales (cf.\ e.g.~\cite{DiLuzio:2018zxy,Cornella:2019hct}).
In order to reduce the number of parameters, we consider only a single breaking scale (as predicted by the model in~\cite{Fuentes-Martin:2020bnh}), and we do not specify the exact mechanism that triggers the spontaneous breaking.
} $v_{\rm LQ}$, the spontaneous breaking
\be
\label{eq:431breaking}
 SU(4)\times SU(3)'\times U(1)_X \to SU(3)_c\times U(1)_Y
\ee
yields the SM color times hypercharge factors.
Given the $SU(3)_{4}\times U(1)_{4}$ subgroup of $SU(4)$, the spontaneous breaking proceeds such that $SU(3)_c$ is the diagonal subgroup of $SU(3)_{4}\times SU(3)'$ and $U(1)_Y$ is the diagonal subgroup of $U(1)_{4}\times U(1)_X$.

\subsection{Vector bosons}
Following the notation of~\cite{Cornella:2019hct}, we denote the gauge fields of $SU(4)$, $SU(3)'$, and $U(1)_X$ by $H_\mu^\alpha$, $C_\mu^a$, and $B^\prime_\mu$, respectively.
The spontaneous breaking yields a massive $U_1$ LQ
\be
 {U_\mu^\pm}^{1,2,3}
 =
 \tfrac{1}{\sqrt{2}}(H_\mu^{9,11,13}\mp i H_\mu^{10,12,14}),
\ee
as well as the massive $Z'_\mu$ and the massive gluon-like `coloron' fields $G^{\prime a}_\mu$, given by the linear combinations
\be
\label{eq:Z'G'defs}
 Z^\prime_\mu
 =
 H_\mu^{15}\, \cos \theta_{41} - B^\prime_\mu\,\sin \theta_{41},
 \quad\quad
 G^{\prime a}_\mu
 =
 H^a_\mu\, \cos \theta_{43} - C^a_\mu\,\sin \theta_{43},
\ee
whereas the linear combinations orthogonal to $Z'_\mu$ and $G^{\prime a}_\mu$ are the massless hypercharge and QCD gauge bosons $B_\mu$ and $G^a_\mu$.
Denoting the gauge couplings of $SU(4)$, $SU(3)'$, $U(1)_X$, $SU(3)_c$, and $U(1)_Y$ by $g_4$, $g_3$, $g_1$, $g_s$, and $g_Y$, the angles $\theta_{41}$ and $\theta_{43}$ are defined analogously to the weak-mixing angle by
\begin{equation}\label{eq:gauge_mixing}
 \cos \theta_{41}
 =
 \frac{g_4}{\sqrt{g_4^2+\tfrac{2}{3} g_1^2}}
 =
 \frac{g_Y}{g_1},
 \quad\quad
 \cos \theta_{43}
 =
 \frac{g_4}{\sqrt{g_4^2+g_3^2}}
 =
 \frac{g_s}{g_3}.
\end{equation}
Since the couplings $g_s$ and $g_Y$ are the known QCD and hypercharge couplings, eq.~\eqref{eq:gauge_mixing} implies that for a given value of $g_4$, the other two couplings $g_1$ and $g_3$ are fixed.
Consequently, the gauge sector of the model can be parameterized by only the two independent parameters
\begin{equation}\label{eq:vLQ_g4}
 v_{LQ}\,,
 \quad
g_4\,.
\end{equation}
The masses of $U_\mu$, $Z^\prime_\mu$, and $G^{\prime}_\mu$ are given by
\be
 \begin{aligned}
  M_U^2 &= \tfrac{1}{4}\, g_4^2\, v_{\rm LQ}^2,
  \\
  M_{Z^\prime}^2 &= \tfrac{1}{4}\, \left(g_4^2+\tfrac{2}{3}\,g_1^2\right)\, v_{\rm LQ}^2,
  \\
  M_{G^\prime}^2 &= \tfrac{1}{4}\, \left(g_4^2+g_3^2\right)\, v_{\rm LQ}^2,
 \end{aligned}
\ee
and they are related through the angles $\theta_{41}$ and $\theta_{43}$ as\footnote{%
Note that this relation can be modified if the 4321 symmetry is broken by different scalars at slightly different scales, cf.\ e.g.~\cite{DiLuzio:2018zxy,Cornella:2019hct}.
}
\be
 M_U = M_{Z^\prime}\, \cos \theta_{41} = M_{G^\prime}\, \cos \theta_{43}~,
\ee
implying that, at tree level, the $U_1$ is expected to be the lightest new vector boson.

\subsection{Fermions}
Among the different possibilities for implementing the SM fermions in a 4321 model (see Appendix~\ref{app:fermions}), a well-motivated and phenomenologically successful variant corresponds to a unification of third-family quarks and leptons~\cite{Bordone:2017bld,Greljo:2018tuh,Bordone:2018nbg,Cornella:2019hct,Fuentes-Martin:2020bnh}.
In this case, the first and second families of SM-like fermions transform under the $SU(3)'\times SU(2)_L\times U(1)_X$ subgroup of the 4321 symmetry like the usual SM fermions, whereas the third-family quarks and leptons are unified into
$\Psi_L^{\prime\,3}\equiv(q_L^{\prime\,3}\; \ell_L^{\prime\,3})^\intercal$, $\Psi_R^{\prime+\,3}\equiv(u_R^{\prime\,3}\; \nu_R^{\prime\,3})^\intercal$, and $\Psi_R^{\prime-\,3}\equiv(d_R^{\prime\,3}\; e_R^{\prime\,3})^\intercal$, which transform under the 4321 symmetry as shown in table~\ref{tab:fermions}.
\begin{table}[t]
\begin{center}
\def\arraystretch{1.4}
\begin{tabular}{|c|c|c|c|c|}
\hline
Field & $SU(4)$ & $SU(3)'$ & $SU(2)_L$ & $U(1)'$ \\
\hline
\hline
$\ell_L^{\prime\,1,2}$&${\bf 1}$ & ${\bf 1}$ & ${\bf 2}$ & $-1/2$ \\
$e_R^{\prime\,1,2}$&${\bf 1}$ & ${\bf 1}$ & ${\bf 1}$ & $-1$ \\
$q_L^{\prime\,1,2}$&${\bf 1}$ & ${\bf 3}$ & ${\bf 2}$ & $+1/6$ \\
$u_R^{\prime\,1,2}$&${\bf 1}$ & ${\bf 3}$ & ${\bf 1}$ & $+2/3$ \\
$d_R^{\prime\,1,2}$&${\bf 1}$ & ${\bf 3}$ & ${\bf 1}$ & $-1/3$ \\
$\Psi_L^{\prime\,3}$&${\bf 4}$ & ${\bf 1}$ & ${\bf 2}$ & $0$ \\
$\Psi_R^{\prime+\,3}$&${\bf 4}$ & ${\bf 1}$ & ${\bf 1}$ & $+1/2$ \\
$\Psi_R^{\prime-\,3}$&${\bf 4}$ & ${\bf 1}$ & ${\bf 1}$ & $-1/2$ \\
\hline
\hline
$\Psi_{\rm DM}$&${\bf 4}$ & ${\bf 1}$ & ${\bm N}$ & $+1/2$ \\
\hline
\end{tabular}
\end{center}
\caption{Quantum numbers of SM-like fermions (upper block) and the vector-like DM multiplet $\Psi_{\rm DM}$ (last row).
First and second generation fermions transform under $SU(3)'\times SU(2)_L\times U(1)_X$ like the usual SM fermions; the third generation quarks and leptons are unified into $\Psi_L^{\prime\,3}\equiv(q_L^{\prime\,3}\; \ell_L^{\prime\,3})^\intercal$, $\Psi_R^{\prime+\,3}\equiv(u_R^{\prime\,3}\; \nu_R^{\prime\,3})^\intercal$, and $\Psi_R^{\prime-\,3}\equiv(d_R^{\prime\,3}\; e_R^{\prime\,3})^\intercal$.
}
\label{tab:fermions}
\end{table}
Due to their quantum numbers, the light SM fermions cannot directly couple to the $U_1$.
However, small but non-vanishing couplings between the $U_1$ and light SM fermions are required to explain the $B$-meson anomalies.
To realize this, we introduce two massive fermions that couple to the $U_1$ and mix with the left-handed first and second generation SM-like fermions.
In addition to couplings between light fermions and the $U_1$, whose sizes are controlled by the mixing, this construction also generates the 2-3 entries in the CKM matrix.
The new heavy fermions transform in the same way as $\Psi_L^{\prime\,3}$ (cf. table~\ref{tab:fermions}) and we denote their left-handed components by $\Psi_L^{\prime\,1,2}$.
While the mixing is important for the couplings of the SM fermions to $U_1$, $Z'$, and $G'$, we do not further discuss the new heavy fermion mass eigenstates since they are not relevant for the DM dynamics as long as their masses are larger than $M_\chi + M_\psi$, which we assume in the following.
Due to the mixing, the first and second generation SM $SU(2)_L$ doublets are in general linear combinations of the SM-like fields $q_L^{\prime\,1,2}$, $\ell_L^{\prime\,1,2}$ and the new heavy fields $\Psi_L^{\prime\,1,2}$.
To avoid large flavour violating effects, we align the mixings between SM fermions and new heavy fermions in the basis in which the down-quark mass matrix is diagonal~(cf.\ e.g.~\cite{DiLuzio:2018zxy}) such that the mixings are flavour-diagonal for the fields
\begin{equation}
  q^i = \begin{pmatrix} V_{ji}^*\,u^j_L \\ d^i_L \end{pmatrix} \,, \qquad
  \ell^j =  \begin{pmatrix} \nu^j_L \\ e^j_L \end{pmatrix} \,, \qquad
  u^i_R \,, \qquad d^i_R \,, \qquad e^i_R \,, \qquad \nu^i_R~,
\end{equation}
where $V$ is the CKM matrix and $u^i$, $d^i$, $e^i$, and $\nu^i$ are mass eigenstates.
A possible misalignment between the quark and lepton components of the fields $\Psi_L^{\prime\,i}$ is parameterized by embedding the quark and lepton components $\Psi_{q\,L}^{\prime\,1,2}$ and  $\Psi_{\ell\,L}^{\prime\,1,2}$ that have a flavour-diagonal mixing with $q_L^{\prime\,1,2}$ and $\ell_L^{\prime\,1,2}$, respectively, as
\begin{equation}
 \Psi_{L}^{\prime\,i}
  = \begin{pmatrix} \Psi_{q\,L}^{\prime\,i} \\ W_{ij}\, \Psi_{\ell\,L}^{\prime\,j}  \end{pmatrix}\,,
\end{equation}
where $W$ is a unitary matrix parameterizing the misalignment.
This matrix is usually chosen to be $C\!P$-conserving and to mix only the second and third generation, i.e.\ we use
\begin{equation}
 W=
 \begin{pmatrix}
  1 & 0 & 0 \\
  0 & \cos \theta_{LQ} & \sin \theta_{LQ} \\
  0 & -\sin \theta_{LQ} & \cos \theta_{LQ} \\
 \end{pmatrix}.
\end{equation}
In the absence of additional new heavy fermions that mix with right-handed SM fermions, a possible quark-lepton misalignment in $\Psi_R^{\prime+\,3}$ and $\Psi_R^{\prime-\,3}$ corresponds to only a phase difference, which we parameterize as
\begin{equation}
 \Psi_R^{\prime+\,3}
  = \begin{pmatrix} \Psi_{u\,R}^{\prime\,3} \\ e^{i \phi_{\nu}}\,\Psi_{\nu\,R}^{\prime\,3}  \end{pmatrix}\,,
  \qquad
 \Psi_R^{\prime-\,3}
  = \begin{pmatrix} \Psi_{d\,R}^{\prime\,3} \\ e^{i \phi_{e}}\,\Psi_{e\,R}^{\prime\,3}  \end{pmatrix}\,.
\end{equation}
Consequently, the SM fields in the basis where the down-quark mass matrix is diagonal can be expressed as
\begin{equation}
\begin{aligned}
 q_L^{1,2} &= q_L^{\prime\,1,2}\,\cos\theta_{q_{1,2}} + \Psi_{q\,L}^{\prime\,1,2}\,\sin\theta_{q_{1,2}}\,,
 \qquad&
 q_L^{3} &= \Psi_{q\,L}^{\prime\,3}\,,
 \\
 \ell_L^{1,2} &= \ell_L^{\prime\,1,2}\,\cos\theta_{\ell_{1,2}} + \Psi_{\ell\,L}^{\prime\,1,2}\, \sin\theta_{\ell_{1,2}}\,,
 \qquad&
 \ell_L^{3} &= \Psi_{\ell\,L}^{\prime\,3}\,,
 \\
 u_R^{1,2} &= u_R^{\prime\,1,2}\,,
 \qquad&
 u_R^{3} &= \Psi_{u\,R}^{\prime\,3}\,,
 \\
 d_R^{1,2} &= d_R^{\prime\,1,2}\,,
 \qquad&
 d_R^{3} &= \Psi_{d\,R}^{\prime\,3}\,,
 \\
 e_R^{1,2} &= e_R^{\prime\,1,2}\,,
 \qquad&
 e_R^{3} &= \Psi_{e\,R}^{\prime\,3}\,,
 \\
 &&
 \nu_R^{3} &= \Psi_{\nu\,R}^{\prime\,3}\,.
\end{aligned}
\end{equation}
In this basis, the couplings of the new vector bosons $U_1$, $Z'$, and $G'$ to the SM fermions and to the DM-sector fields $\psi$ and $\chi$ are given by
\begin{equation}\label{eq:couplings_fermions_vectors}
\begin{aligned}
 \mathcal{L}_{Z'}
 &\supset
 \frac{g_{4}}{2\sqrt{6}\cos\theta_{41}}\,Z_\mu^\prime
 \begin{aligned}[t]
 \Big(&
   \xi_q^i\,\bar{q}_L^i\gamma^\mu q_L^i
   +
   \xi_u^i\,\bar{u}_R^i\gamma^\mu u_R^i
   +
   \xi_d^i\,\bar{d}_R^i\gamma^\mu d_R^i
   +
   \xi_\psi\,\bar{\psi} \gamma^\mu \psi
   \\
  & -3\left(
   \xi_\ell^i\,\bar{\ell}_L^i\gamma^\mu \ell_L^i
   +
   \xi_e^i\,\bar{e}_R^i\gamma^\mu e_R^i
   +
   \xi_\nu\,\bar{\nu}_R^3\gamma^\mu \nu_R^3
   +
   \xi_\chi\,\bar{\chi} \gamma^\mu \chi
   \right)\!\Big),
 \end{aligned}
   \\
 \mathcal{L}_{G'}
 &\supset
 \frac{g_{4}}{\cos\theta_{43}}\,G_\mu^{\prime a}
 \Big(
   \kappa_q^i\,\bar{q}^i\gamma^\mu T^a q^i
   +
   \kappa_u^i\,\bar{u}_R^i\gamma^\mu T^a u_R^i
   +
   \kappa_d^i\,\bar{d}_R^i\gamma^\mu T^a d_R^i
   +
   \kappa_\psi\,\bar{\psi} \gamma^\mu T^a \psi
   \Big),
   \\
 \mathcal{L}_{U_1}
 &\supset
 \frac{g_4}{\sqrt{2}}\,U_\mu^+
 \Big(
   \beta_{q\ell}^{ij}\,\bar{q}^i\gamma^\mu \ell^j
   +
   \beta_{de}\,\bar{d}_R^3\gamma^\mu e_R^3
   +
   \beta_{u\nu}\,\bar{u}_R^3\gamma^\mu \nu_R^3
   +
   \bar{\psi} \gamma^\mu \chi
   \Big)
   +h.c.\,,
\end{aligned}
\end{equation}
where the constants $\kappa$ and $\xi$ that appear in the $G'$ and $Z'$ couplings are collected in table~\ref{tab:couplings} and the constants $\beta$ that appear in the $U_1$ couplings are given by
\begin{table}[t]
\begin{center}
\def\arraystretch{1.4}
\begin{tabular}{|c|c|c|c|c|}
\hline
Field & \multicolumn{2}{c|}{$\kappa$} & \multicolumn{2}{c|}{$\xi$} \\
\hline
\hline
&$i=1,2$ & $i=3$ &$i=1,2$ & $i=3$ \\
\cline{2-5}
$q_L^i$
    & $\sin^2\theta_{q_{1,2}} - \sin^2\theta_{43}$
    & $\cos^2\theta_{43}$
    & $\sin^2\theta_{q_{1,2}}-\sin^2\theta_{41}$
    & $\cos^2\theta_{41}$ \\
$u_R^i$
    & $-\sin^2\theta_{43}$
    & $\cos^2\theta_{43}$
    & $-4\,\sin^2\theta_{41}$
    & $1-4\,\sin^2\theta_{41}$ \\
$d_R^i$
    & $-\sin^2\theta_{43}$
    & $\cos^2\theta_{43}$
    & $2\,\sin^2\theta_{41}$
    & $1+2\,\sin^2\theta_{41}$ \\
$\ell_L^i$
    &&
    & $\sin^2\theta_{\ell_{1,2}}-\sin^2\theta_{41}$
    & $\cos^2\theta_{41}$ \\
$e_R^i$
    &&
    & $-2\,\sin^2\theta_{41}$
    & $1-2\,\sin^2\theta_{41}$ \\
$\nu_R^i$
    &&
    &
    & $1$ \\
    \hline
    \hline
$\psi$
    & \multicolumn{2}{c|}{$\cos^2\theta_{43}$}
    & \multicolumn{2}{c|}{$1-4\,\sin^2\theta_{41}$}\\
$\chi$
    & \multicolumn{2}{c|}{}
    & \multicolumn{2}{c|}{$1$}\\
    \hline
\end{tabular}
\end{center}
\caption{Constants $\kappa$ and $\xi$ entering the couplings of fermions to $G'$ and $Z'$ (cf.\ eq.~\eqref{eq:couplings_fermions_vectors}).}
\label{tab:couplings}
\end{table}
\begin{equation}\label{eq:beta_couplings}
 \beta_{q\ell}
 =
 \begin{pmatrix}
  \sin\theta_{q_1}\,\sin\theta_{\ell_1} & 0 & 0 \\
  0 & \sin\theta_{q_2}\,\sin\theta_{\ell_2}\,\cos \theta_{LQ} & \sin\theta_{q_2}\,\sin \theta_{LQ} \\
  0 & -\sin\theta_{\ell_2}\,\sin \theta_{LQ} & \cos \theta_{LQ} \\
 \end{pmatrix}\,,
   \quad
 \beta_{de}
 =
 e^{i \phi_{e}}\,,
   \quad
 \beta_{u\nu}
 =
 e^{i \phi_{\nu}}\,.
\end{equation}
Note that in the limit of large $g_4$, where $\cos\theta_{41}\approx\cos\theta_{43}\approx1$ and $\sin\theta_{41}\approx\sin\theta_{43}\approx0$, the constants $\kappa$ and $\xi$ are approximately
\begin{equation}\label{eq:parameters_approx}
\begin{aligned}
 \kappa_q^{1,2} &\approx \xi_q^{1,2} \approx \sin^2\theta_{q_{1,2}}\,,
 \qquad
 \xi_\ell^{1,2}\approx \sin^2\theta_{\ell_{1,2}}\,,
 \qquad
 \kappa_u^{1,2}=\kappa_d^{1,2}\approx\xi_u^{1,2}\approx\xi_d^{1,2}\approx\xi_e^{1,2}\approx0\,,
 \\
 \kappa_q^{3}&=\kappa_u^{3}=\kappa_d^{3}\approx\xi_q^{3}\approx\xi_u^{3}\approx\xi_d^{3}\approx\xi_\ell^{3}\approx\xi_e^{3}\approx\xi_\nu^{3}\approx\kappa_\psi\approx\xi_\chi\approx\xi_\psi\approx1\,,
\end{aligned}
\end{equation}
i.e.\ the couplings of left-handed light fermions are proportional to their mixings with the new heavy fermions, the couplings of right-handed light fermions vanish, and all couplings of third-generation SM fermions and DM sector fields satisfy $\kappa\approx\xi\approx1$.
The constants $\beta$, on the other hand, are independent of the value of $g_4$ and only depend on fermion mixing angles and phases.

While the parameterization described above allows explaining the $B$-meson anomalies and avoids strong constraints from large flavour violating effects, the number of parameters can be further reduced by the following phenomenologically motivated assumptions:
\begin{itemize}
 \item To maximize the agreement with the $B$-decay measurements that deviate from the SM, one can take $\beta_{de}=-1$~\cite{Cornella:2019hct}, which fixes the phase $\phi_e=\pi$.
 \item Since the phase $\phi_\nu$ is currently not constrained by any measurement one can use $\phi_\nu=0$ for simplicity.
 \item An approximate $U(2)$ symmetry in the quark sector, i.e.\ $\theta_{q_1}\approx\theta_{q_2}$, can be employed to suppress tree-level FCNCs in the up-quark sector that are mediated by the $Z'$ and $G'$~\cite{DiLuzio:2018zxy}.
 Without such a $U(2)$ protection, excessive contributions to $\Delta C=2$ observables would be possible.
 \item The first-generation lepton doublet can be taken to be purely a singlet of $SU(4)$, i.e.\ $\theta_{\ell_1}=0$, to be safe from LFV due to $U_1$ couplings involving the electron.
\end{itemize}
Making all of the above assumptions and defining $\theta_{q_{12}}=\theta_{q_1}=\theta_{q_2}$, the only remaining free parameters in the fermion sector are
\begin{equation}\label{eq:flavour_DM_pars}
 \theta_{q_{12}}\,,
 \quad
 \theta_{\ell_2}\,,
 \quad
 \theta_{LQ}\,,
 \quad
 M_{\chi}\,,
 \quad
 N\,,
\end{equation}
where $M_\chi$ denotes, here and henceforth, the mass of the DM-candidate\footnote{The mass $M_\chi$ is related to the tree-level mass of the multiplet $\Psi_{\rm DM}$ as discussed in Appendix~\ref{app:mass_splitting}.} $\chi$ and $N$ is the dimension of its $SU(2)_L$ representation.

\subsection{Parameter ranges}\label{sec:parameter_ranges}

As summarised in eqs.~(\ref{eq:vLQ_g4}) and (\ref{eq:flavour_DM_pars}), within reasonable assumptions the `effective' model parameters are the following:
\be
\label{eq:parameter_space}
g_4~,~~~~v_{LQ}~,~~~~\theta_{q_{12}}~,~~~~\theta_{\ell_{2}}~,~~~~\theta_{LQ}~,~~~~M_{\chi}~,~~~~N~.
\ee
In this section we would like to collect the non-negligible information available on these parameters from a fit to flavour data as well as from constraints due to direct searches. In Sec.~\ref{sec:results} we will then address the question to what extent these constraints are compatible with those coming from cosmological and direct-detection information about Dark Matter.

The above parameters can be grouped into three classes according to their impact on the DM phenomenology:
\begin{itemize}
 \item $g_4$, $v_{LQ}$, $M_{\chi}$, $N$: The DM phenomenology depends crucially on these parameters.
 \item $\theta_{q_{12}}$: This parameter is important only for DM direct detection.
 \item $\theta_{\ell_{2}}$, $\theta_{LQ}$: The DM phenomenology is essentially independent of these parameters.
\end{itemize}
Within our model, a combination of the parameters $v_{LQ}$, $\theta_{q_{12}}$, $\theta_{\ell_{2}}$, and $\theta_{LQ}$ is constrained by $B$-physics data alone, in particular by the $R(D^{(*)})$ discrepancies.
A global fit of $v_{LQ}$ and the constants $\beta$ (cf.\ eq.~\eqref{eq:beta_couplings}) was performed in Ref. \cite{Cornella:2019hct}.
Expressed in our notation, the fit prefers values for $v_{LQ}$ in the range $v_{LQ}/\cos\theta_{LQ} \in [3.1, 4.6 ]~{\rm TeV}$, with $\cos\theta_{LQ}\approx0.8-0.9$.
While the preferred value of $v_{LQ}$ is correlated with the preferred values of $\theta_{\ell_{2}}$ and $\theta_{LQ}$, our DM phenomenology is essentially independent of the latter two parameters.
Consequently, for any reasonable value of $v_{LQ}$, we can set the parameters $\theta_{\ell_{2}}$ and $\theta_{LQ}$ to comply with the fit in \cite{Cornella:2019hct}, while fulfilling all DM constraints.
The fit to $B$-physics data leaves some freedom for $\theta_{q_{12}}$, which, however, is constrained by direct searches (see below).
In short, for definiteness we take
\be
\label{eq:vLQrange}
v_{LQ} \in [3, 5]~{\rm TeV}
\ee
as our fiducial range for $v_{LQ}$.

\begin{figure}[h!]
  \centering
  \includegraphics[width=.294\textwidth]{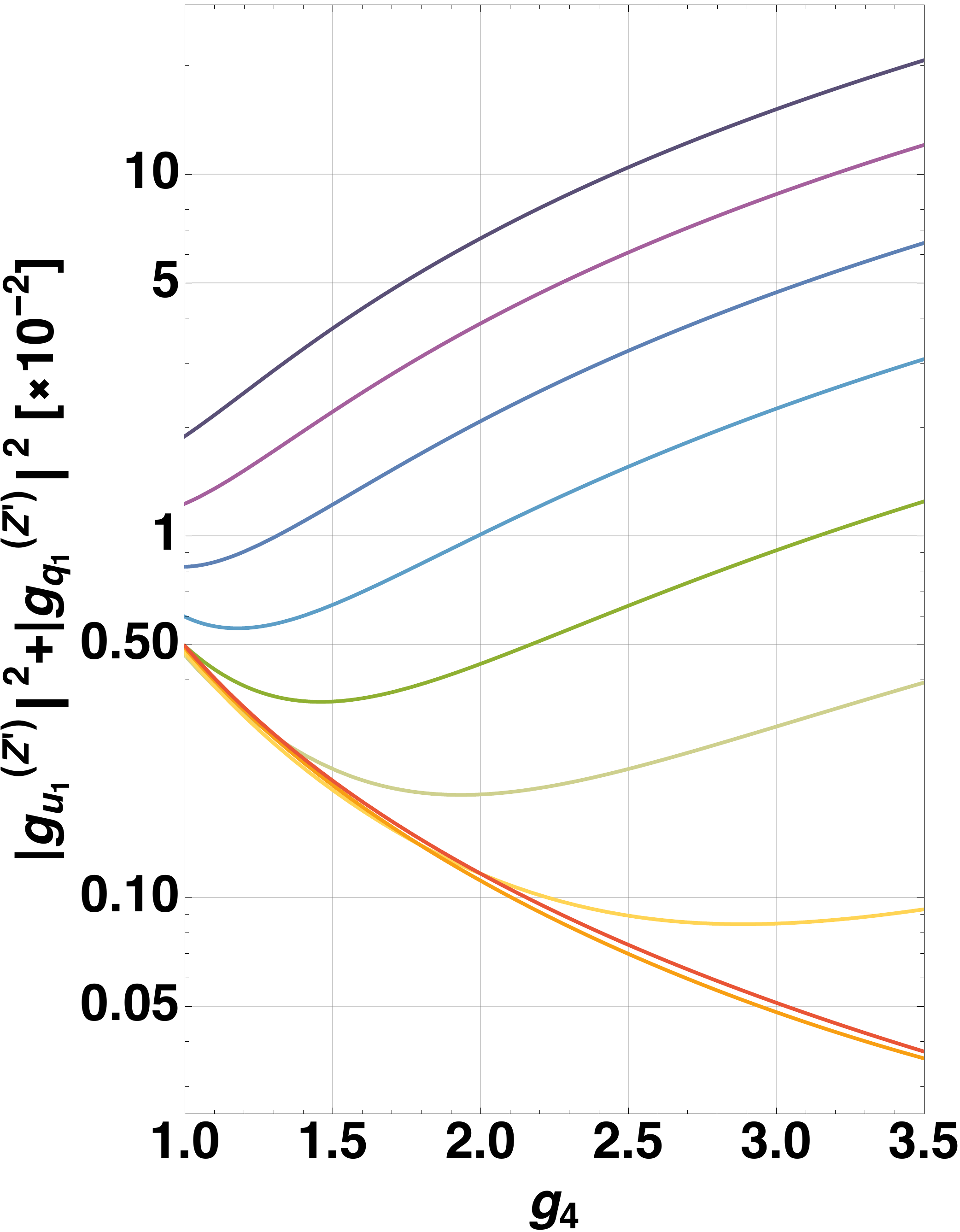}
  \includegraphics[width=.294\textwidth]{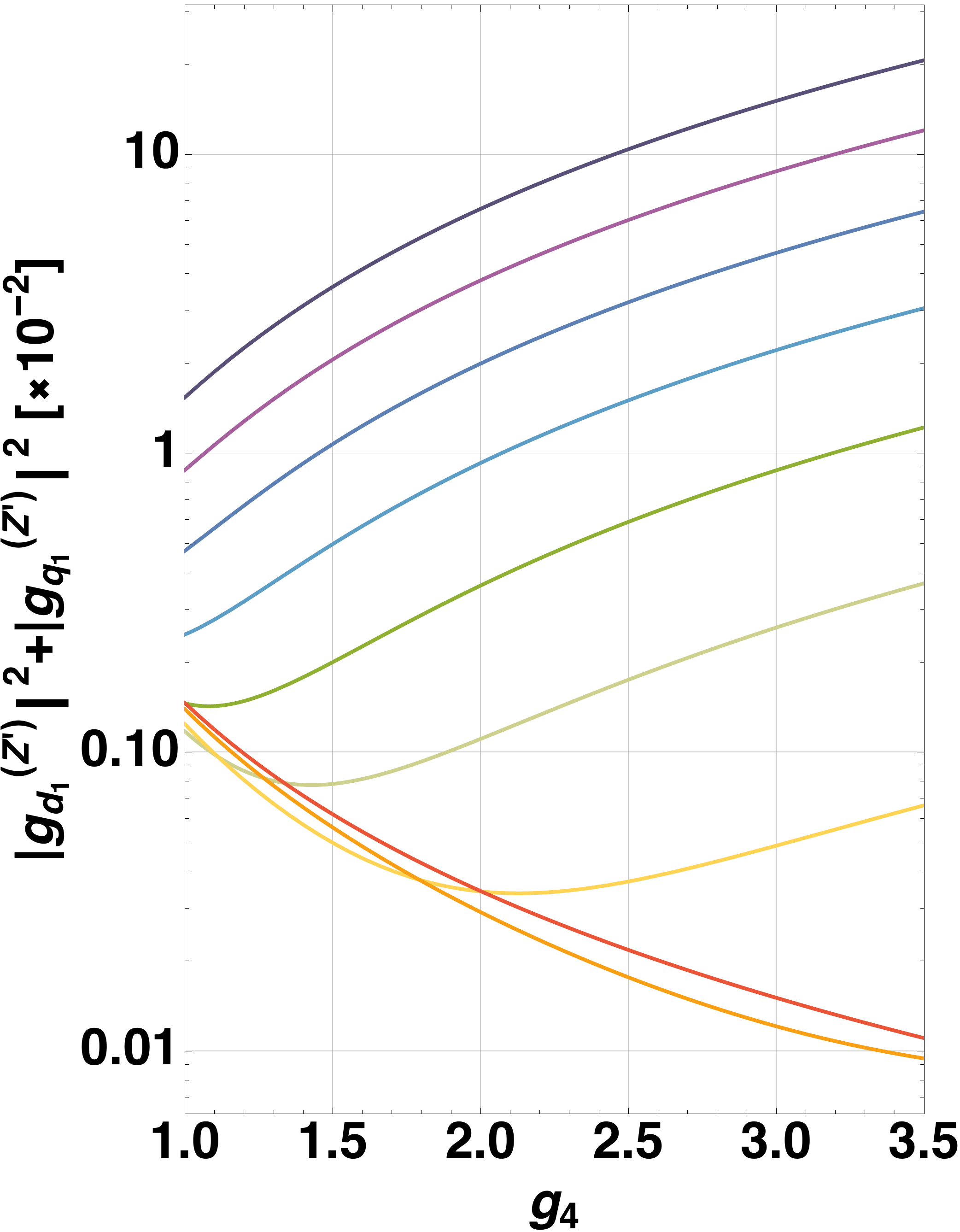}
  \includegraphics[width=.396\textwidth]{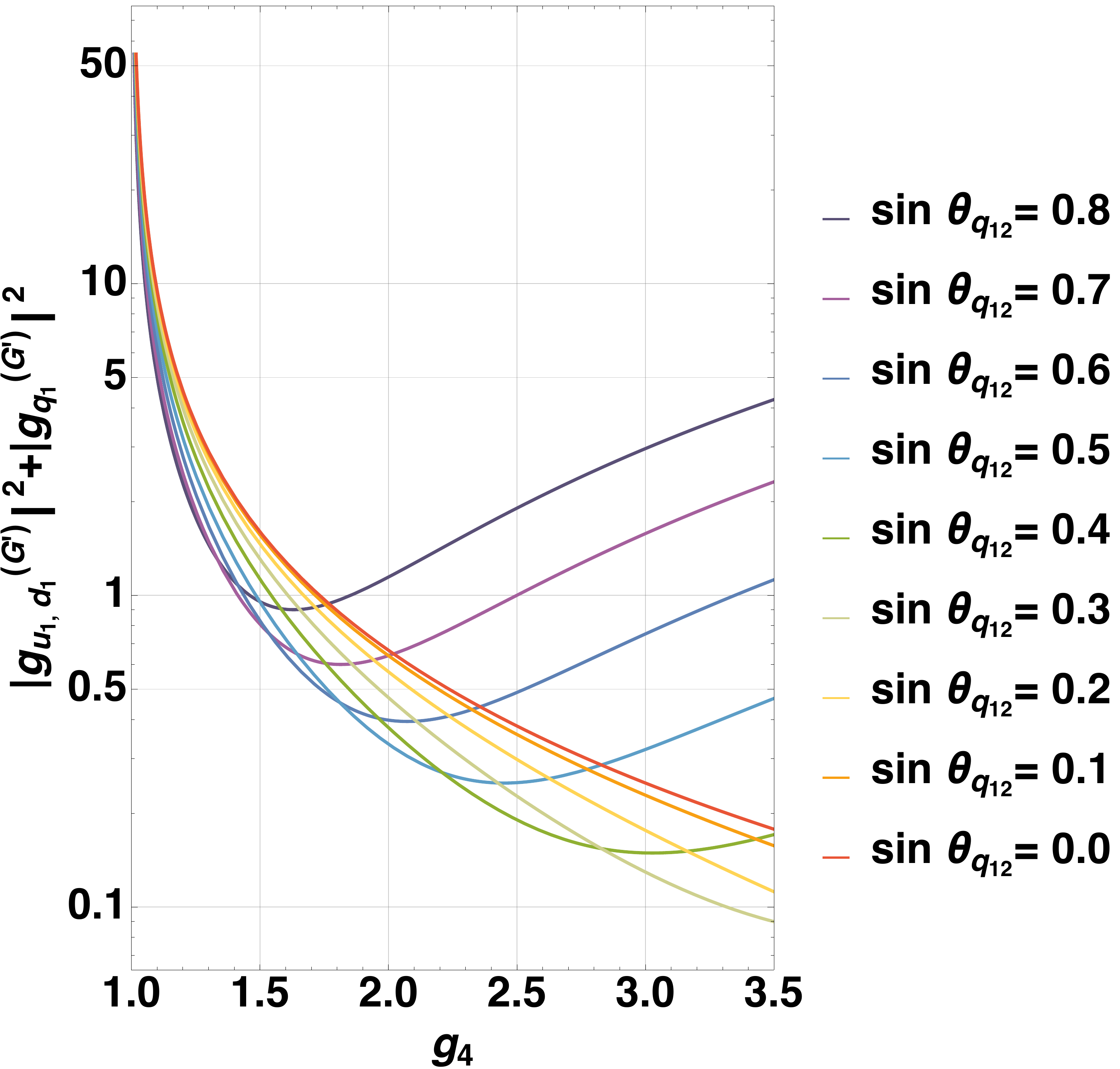}
  \caption{First-generation coupling combinations relevant for $Z^\prime$ searches through the vertex $Z^\prime \bar u u$ (left), $Z^\prime \bar d d$ (center), or for $G^\prime$ searches through the vertices $G^\prime \bar u u$ or $G^\prime \bar d d$ (right).}
  \label{fig:ZpGpcouplings}
\end{figure}

The parameters $g_4$ and $\theta_{q_{12}}$ enter
the definition of the fermionic-currents' couplings to the $U_1$, the $Z^\prime$ and the $G^\prime$, which are constrained by direct searches.
In fig.~\ref{fig:ZpGpcouplings}, we show the $g_4$ dependence of light-quark couplings to the $Z^\prime$ as well as to the $G^\prime$, for different values of $\sin \theta_{q_{12}}$.
This dependence displays transparently the $g_4$ and $\sin \theta_{q_{12}}$ ranges preferred by direct searches. The figure shows at a glance that an efficient suppression of these coupling combinations is achieved for small $\sin \theta_{q_{12}}$ {\em and} large $g_4$. Representative ranges are
\be
\label{eq:sinq12_g4_ranges}
\sin \theta_{q_{12}} \lesssim 0.2~~~\mbox{and}~~~g_4 \gtrsim 3~.
\ee
These two requirements suppress respectively the two terms entering the coupling constants $\kappa_{q}^{1,2}$, $\xi_{q}^{1,2}$, and $\xi_{\ell}^{1,2}$ (cf.\ table~\ref{tab:couplings}).\footnote{The figure also shows that, for $G^\prime$ couplings, a cancellation between these two terms can be engineered for $g_4 \approx 1.5$ - $2$ and $\sin \theta_{q_{12}} \ge 0.5$. However, the resulting light-quark - $G^\prime$ coupling is not nearly as suppressed as in the case of $\sin \theta_{q_{12}} \lesssim 0.2$ and $g_4 \gtrsim 3$.}

Bounds from direct searches have been extensively studied in the literature, see \cite{Faroughy:2016osc,Diaz:2017lit,DiLuzio:2017vat,Schmaltz:2018nls,Angelescu:2018tyl,Greljo:2018tzh,DiLuzio:2018zxy,Baker:2019sli,Cornella:2019hct,Fuentes-Martin:2019ign}.
It is straightforward to verify that the ranges in eqs.~(\ref{eq:vLQrange})-(\ref{eq:sinq12_g4_ranges}) yield $U_1$, $Z^\prime$ and $G^\prime$ masses and couplings in accord with these bounds.
In particular, for $g_4 = 3$ and $v_{LQ} \ge 3$ TeV we have $M_{U}\approx M_{Z'} \ge 4.5$ TeV and $M_{G^\prime} \ge 4.8$ TeV, such that the bounds found in~\cite{Baker:2019sli} can be comfortably satisfied.

We conclude that agreement with $B$-decay discrepancies and with the constraints coming from direct searches select the parameter regions in eqs.~(\ref{eq:vLQrange})-(\ref{eq:sinq12_g4_ranges}). In Sec.~\ref{sec:results} we will confront this parameter space with the constraints imposed by Dark Matter.

\section{Dark matter relic abundance} \label{sec:Omega_DM}

In this section we shall address the question whether our setup, as introduced around eq.~(\ref{eq:chi-psi-SU2L}) and discussed in Sec.~\ref{sec:model}, can accommodate the relic abundance of DM observed today, $\Omega_0 h^2$.

In addition to the DM candidate $\chi_0$, the DM sector of our model includes also several co-annihilation partners: all the other components, charged under weak isospin, of the $\chi$ and $\psi$ $SU(2)_L$ multiplets.
As shown in classic work, even in the presence of co-annihilators an estimate of $\Omega_0 h^2$ accurate to about $10\%$ may be obtained analytically \cite{Griest:1990kh} (see also \cite{Kolb:1990vq}).
This accuracy is satisfactory in our case, in view of several uncertainties inherent in the problem and likewise discussed in the above works.

The first main step towards the estimate of $\Omega_0 h^2$ is the determination of $\Omega h^2$ at the `freeze-out' temperature $T_f$.
It is convenient to introduce the variable $x$ denoting the inverse temperature in units of the DM mass, i.e.\ $x\equiv M_{\chi_0} / T$, and to define $x_f\equiv x|_{T=T_f}$.
In the case -- like ours -- where co-annihilators are present, $x_f$ is determined iteratively from the relation \cite{Griest:1990kh}
\be
\label{eq:Tf}
x_f ~=~ \ln
\frac{0.038\, g_{\rm eff}\, M_{\rm Pl}\, M_{\chi_0}\, \< \sigma_{\rm eff}\,v\>}{g_*^{1/2}\, x_f^{1/2}}\,,
\ee
where $M_{\rm Pl}=1.22\times10^{19}$ GeV, $g_\ast$ denotes the total number of effectively relativistic d.o.f. at freeze-out, $g_{\rm eff}$ denotes the number of effective d.o.f. within the DM sector, and the thermally averaged annihilation cross section $\< \sigma_{\rm eff}\,v\>$ is the main dynamical quantity.
After freeze-out, the relic abundance is subject to post-freeze-out annihilation processes.
The efficiency of this post-freeze-out annihilation is given by~\cite{Griest:1990kh,Jungman:1995df}
\be
\label{eq:J}
J \equiv \int_{x_f}^\infty \frac{\< \sigma_{\rm eff} \, v\>}{x^2} dx\,.
\ee
The present-day DM abundance can then be estimated as\footnote{One can derive this relation by using $H(T) = \sqrt{8 \pi^3 g_* / 90} \,\, T^2 / M_{\rm Pl}$, $s = 2 \pi^2 g_* T^3 / 45$ and $\rho_c = 3 H_0^2 / (8 \pi G_N)$, with $H(T_0) \equiv H_0 = 100 h$ km$/$ (s Mpc), as customary.}
\begin{equation}\label{eq:Omega_DM_J}
\Omega_0 h^2
=
\sqrt{\frac{45}{\pi}}
\frac{s_0}{\rho_c}
\frac{1}{g_*^{1/2}\,M_{\rm Pl}\,J}
\simeq
\frac{1.07 \times 10^9\,{\rm GeV}^{-1}}{g_*^{1/2}\,M_{\rm Pl}\,J}\,.
\end{equation}
The crucial ingredient in the determination of $\Omega_0 h^2$ is the calculation of the thermally averaged annihilation cross section $\< \sigma_{\rm eff}\,v\>$, which enters in $\Omega_0 h^2$ through the post-freeze-out annihilation efficiency $J$.
To this end, it is also necessary to determine $g_{\rm eff}$, which enters in both $\< \sigma_{\rm eff}\,v\>$ and $x_f$.
In turn, $g_{\rm eff}$ and $\sigma_{\rm eff}$ depend in an important way on the mass differences between the DM candidate and its co-annihilation partners~\cite{Griest:1990kh}. In the following, we discuss in great detail the mass differences, the effective degrees of freedom $g_{\rm eff}$, and therewith proceed to the estimate of the thermally averaged annihilation cross section $\< \sigma_{\rm eff}\,v\>$.

\subsection{Mass differences} \label{sec:dmi}

For mass differences between the DM candidate and its co-annihilation partners comparable to the freeze-out temperature (which quantifies the average amount of kinetic energy available in the collisions), the co-annihilation partners become nearly as kinematically accessible as the DM candidate.
The mass differences are therefore an important ingredient in the determination of the annihilation cross section.
In our model, the DM candidate and its co-annihilation partners are components of VL fermion multiplets that transform non-trivially under the $SU(2)_L$ and $SU(4)$ gauge groups.
The relevant mass differences in this case correspond to the mass splitting inside these multiplets, which are generated after the spontaneous breaking of the $SU(2)_L$ and $SU(4)$ symmetries.
While it is possible to generate a mass splitting at tree level, e.g.\ by coupling the VL multiplets to the scalar operators responsible for the spontaneous symmetry breaking, a mass splitting is generated even in the absence of such tree-level terms.
At the one-loop order, the gauge bosons associated with the spontaneously broken symmetries, of which some become massive due to the breaking, induce corrections to the fermion masses.
Since the components of the VL multiplets correspond to different irreducible representations of the unbroken gauge group, each of them couples differently to the gauge bosons and thus receives a different contribution to its mass.

It is possible to obtain a generic result for the one-loop mass splitting among components of a VL multiplet that is applicable to a large set of spontaneously broken gauge groups (see appendix~\ref{app:mass_splitting}).
Applying our generic result, eq.~\eqref{eq:generic_mass_splitting}, to the EW gauge group, we can determine the relative mass difference
\begin{equation}
 \Delta_{\xi\eta} = \frac{M_\xi-M_\eta}{\hat{M}}
\end{equation}
 between components $\xi$ and $\eta$ of a VL multiplet of hypercharge $Y$ and mass $\hat{M}$.
 We find
 \begin{equation} \label{eq:mass_splitting_EW}
 \begin{aligned}
 \Delta_{\xi\eta}^{\rm EW}
 %&
 =
 \frac{g^2}{16\,\pi^2}
 \bigg\{
 %&\!\!\!\!
 \left(
 (Q_\xi-Y)^2
 -
 (Q_\eta-Y)^2
 \right)
 \left[
 f\left(\tfrac{M_{W}}{\hat{M}}\right)
 -
 f\left(\tfrac{M_{Z}}{\hat{M}}\right)
 \right]
 %&
 %\\
 %&&
 +
 s_W^2
 \,
 (
 Q_\xi^2
 -
 Q_\eta^2
 )
 \,f\left(\tfrac{M_{Z}}{\hat{M}}\right)
 %&
 \bigg\}\,,
 \end{aligned}
 \end{equation}
where $Q_\xi$ and $Q_\eta$ are the electric charges of $\xi$ and $\eta$, and $f(r)$ is a finite loop function given in eq.~\eqref{eq:f_loop_function}.
This reproduces the well-known result for the mass splitting in EW VL multiplets (cf.\ e.g.~\cite{Cirelli:2005uq}).
For reference, the relative mass splitting within the $\psi$ and $\chi$ $SU(2)_L$ multiplets of our DM sector is between $\mathcal{O}(10^{-3})$ and $\mathcal{O}(10^{-4})$ for $\hat{M}=\mathcal{O}(1\ {\rm TeV})$.

\begin{figure}[t]
  \centering
  \includegraphics[width=.70\textwidth]{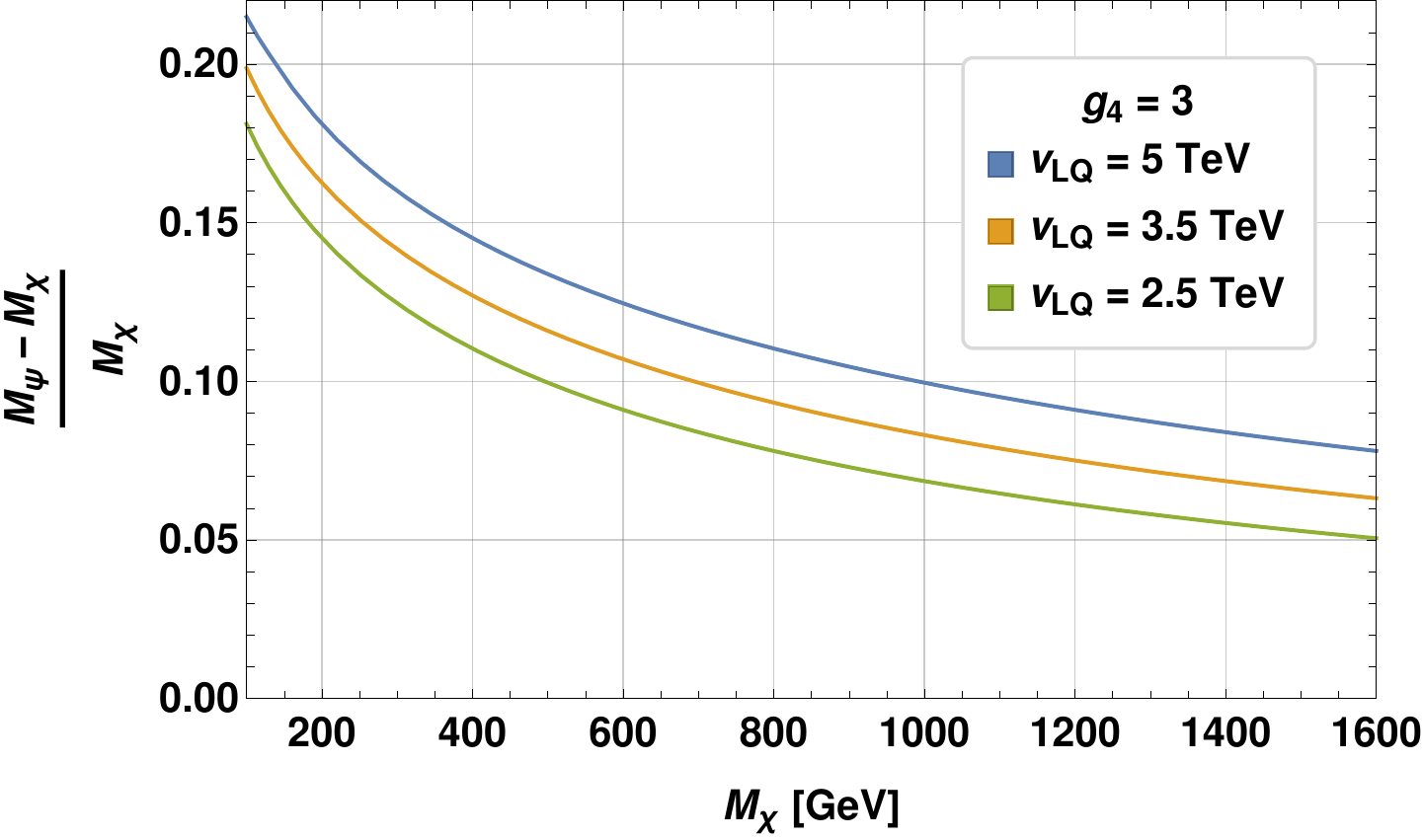}
  \caption{Mass splitting between $\psi$ and $\chi$ induced by the 4321 gauge bosons.}
  \label{fig:mass-splitting}
\end{figure}

Having at hand the generic result, eq.~\eqref{eq:generic_mass_splitting}, it is straightforward to determine the relative mass splitting between our DM candidate $\chi$ and its colored co-annihilation partner $\psi$.
This mass splitting is induced by the vector bosons associated with the 43(2)1 symmetry breaking and is given by
 \begin{equation}
 \label{eq:Deltapsichi4321}
 \Delta_{\psi\chi}^{4321}
 =
 \frac{g_4^2}{16\,\pi^2}
 \bigg\{
 f\left(\tfrac{M_{U}}{\hat{M}}\right)
 +
 \frac{1}{3}
 (2\,\sin^2 \theta_{41}+1)\,
 f\left(\tfrac{M_{Z'}}{\hat{M}}\right)
 +
 \frac{4}{3}(\sin^2 \theta_{43}-1)\,
 f\left(\tfrac{M_{G'}}{\hat{M}}\right)
 \bigg\}\,.
 \end{equation}
The value of $\Delta_{\psi\chi}^{4321}$ is around $8-15\%$ for the parameter region of interest (see fig.~\ref{fig:mass-splitting}) and its significance is further discussed in sections~\ref{sec:g_eff} and~\ref{sec:sig_eff_v}.

Additional mass splittings are induced by the non-zero temperature at which our processes of interest take place.
We estimate these mass splittings to be of order $\Delta^{\rm EW}({T}) \sim (g\, T / \hat{M})^2$~\cite{Cirelli:2005uq} within $SU(2)_L$ multiplets and $\Delta^{4321}({T}) \sim (g_4\, T / \hat{M})^2$ within $SU(4)$ multiplets.
Recalling that $T_f /\hat{M} \simeq 3\%$, and that such ratio enters quadratically in $\Delta({T})$, we estimate $\Delta^{\rm EW}({T})$ to lie between $\mathcal{O}(10^{-3})$ and $\mathcal{O}(10^{-4})$ and $\Delta^{4321}({T})\lesssim 1\%$.
Since $\Delta^{4321}({T})$ is small compared to the splitting induced by eq.~\eqref{eq:Deltapsichi4321}, we neglect it in the following.
The size of $\Delta^{\rm EW}({T})$ is similar to the splitting induced by eq.~\eqref{eq:mass_splitting_EW} and we estimate the combination of both contributions to lie between $\mathcal{O}(10^{-3})$ and $\mathcal{O}(10^{-4})$.

\subsection{Effective degrees of freedom}\label{sec:g_eff}
In our case, the number of effective d.o.f. in the DM sector $g_{\rm eff}$ is given by~\cite{Griest:1990kh}
\begin{equation}\label{eq:g_eff}
g_{\rm eff}
=
\sum_i
\left(
g_\chi\, (1 + \Delta_{\chi_i})^{3/2} \exp(-x\, \Delta_{\chi_i})
+
g_\psi\, (1 + \Delta_{\psi_i})^{3/2} \exp(-x\, \Delta_{\psi_i})
\right)
\,,
\end{equation}
where the index $i$ runs over the $N$ components of the $SU(2)_L$ multiplets $\chi$ and $\psi$.
Besides $g_\chi=4$ and $g_\psi=12$ denote the internal (spin, color, ...) d.o.f. of the components of these multiplets.\footnote{Since $\chi$, $\psi$ belong to complex VL representations of the gauge group, they are Dirac fermions.}
The relative mass splittings $\Delta_{\chi_i}$ and $\Delta_{\psi_i}$ are defined as
\begin{equation}
 \Delta_{\chi_i} = (M_{\chi_i} - M_{\chi_0}) / M_{\chi_0}\,,
 \qquad
 \Delta_{\psi_i} = (M_{\psi_i} - M_{\chi_0}) / M_{\chi_0}\,.
\end{equation}
The relative mass differences within the $\chi$ and $\psi$ multiplets are only at the per mil level (see discussion below eq.~(\ref{eq:mass_splitting_EW})).
We thus neglect them in the following and use a common mass and relative mass splitting for each multiplet,
\begin{equation}\label{eq:approx_splittings}
 M_{\chi_i} \to M_{\chi}\,,
 \qquad
 M_{\psi_i} \to M_{\psi}\,,
 \qquad
 \Delta_{\chi_i} \to 0\,,
 \qquad
 \Delta_{\psi_i} \to \Delta_{\psi}\,,
\end{equation}
such that $M_\psi = (1 + \Delta_\psi)\, M_\chi$.
Employing this approximation, the number of effective d.o.f. simplifies to
\begin{equation}\label{eq:g_eff_simpl}
 g_{\rm eff} \approx N\left(
 g_\chi + g_\psi\, (1 + \Delta_\psi)^{3/2} \exp(-x\, \Delta_\psi)
 \right)\,.
\end{equation}
We see that, at the freeze-out temperature, $g_{\rm eff}$ departs appreciably from $N(g_\chi + g_\psi)$ unless $x_f\,\Delta_\psi\ll 1$, i.e.\ unless the $\psi$-$\chi$ mass splitting is much smaller than the freeze-out temperature.
Combining eq.~(\ref{eq:Tf}) with the value of the $\psi$-$\chi$ mass splitting discussed in Sec.~\ref{sec:dmi}, $\Delta_\psi \approx 0.1$, we find
\begin{equation}\label{eq:approx_psi_splitting}
 (1 + \Delta_\psi)^{3/2} \exp(-x_f\, \Delta_\psi)\approx0.06\,.
\end{equation}
eqs.~(\ref{eq:g_eff_simpl})-(\ref{eq:approx_psi_splitting}) provide an accurate determination of $g_{\rm eff}$ within our setup.

\subsection{Thermally averaged cross section}\label{sec:sig_eff_v}
The main dynamical quantity in eq.~(\ref{eq:Tf}) is $\< \sigma_{\rm eff}\,v\>$, where \cite{Griest:1990kh}
\begin{equation}
\begin{aligned}
 \sigma_{\rm eff}
 = \frac{1}{g_{\rm eff}^2}\, \sum_{i,j}\Big(
 & \sigma_{\chi_i\chi_j}\,g_\chi^2\,(1+\Delta_{\chi_i})^{3/2}\,(1+\Delta_{\chi_j})^{3/2}\, e^{-x(\Delta_{\chi_i}+\Delta_{\chi_j})}
 \\
 +&\sigma_{\psi_i\psi_j}\, g_\psi^2\,(1+\Delta_{\psi_i})^{3/2}\,(1+\Delta_{\psi_j})^{3/2}\, e^{-x(\Delta_{\psi_i}+\Delta_{\psi_j})}
 \phantom{\sum_{i}}
 \\
 +& 2\,\sigma_{\chi_i\psi_j}\, g_\chi\, g_\psi\,(1+\Delta_{\chi_i})^{3/2}\,(1+\Delta_{\psi_j})^{3/2}\, e^{-x(\Delta_{\chi_i}+\Delta_{\psi_j})}
 \Big)
\end{aligned}
\end{equation}
with e.g.
\begin{equation}
\sigma_{\chi_i \psi_j} \equiv \sigma(\chi_i \psi_j \to X X^\prime)\,,
\end{equation}
and the other cross sections defined analogously. Here $X, X^\prime$ denote any particles other than $\chi, \psi$.
Since we assume that the DM sector is lighter than any of the $U_1, Z^\prime, G^\prime$ mediators or new vector-like fermions, for the $X, X^\prime$ we only consider SM particles.
Neglecting again the mass differences within the $\psi$ and $\chi$ multiplets, i.e.\ employing the replacements in eq.~(\ref{eq:approx_splittings}), the effective cross section simplifies to
\begin{equation}\label{eq:general_sigma_eff}
\sigma_{\rm eff}
=
\frac{1}{g_{\rm eff}^2} \, \sum_{i,j}
\left(
\sigma_{\chi_i\chi_j}\,g_\chi^2
+
2\,\sigma_{\chi_i\psi_j}\, g_\chi\, g_\psi\, (1 + \Delta_\psi)^{3/2} \, e^{-x \Delta_\psi}
+
\sigma_{\psi_i\psi_j}\, g_\psi^2\,(1 + \Delta_\psi)^{3} \, e^{-2 x \Delta_\psi}
\right)\,.
\end{equation}
The cross sections $\sigma_{\chi_i\chi_j}$, $\sigma_{\chi_i\psi_j}$, and $\sigma_{\psi_i\psi_j}$ are due to the exchange of either SM bosons or the new heavy gauge bosons $U_1$, $Z'$, and $G'$.
Because of the dependence of the cross sections on the fourth power of the couplings, contributions due to the electroweak sector are negligible compared to those involving the relatively strongly coupled new heavy gauge bosons or gluons.
We find that all the cross sections mediated by $U_1$, $Z'$, $G'$, and gluons are of comparable size.
However, in the effective cross section, $\sigma_{\chi_i\psi_j}$ and $\sigma_{\psi_i\psi_j}$ are multiplied by one and two powers, respectively, of a factor that is suppressed by the $\psi$-$\chi$ mass splitting, cf.\ eq.~(\ref{eq:approx_psi_splitting}).
Consequently, the $Z'$-mediated cross section $\sigma_{\chi_i\chi_j}$ is larger than any other contribution to $\sigma_{\rm eff}$ by one to two orders of magnitude.
In view of the overall 10\% uncertainty in our analytical estimate, we can therefore approximate
\begin{equation}\label{eq:sigma_eff}
 \sigma_{\rm eff}
 \approx \frac{1}{N}\, \sigma(\chi_0\chi_0\to Z'\to XX)\,,
\end{equation}
where we have used eqs.~(\ref{eq:g_eff_simpl})-(\ref{eq:approx_psi_splitting}) and $\sigma_{\chi_i\chi_j} \approx \delta_{ij}\,\sigma(\chi_0\chi_0\to Z'\to XX)$.

The thermal average $\< \sigma_{\rm eff} v\>$ can be determined as an expansion in powers of $1/x$, and this expansion can be related order by order to the expansion of $\sigma_{\rm eff}$ around $s = 4 M_\chi^2$ \cite{Srednicki:1988ce}. To this end, we follow the notation of~\cite{Cannoni:2013bza}. Neglecting the masses of the annihilation products, we define
\begin{equation}
\label{eq:sig0_definition}
 \sigma_0(y) ~=~
2 \sqrt{y^2-y} \, \sigma_{\rm eff}(y)
\end{equation}
where we substituted $s$ by the dimensionless variable $y=s/(4\,M_{\chi}^2)$.
The thermal average $\< \sigma_{\rm eff} \, v\>$ can then be expressed as~\cite{Srednicki:1988ce,Cannoni:2013bza}
\begin{equation}\label{eq:sigmav_expansion}
 \< \sigma_{\rm eff} \, v\>
 = \sigma_0(1) \left(1 + \sum_{k=1}^\infty \frac{c_k}{x^k}\right)\,,
\end{equation}
where the first coefficients are
\begin{equation}
 c_1 = -3+\frac{3}{2}\,\lambda_1\,,
 \qquad
 c_2 = 6-3\,\lambda_1+\frac{15}{8}\,\lambda_2\,,
 \qquad
 c_3 = -\frac{5}{16}\left(30-15\,\lambda_1+3\,\lambda_2-7\,\lambda_3\right)\,,
\end{equation}
and we have defined
\begin{equation}
 \lambda_n= \frac{1}{\sigma_0(y)}\frac{d^n \sigma_0(y)}{dy^n}\Bigg|_{y=1}\,.
\end{equation}
The above relations allow thus to verify that, by including higher powers in the small-velocity expansion of $\sigma_0(y)$, higher powers in the small-temperature expansion of $\< \sigma_{\rm eff} v\>$ are smaller and smaller.

\subsection{Present-day DM abundance}

In order to obtain the present-day DM abundance, one convolutes the calculated $\< \sigma_{\rm eff} v\>$ in the post-freeze-out annihilation efficiency $J$ in eq.~(\ref{eq:J}). Using the expansion in eq.~(\ref{eq:sigmav_expansion}) to order $1/x_f^2$, the present-day DM abundance in eq.~(\ref{eq:Omega_DM_J}) yields
\be
\label{eq:Omega_DM_sig}
\Omega_0 h^2 ~\simeq~ \frac{1.07 \times 10^9\,{\rm GeV}^{-1}}{g_*^{1/2} M_{\rm Pl}} \cdot \frac{x_f}{\sigma_0(1)} \cdot \frac{1}{1 + c_1/(2 x_f) + c_2/(3 x_f^2)}~.
\ee
We will perform a full numerical study in Sec.~\ref{sec:results}, following the discussion of the constraints imposed by direct detection in Sec.~\ref{sec:DD}. Here we would like to make a few qualitative considerations around eq.~(\ref{eq:Omega_DM_sig}).
Using $\sigma_{\rm eff}$ as in eq.~(\ref{eq:sigma_eff}), eq.~(\ref{eq:sig0_definition}) yields
\begin{equation}
\label{eq:sig0_chichi}
 \sigma_0(y) =
 \frac{1}{128\,\pi}\left(\frac{g_4}{\cos\theta_{41}}\right)^4 \frac{M_\chi^2\,(2\,y^2+y)}{(4\,y\,M_\chi^2-M_{Z'}^2)^2} \frac{1}{N} f(\{ \xi^i \})~,
\end{equation}
where $N = 1, 3, 5, ...$ denotes the $SU(2)_L$ size of the $\chi$ and $\psi$ multiplets and for brevity we introduced the flavour function
\be
\label{eq:fxi}
f(\{ \xi^i \}) \equiv \sum_{i=1}^3 \left(2|\xi_q^i|^2 + |\xi_u^i|^2 + |\xi_d^i|^2 + 3(2|\xi_\ell^i|^2 + |\xi_e^i|^2 + |\xi_\nu^i|^2)\right)~.
\ee
Plugging eq.~(\ref{eq:sig0_chichi}) into eq.~(\ref{eq:Omega_DM_sig}), and taking the representative value $g_4 = 3$, we find
\be
\label{eq:Omega_DM_approx}
\Omega_0 h^2 ~ \approx ~ 0.06 \frac{N}{f(\{ \xi^i \})} \left(\frac{v_{LQ}}{5~{\rm TeV}}\right)^2 \left( \frac{v_{LQ}}{M_\chi} \right)^2~.
\ee
A few remarks are in order. First, eq.~(\ref{eq:Omega_DM_approx}) assumes $4 M_\chi^2 \ll M_{Z^\prime}^2$. For $v_{LQ} = 3$~TeV (5~TeV), this approximation implies an error $\lesssim 15\%$ ($\lesssim 40\%$) in eq.~(\ref{eq:Omega_DM_approx}), keeping in mind that the preferred range for $M_\chi$ are $\lesssim 600$ GeV ($\lesssim 1.5$ TeV), see Sec.~\ref{sec:results}. Second, with the above mass ranges at hand, we can discuss the relative size of the corrections due to the $c_{1,2}$ terms in the $1/x_f$ expansion (see eq.~(\ref{eq:Omega_DM_sig})). These terms induce corrections in the per mil ballpark for $g_4 = 3$ and the just mentioned mass ranges for $M_\chi$ and $v_{LQ}$. Besides, these corrections do not depend on the choice of any other of our model's parameters, e.g. $f(\{\xi^i\})$ and $N$, as such dependences cancel in the $\lambda_n$ ratios. The function $f( \{ \xi^i \})$ is typically of O(10) in the region satisfying all constraints. For example, taking table~\ref{tab:couplings} with $g_4 = 3$, $\sin \theta_{q_{12}} =  0.2$,\footnote{This choice of parameters follows from the discussion in Sec.~\ref{sec:parameter_ranges}.} one has $f( \{ \xi^i \}) \approx 16$.

\begin{center}
***
\end{center}

The procedure outlined in this section, and leading to eq.~(\ref{eq:Omega_DM_sig}), with $\sigma_{\rm eff}$ including $U_1$, $Z^\prime$, $G^\prime$ and gluon contributions, will be used in Sec.~\ref{sec:results} to identify the regions of parameter space that are viable in the light of all constraints, including $B$ discrepancies, the relic abundance and also direct-detection constraints, to be discussed in the next section.

The approximate formula in eq.~(\ref{eq:Omega_DM_approx}), and the discussion around it, demonstrate that $\Omega_0 h^2$ of the order of the observed value can be obtained without effort, in compliance with all other constraints.

\section{Dark-Matter Direct Detection}\label{sec:DD}

One of the most straightforward signals one may expect of our model are DM collisions on nuclei. The latter are constrained by a large number of direct-detection experiments, the most stringent bounds for the DM masses of interest to us being Refs. \cite{Akerib:2016vxi,Cui:2017nnn,Aprile:2018dbl}. Actually, it is precisely in the light of these constraints that we restricted our attention to DM multiplets that transform under $SU(2)_L$ representations with {\em odd} dimensions, allowing for a $Y=0$ multiplet member -- the DM candidate -- as discussed in Sec.~\ref{sec:intro}.

Even in our case however, the DM-nucleon cross section receives a tree-level contribution mediated by a $Z^\prime$.
Since DM is non-relativistic, and since $M_{Z^\prime}$ is also much larger than the relevant momentum transfer, the scattering process with the nucleon constituents may be accounted by a local Lagrangian
\be
\label{eq:Lchiq}
\mc L_{\chi q} ~=~ \frac{g_{Z^\prime}^2}{12 \, M_{Z^\prime}^2} \xi_\chi \, (\bar \chi \gamma^\mu \chi)\,(\xi_q^i \, \bar q^i \gamma_\mu q^i + \xi_u^i \, \bar u_R^i \gamma_\mu u_R^i + \xi_d^i \, \bar d_R^i \gamma_\mu d_R^i)~.
\ee
Furthermore, if we are able to neglect corrections due to the finite momentum transfer between the DM and the nucleons, we may parametrize the matrix elements between vector or axial-vector quark $q$ currents and the external-state nucleons $N$ as
\bea
\label{eq:NNqq}
\< N(p') | \bar q \gamma^\mu q | N(p) \>|_{\vec{p} = \vec{p}^{\,\prime} = 0} &=& F_1^{q/N}(0) \,\,  \bar u_{N}(p') \gamma^\mu u_{N}(p)~,\nn \\
\< N(p') | \bar q \gamma^\mu \gamma_5 q | N(p) \>|_{\vec{p} = \vec{p}^{\,\prime} = 0} &=& F_A^{q/N}(0) \,\, \bar u_{N}(p') \gamma^\mu \gamma_5 u_{N}(p)~.
\eea
For the two form factors at zero momentum transfer we follow conventions common in the literature.
In particular, $F_1^{q/N}(0)$ counts the number of valence quarks $q$ in the nucleon $N$, e.g. $F_1^{d/n}(0) = 2$.
eqs.~(\ref{eq:Lchiq})-(\ref{eq:NNqq}) yield the following spin-independent cross section for elastic scattering between DM and a single nucleon $N = p$ or $n$
\be
\label{eq:sigSI^N}
\sigma_{\rm SI}^N ~=~ \frac{g_{Z'}^4 \, \xi^2_\chi \, M_N^2}{144 \pi \, M_{Z'}^4} |C_V^N|^2~,
\ee
where
\be
C_V^p = 2 C_V^u + C_V^d~,~~~~~C_V^n = C_V^u + 2 C_V^d~,
\ee
and
\be
C_V^u = \frac{\xi_q^1 + \xi_u^1}{2}~,~~~~~C_V^d = \frac{\xi_q^1 + \xi_d^1}{2}~.
\ee
Starting from eq.~(\ref{eq:sigSI^N}), in order to estimate the matrix element on a nucleus $\mc N$ with mass number $A$ and atomic number $Z$, one may assume (see e.g. \cite{Salati:2007zz,Anand:2013yka}) that DM scatters coherently on the $A$ nucleons of the target. In the static limit, the DM - nucleon cross section measured by experiments operating with nuclei $\mc N$ as target material can thus be estimated from eq.~(\ref{eq:sigSI^N}) with the replacement
\be
\label{eq:CVN}
|C_V^N|^2 \rightarrow \frac{|Z C_V^p + (A-Z) C_V^n|^2}{A^2}~.
\ee

The above procedure is crude in a number of ways, that have been amply discussed in the literature \cite{Fan:2010gt,Fitzpatrick:2012ix,Fitzpatrick:2012ib,Anand:2013yka,Cirigliano:2012pq,DelNobile:2013sia,Barello:2014uda,Hill:2014yxa,Hoferichter:2015ipa,Catena:2014uqa,Hill:2013hoa,Hill:2011be,Hoferichter:2016nvd,Kurylov:2003ra,Pospelov:2000bq,Bagnasco:1993st,Bishara:2016hek,Bishara:2017pfq}.
A first outstanding limitation is the fact that the matching scale for the interactions in eq.~(\ref{eq:Lchiq}) is well above the effective scale for the hadronic matrix elements in eq.~(\ref{eq:NNqq}), hence renormalization-group effects are in general non-negligible, and the relativistic-operator basis of eq.~(\ref{eq:Lchiq}) has to be matched onto the non-relativistic basis relevant for the interaction with the nucleus.
A second crucial limitation occurs if the DM momentum is large enough that the pointlike-nucleon approximation inherent in eq.~(\ref{eq:NNqq}) loses validity.

The impact of the above approximations may be explored using the public codes {\tt DirectDM} \cite{Bishara:2016hek,Bishara:2017pfq,Bishara:2017nnn} -- that accounts for renormalization-group effects from the UV scale to the scale of the (non-relativistic) interaction with the nucleons -- and {\tt DMFormFactor} \cite{Fitzpatrick:2012ix,Fitzpatrick:2012ib,Anand:2013yka} -- which estimates the non-trivial dynamics due to non-negligible momentum transfers between the DM and the nucleon.

We performed a detailed comparison between the prediction obtained within the analytic approach of eqs.~(\ref{eq:sigSI^N})-(\ref{eq:CVN}) and the numerical estimate obtained within the {\tt DirectDM} and {\tt DMFormFactor} codes. The analytic approach of eqs.~(\ref{eq:sigSI^N})-(\ref{eq:CVN}) yields a prediction in agreement to $\lesssim 25\%$ with the numerical estimate, provided $M_\chi \lesssim 400$ GeV, and for any $M_\chi \gtrsim 200$ GeV the numerical prediction is {\em lower} with respect to the analytic result by a factor of $\lesssim 2$ for $M_\chi \lesssim 1.5$ TeV. This conclusion holds for any choice of $v_{LQ}$ within our fiducial range (see eq. \ref{eq:vLQrange}).

This comparison is in agreement with the expectation that, for $M_\chi$ light enough, the DM Compton wavelength -- of order $v_0/c \times M_\chi \approx 10^{-3} M_\chi$, where $v_0$ is the typically assumed RMS velocity of the DM halo distribution  -- is not sufficiently large to resolve the inner nucleon structure, so that the pointlike-nucleon approximation is tenable.

We conclude that, for $M_\chi$ masses in the range required by the relic-density constraint, $300$~GeV$ - 1.5$ TeV, use of the analytic prediction of eqs.~(\ref{eq:sigSI^N})-(\ref{eq:CVN}) will produce direct-detection bounds that are somewhat stronger -- yet quite realistic -- than those produced with numerical codes. In Sec.~\ref{sec:results} we will compare this analytic prediction with the latest bound obtained by the Xenon1T experiment \cite{Aprile:2018dbl}.

\section{Results}\label{sec:results}

The relic-density and direct-detection constraints discussed in the previous two sections represent significant phenomenological input for our model. We summarised our parameter space in eq.~(\ref{eq:parameter_space}) and discussed how $B$-decay discrepancies and collider constraints lead to the preferred ranges in eqs.~(\ref{eq:vLQrange})-(\ref{eq:sinq12_g4_ranges}). In this section we discuss to what extent such ranges are compatible with those imposed by the DM relic-density and direct-detection constraints.

\begin{figure}[h!]
  \centering
  \includegraphics[width=.75\textwidth]{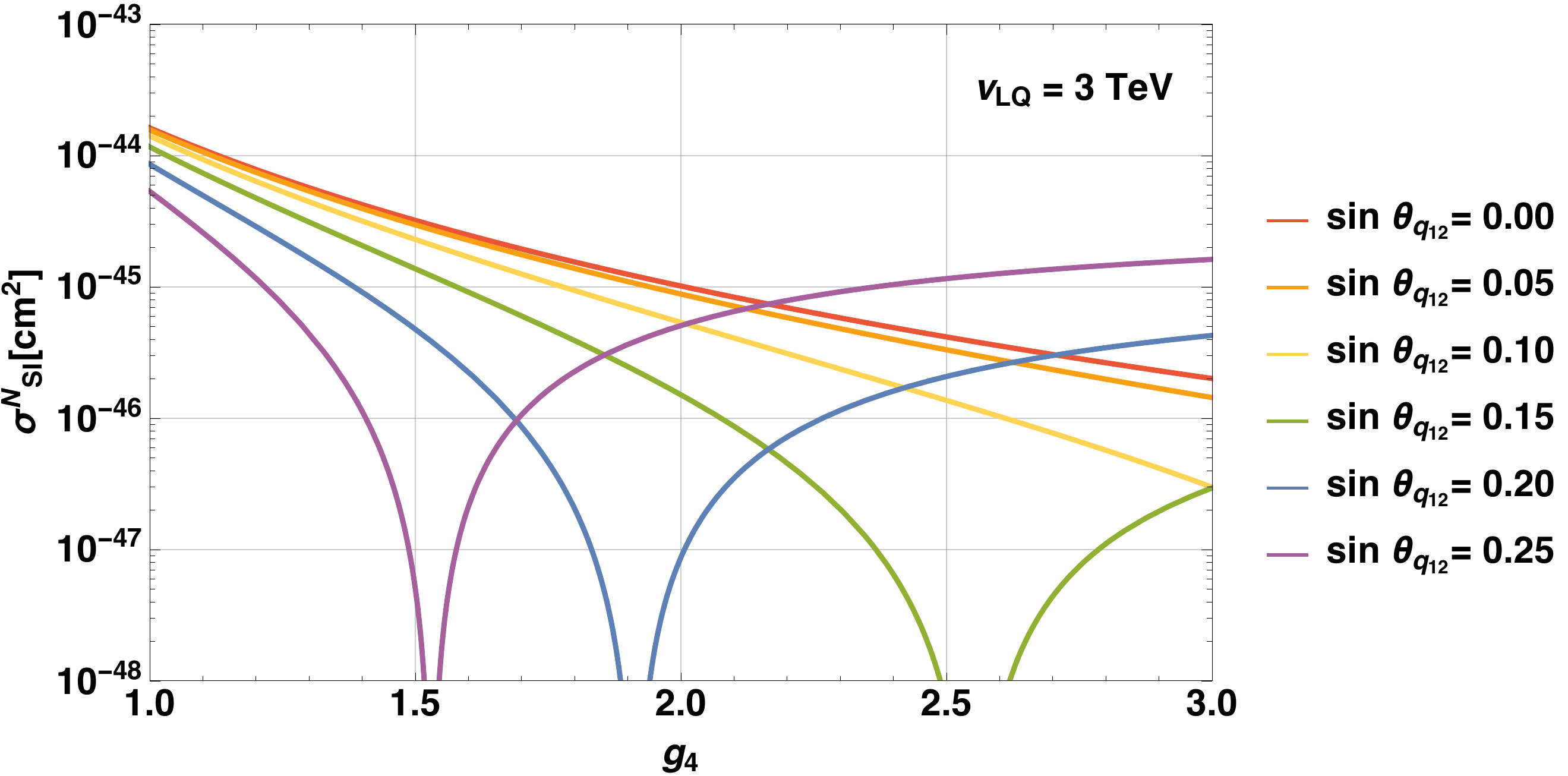}
  \caption{The DM - nucleon cross section (see eq.~(\ref{eq:sigSI^N})) as a function of $g_4$ for different choices of the mixing angle $\sin \theta_{q_{12}}$ and $v_{LQ} = 3$ TeV. See text for details.}
  \label{fig:sigmaN_vs_g4}
\end{figure}

Quite remarkably, the $g_4$ and $\sin \theta_{q_{12}}$ ranges in eq.~(\ref{eq:sinq12_g4_ranges}) are also favoured by direct-detection constraints, as illustrated in fig.~\ref{fig:sigmaN_vs_g4}. Here we show the DM - nucleon cross section $\sigma_{\rm SI}^N$ discussed around eq.~(\ref{eq:sigSI^N}), as a function of $g_4$ for increasing values of $\sin \theta_{q_{12}} \ge 0$. As discussed in Sec.~\ref{sec:DD}, experiments yield severe limits, as strong as $\sigma_{\rm SI}^N < 10^{-45} {\rm cm}^2$. As the figure shows, these limits can be comfortably satisfied with the choice $\sin \theta_{q_{12}} \lesssim 0.2$ and $g_4 \gtrsim 3$.\footnote{As an alternative, one may advocate $\sin \theta_{q_{12}} \ge 0.25$ and $g_4 \lesssim 2$, but the constraint would be satisfied in (or very close to) a fine-tuned region.}

Therefore, the constraints from $Z^\prime$ and $G^\prime$ direct searches on the one side, and from DM direct detection on the other side, identify one and the same region for $g_4$ and $\sin \theta_{q_{12}}$. Although a correlation between the suppression required to DY-produced $Z^\prime$ and the suppression required to $Z^\prime$-mediated DM - nucleon scattering may be expected just by crossing symmetry, the coupling combinations involved in the two processes are entirely different. Besides, it is non-trivial that couplings in compliance with $Z^\prime$ searches would {\em also} yield a DM cross section on nucleons as small as $10^{-45}$ cm$^2$.

\begin{figure}[bh]
  \centering
  \includegraphics[width=.70\textwidth]{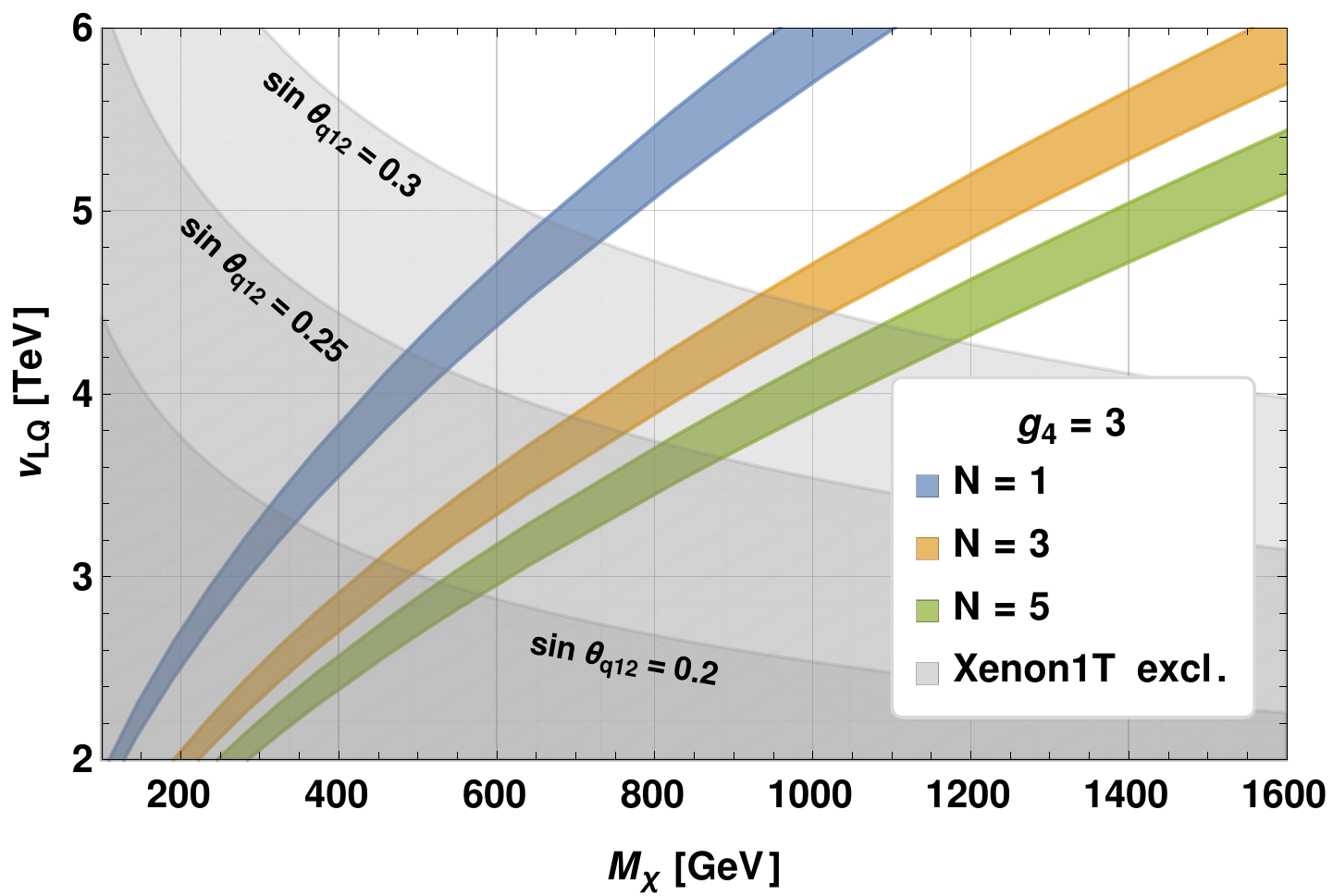}
  \caption{DM constraints in the $M_\chi$ vs. $v_{LQ}$ plane. Coloured rays fulfil the $\Omega_0 h^2$ constraint within $\pm 15\%$. Grey regions, with $\sin \theta_{q_{12}}$ fixed at the displayed value, are excluded by Xenon1T \cite{Aprile:2018dbl}. See text for details.
}
  \label{fig:M_vs_vLQ}
\end{figure}
We next discuss the behaviour of $\sin \theta_{q_{12}}$, $g_4$ and $v_{LQ}$ as a function of the DM mass $M_\chi$, when the relic-density and direct-detection constraints are imposed. In fig.~\ref{fig:M_vs_vLQ} we show as colored rays the regions selected by the constraint $\Omega_0 h^2$ in the $M_\chi$ vs. $v_{LQ}$ plane. The constraint is imposed with $\pm 15\%$ accuracy, corresponding to the error we attach to its calculation -- see discussion in Sec.~\ref{sec:Omega_DM}. The different rays refer to different choices of $N$, whereas $g_4 = 3$ following our above discussion.
We see that the $v_{LQ}$ range in eq.~(\ref{eq:vLQrange}), plus the $\Omega_0 h^2$ constraint, allow to identify the following, indicative $M_\chi$ ranges
\renewcommand{\arraystretch}{1.5}
\be
\label{eq:Mchi_ranges}
M_\chi [{\rm GeV}] \in
\begin{array}{ll}
{[}260, 720] & (N = 1)\\
{[}450, 1190] & (N = 3)\\
{[}570, 1460] & (N = 5)
\end{array}~.
\ee
\renewcommand{\arraystretch}{1.0}
Also shown in different shades of grey depending on $\sin \theta_{q_{12}} \in \{ 0.2, 0.25, 0.3 \}$ are regions excluded by direct detection. It is clear that, for $\sin \theta_{q_{12}} = 0.2 $ and $v_{LQ} \ge 3$ TeV (see eq.~(\ref{eq:vLQrange})), only a tiny fraction of the parameter space is excluded by direct detection, and mostly for $N = 1$.

\begin{figure}[h!]
  \centering
  \includegraphics[width=.49\textwidth]{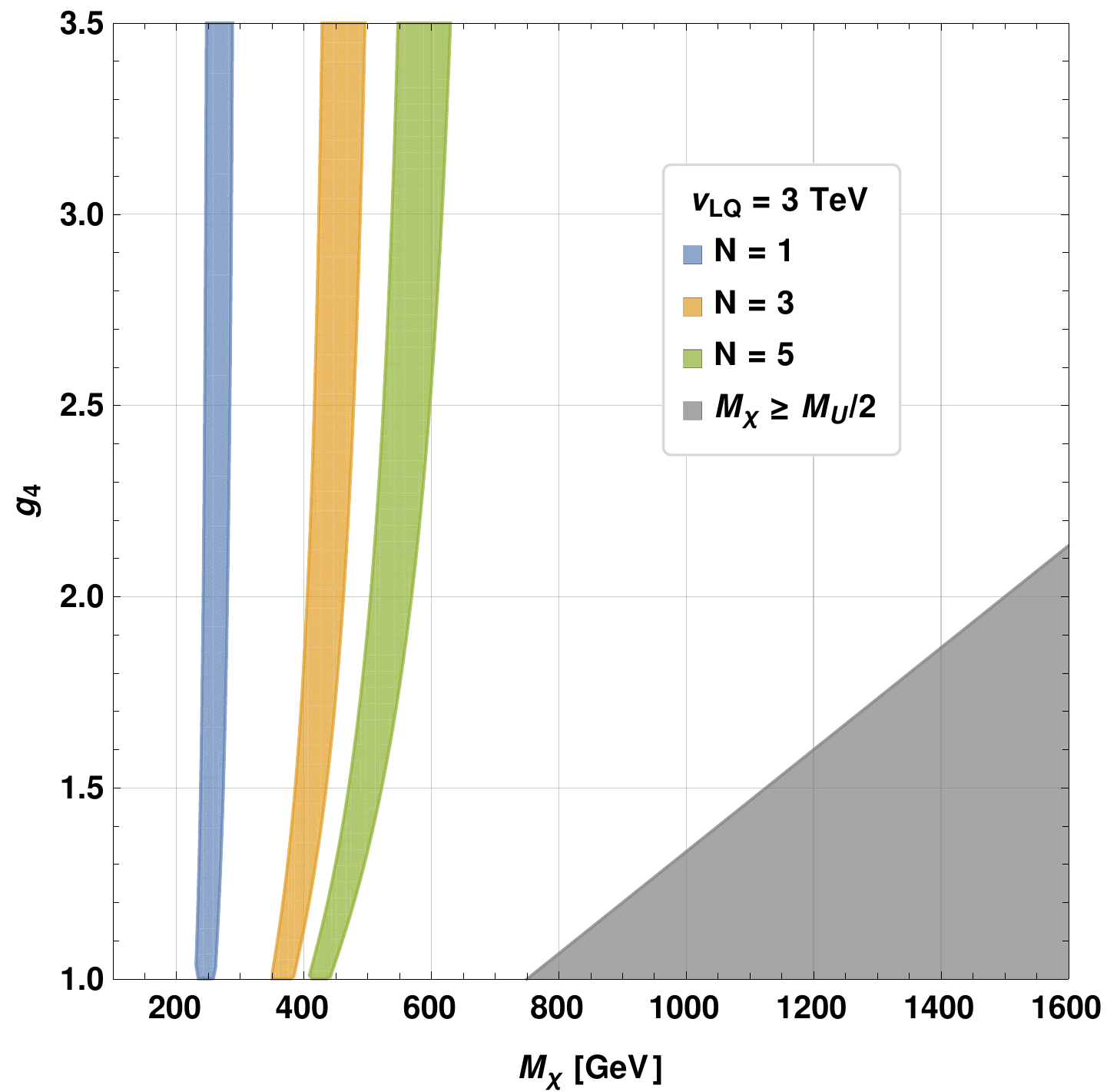}
  \hfill
  \includegraphics[width=.49\textwidth]{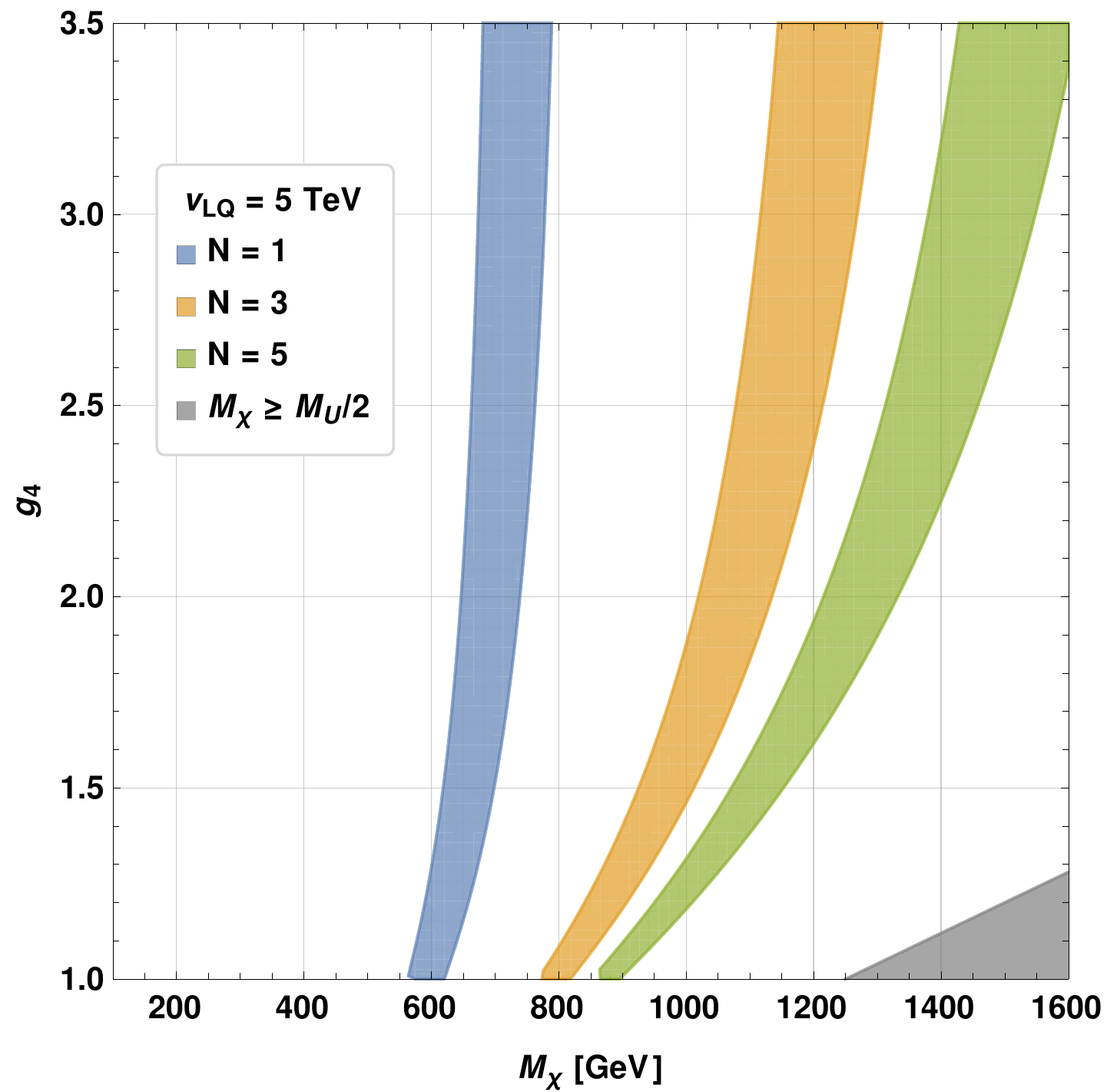}
  \caption{Coloured rays fulfil the $\Omega_0 h^2$ constraint within $\pm 15\%$. The grey region corresponds to an on-shell intermediate $U_1$ leptoquark (see text for details).
  }
  \label{fig:M_vs_g4}
\end{figure}
It is interesting to also test the dependence of the region selected by the $\Omega_0 h^2$ constraint on $g_4$, rather than on $v_{LQ}$. We show such dependence in the two panels of fig.~\ref{fig:M_vs_g4}, corresponding to $v_{LQ}$ set to 3 and 5 TeV, respectively. We see that the dependence on $g_4$ is actually very weak, especially if this coupling is large. The figure also shows the region where $M_\chi \ge M_U/2$, that we excluded for simplicity. In fact, as the $U_1$ becomes on-shell, $2 \to 3$ and $2 \to 4$ decay channels open up, whereas we restricted to $2 \to 2$ processes in the calculation of $\sigma_{\rm eff}$.

\begin{figure}[h!]
  \centering
  \includegraphics[width=.70\textwidth]{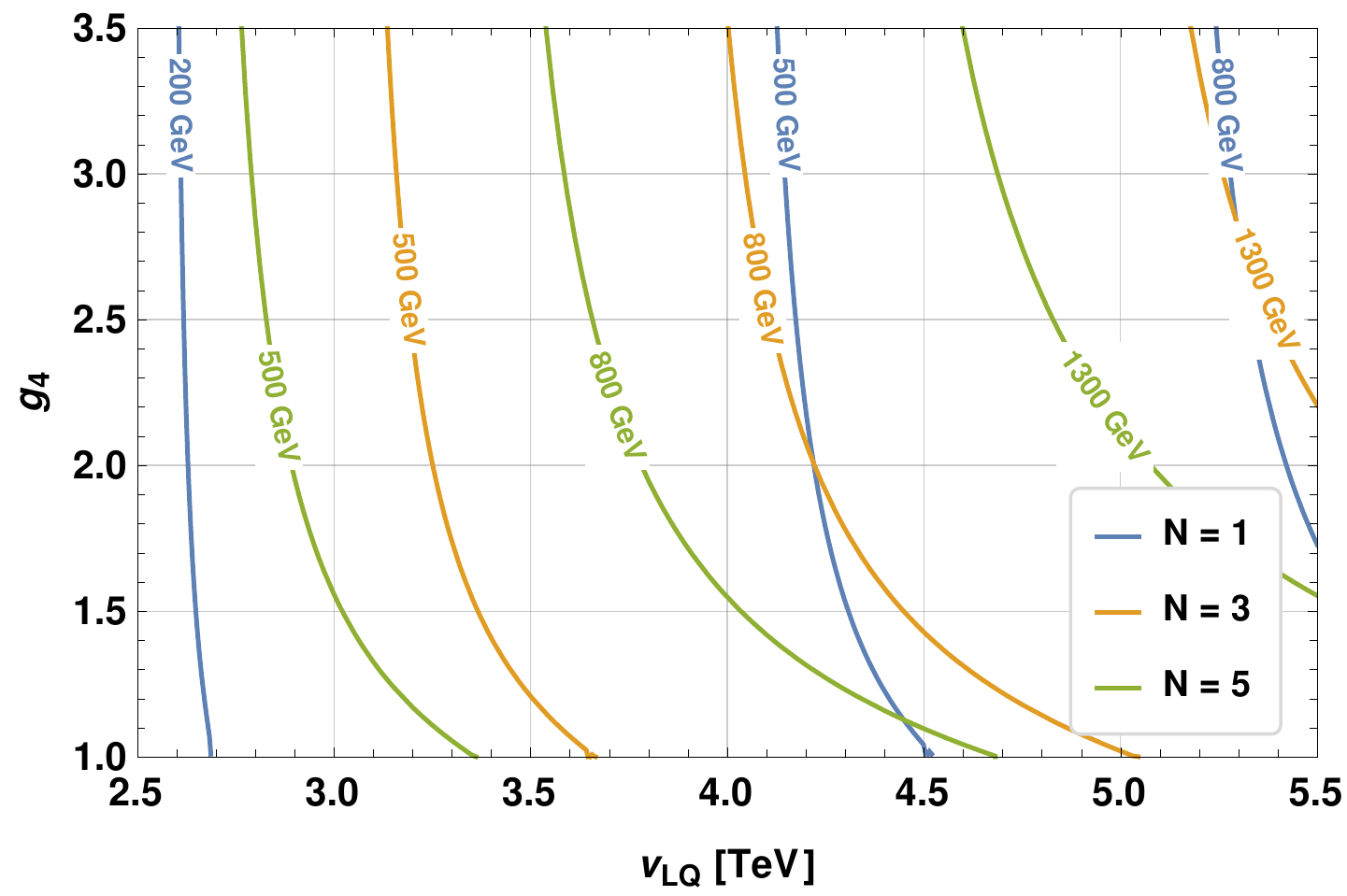}
  \caption{The $\Omega_0 h^2$ in the $v_{LQ}$ vs. $g_4$ plane. Along the isolines, the $\Omega_0 h^2$ constraint is fulfilled for the  value of $M_\chi$ specified along the line itself.}
  \label{fig:vLQ_vs_g4_isoMchi}
\end{figure}
The dependence of the $\Omega_0 h^2$ constraint on $M_\chi$ vs. $v_{LQ}$ or $g_4$ shown in figs.~\ref{fig:M_vs_vLQ} and \ref{fig:M_vs_g4} can be captured simultaneously in fig.~\ref{fig:vLQ_vs_g4_isoMchi}, that shows this constraint in the plane $v_{LQ}$ vs. $g_4$ for a few reference values of $M_\chi$ and of $N$, represented as isolines. The figure shows at a glance that within the fiducial $v_{LQ}$ range of eq.~(\ref{eq:vLQrange}), the relic-density constraint can be comfortably satisfied whatever the choice of $N$, and also nearly irrespective of the choice of $g_4$ -- which, as we discussed, is instead constrained by direct detection.

\begin{figure}[h!]
  \centering
  \includegraphics[width=.70\textwidth]{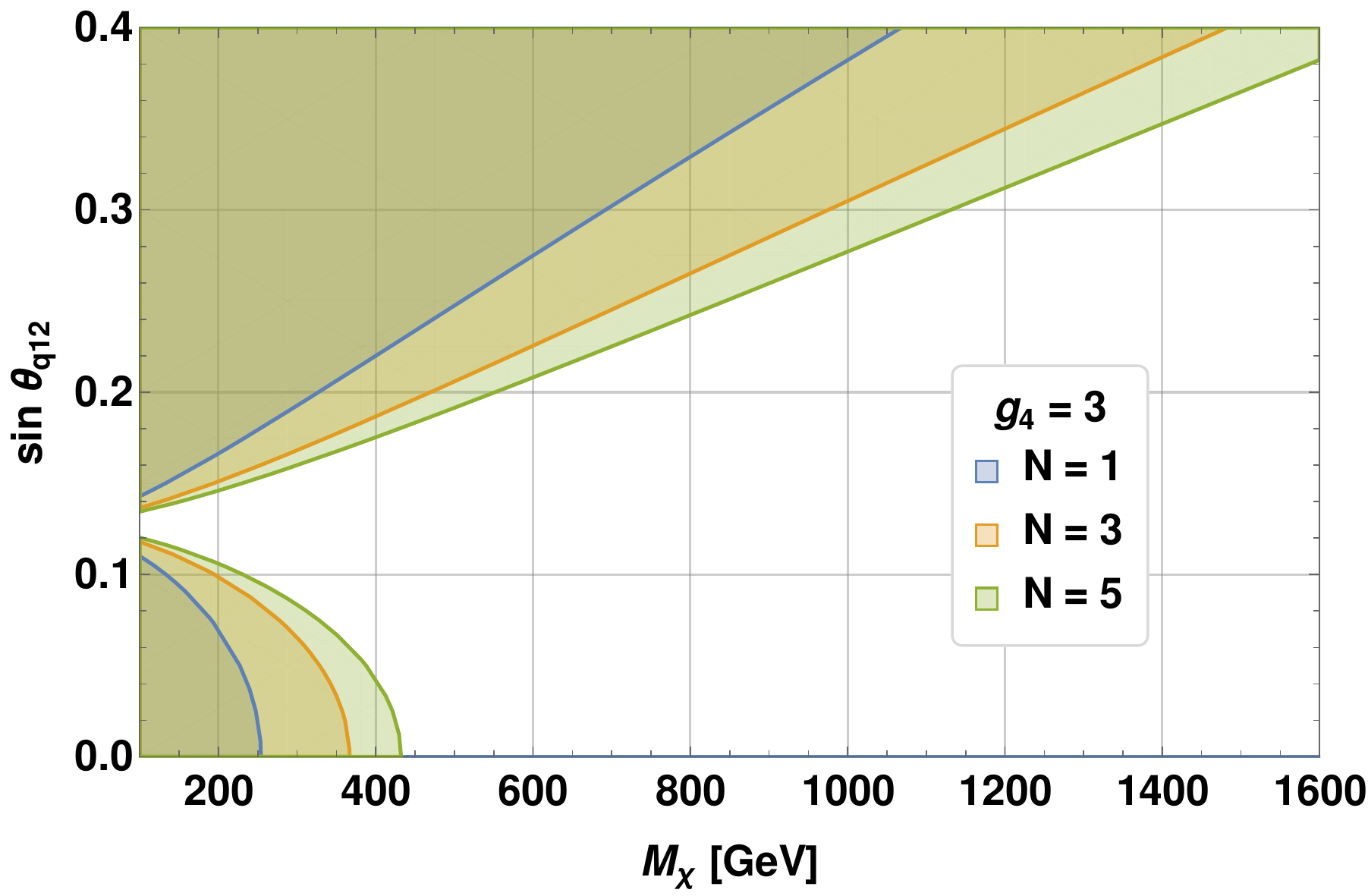}
  \caption{DM constraints in the $M_\chi$ vs. $\sin \theta_{q_{12}}$ plane, with $g_4 = 3$. The coloured `curtains' denote regions excluded by Xenon1T \cite{Aprile:2018dbl}.
  }
  \label{fig:M_vs_SinTheta}
\end{figure}
Finally, fig.~\ref{fig:M_vs_SinTheta} displays the allowed parameter space in the $M_\chi$ vs. $\sin \theta_{q_{12}}$ plane, with $g_4 = 3$. The coloured `curtains' show the region excluded by direct detection, depending on the choice of $N$.
Similarly as fig.~\ref{fig:sigmaN_vs_g4} and the ensuing discussion, this plot shows that direct detection tends to prefer small $\sin \theta_{q_{12}}$, the upper bound for a given $M_\chi$ becoming stronger as $N$ increases.
As discussed earlier, imposing the relic-density constraint plus the $v_{LQ}$ range suggested by $B$ discrepancies yields the {\em indicative} $M_\chi$ ranges in eq.~(\ref{eq:Mchi_ranges}). These ranges are not reported in fig.~\ref{fig:M_vs_SinTheta} to limit clutter.

\section{Conclusions} \label{sec:conclusions}

We investigate a possible common description of the only hint of
physics beyond the SM in collider searches -- the $B$-decay discrepancies -- and of one of the strongest phenomenological indications of new physics -- the existence of Dark Matter.\footnote{Such common description has also been studied elsewhere in the literature, see \cite{Sierra:2015fma,Belanger:2015nma,Allanach:2015gkd,Bauer:2015boy,Celis:2016ayl,Altmannshofer:2016jzy,Ko:2017quv,Ko:2017yrd,Cline:2017lvv,Sala:2017ihs,Ellis:2017nrp,Kawamura:2017ecz,Baek:2017sew,Cline:2017aed,Cline:2017qqu,Dhargyal:2017vcu,Chiang:2017zkh,Vicente:2018xbv,Falkowski:2018dsl,Arcadi:2018tly,Baek:2018aru,Azatov:2018kzb,Barman:2018jhz,Cerdeno:2019vpd,Trifinopoulos:2019lyo,DaRold:2019fiw,Han:2019diw}.}

We adopt the 4321 gauge ansatz, a well-motivated, UV-complete, calculable setup for explaining the $B$ anomalies based on the gauge group
$SU(4)\times SU(3)'\times SU(2)_L\times U(1)_X$.
To this ansatz, we add a minimal DM sector, represented by a fermionic multiplet sitting in the fundamental {\bf 4} of $SU(4)$ and in a representation of $SU(2)_L$ with odd dimension. After the breaking to the SM group, this multiplet gives rise to a DM candidate and a coloured co-annihilation partner with mass $\approx 10\%$ larger.

We find that the parameter space selected by collider and $B$-physics data is also the one favoured by DM phenomenology, in particular by the constraints imposed by direct detection. These parameter choices include the 43(2)1 symmetry breaking scale ${v_{LQ} \in [3, 5]}$~TeV, large $SU(4)$ gauge coupling $g_4 \approx 3$ and small fermion mixing parameters of light quarks ${\sin \theta_{q_{12}} \lesssim 0.2}$. Within this parameter space, direct-detection signals happen to be below the severe bounds imposed in particular by Xenon1T. For $v_{LQ}$ in the above mentioned range, the requirement of the correct DM relic density is easily fulfilled with a DM mass between about 250 GeV and 1.5 TeV, depending on the $SU(2)_L$ representation, and on the $v_{LQ}$ value.

In short, we find our setup neatly compatible with the most accurate DM observations -- the relic density and the limits imposed by direct detection. Interestingly, while the new particle in 4321 models that is mainly responsible for $B$-physics discrepancies is the $U_1$ LQ, it is the $Z^\prime$ that plays the leading role in DM phenomenology.

The study of this class of models thus warrants further scrutiny, especially if the $B$ anomalies will be consolidated by forthcoming measurements.

\section*{Acknowledgments}

DG would like to thank Marco Cirelli for very useful feedback, and Eugenio Del Nobile for critical comments on Sec.~\ref{sec:DD}. Exchanges with Fady Bishara, Jacopo Ghiglieri and Liam Fitzpatrick are also acknowledged. MR thanks the Universit\'e de Montr\'eal for kind hospitality in Spring 2019, and for discussions with David London and Jacky Kumar on related topics. This work is supported by an ANR PRC (contract n. 202650) and by the Labex Enigmass.

\appendix

%:%% BEGIN mass splitting app
\section{Mass splitting}\label{app:mass_splitting}

In order to determine the one-loop mass splitting, we compute the pole masses of the components of a VL fermion multiplet.
It is possible to obtain
these pole masses
in a way that is applicable to a large set of different spontaneously broken gauge groups.
To this end, we consider a gauge group $G\times H'$ that is spontaneously broken to its subgroup $H$,
\begin{equation}
 G\times H' \to H,
\end{equation}
where $G$ has a subgroup $H^G\subseteq G$ that is isomorphic to both $H'$ and the unbroken $H$, i.e. $H'\cong H\cong H^G$, and $H$ is the diagonal subgroup of $H^G\times H'$.
We take $G$ to be a simple group and $H'$ to be semi-simple and given by the direct product of simple groups $H'_i$ as $H'=H'_1 \times H'_2 \times \dots \times H'_n$.\footnote{%
In case of a semi-simple $G=G_1 \times G_2 \times \dots \times G_m$, one can treat each simple group factor $G_i$ separately.
The results given here for a simple group $G$ can therefore be generalized easily to a semi-simple $G$.
}
An analogous decomposition into simple groups $H_i$ and $H^G_i$ applies to $H$ and $H^G$.
We denote the gauge couplings of $G$, $H'_i$, and $H_i$ by $g_G$, $g_{H'_i}$, and $g_{H_i}$, respectively, and we note that the gauge couplings of all simple group factors of $H^G$ are equal to $g_G$.
It is convenient to define mixing angles $\theta_i$ by
\begin{equation}
 c_i\equiv \cos \theta_i = \frac{g_{H_i}}{g_{H'_i}},
 \quad
 s_i\equiv \sin \theta_i = k_i\,\frac{g_{H_i}}{g_G},
\end{equation}
where $k_i$ is a normalization factor relevant in case of an abelian group $H_i=U(1)$ and corresponds to the normalization of the $U(1)$ charges.
For non-abelian $H_i$, we set $k_i=1$.
After the spontaneous symmetry breaking $G\times H' \to H$, there are $\dim(H)$ massless vector bosons that are linear combinations of the $H'$ and $H^G$ gauge bosons.
The orthogonal linear combinations constitute $\dim(H)$ massive vector bosons with masses $M_{H_i}$.
In addition, all vector bosons associated with the coset $G/H$ become massive and have the common mass $M_{G/H}$.

We find it illustrative to show how the above described spontaneously broken gauge group is a generalization both of the EW group in the SM and of the 421 part in 4321 models.
\begin{itemize}
 \item For EW symmetry breaking, $G=SU(2)_L$, $H'=U(1)_Y$, and $H=U(1)_{\rm em}$. The couplings are $g_G = g$, $g_{H'}=g_Y$, and $g_H=e$. In this case, there is only a single mixing angle that can be identified with the weak-mixing angle, $\theta=\theta_W$.
 There is $\dim(H)=1$ massless vector boson, which can be identified with the photon, $\dim(H)=1$ massive vector boson, which can be identified with the $Z$, and $\dim(G/H)=2$ massive coset vector bosons that can be identified with $W^{\pm}$.
 If one defines the electric charge $Q$ as $Q=T_3+Y$, where $T_3$ is the diagonal generator of $SU(2)_L$ normalized as $\text{tr} [T_3 T_3]=1/2$ for the fundamental representation, then $k=1$.

 \item For the 4321 breaking, $G=SU(4)$, $H'_1=U(1)'$, $H'_2=SU(3)'$, $H_1=U(1)_Y$, and $H_2=SU(3)_c$.
 The couplings are $g_G=g_4$, $g_{H'_1}=g_1$, $g_{H'_2}=g_3$, $g_{H_1}=g_Y$, and $g_{H_2}=g_s$. In this case, there are two mixing angles $\theta_1=\theta_{41}$ and $\theta_2=\theta_{43}$.
 There are $\dim(H_1)=1$ plus $\dim(H_2)=8$ massless vector bosons, which can be identified with the $B$ and the gluons, $\dim(H_1)=1$ plus $\dim(H_2)=8$ massive vector bosons, which can be identified with $Z'$ and $G'$, and $\dim(G/H)=6$ massive coset vector bosons that can be identified with colored $U^{\pm}$.
 In order to be able to use the conventional normalization of the electric charge defined by $Q=T_3+Y$, one has to set $k_1=\sqrt{2/3}$.
\end{itemize}

We now consider a VL fermion multiplet that transforms under a representation $R_G$ of $G$ and under representations $R_{H'_i}$ of the $H'_i$.
Note that since the fermion multiplet is VL under $G\times H'$, all of its components transform under $G$ and $H'$ in the same way.
However, the components transform differently under the subgroup $H^G$ and under the unbroken group $H$.
A given multiplet component $\xi$ transforms under representations $r^\xi_{H^G_i}$ of the $H^G_i$ and under representations $r^\xi_{H_i}$ of the $H_i$.
The one-loop pole mass can then be expressed in terms of the quadratic Casimir invariants of the different groups and representations involved.
In particular, we denote them as follows:
\begin{itemize}
 \item $C_2^G(R_G)$: quadratic Casimir of the $R_G$ representation of $G$,
 \item $C_2^{H'_i}(R_{H'_i})$: quadratic Casimir of the $R_{H'_i}$ representation of $H'_i$,
 \item $C_2^{H^G_i}(r^\xi_{H^G_i})$: quadratic Casimir of the $r^\xi_{H^G_i}$ representation of $H^G_i$,
 \item $C_2^{H_i}(r^\xi_{H_i})$: quadratic Casimir of the $r^\xi_{H_i}$ representation of $H_i$.
\end{itemize}
Each multiplet component $\xi$ receives contributions to its pole mass from three different kinds of one-loop diagrams.
\begin{enumerate}
 \item From the massive vector bosons that transform in the adjoint representation of $H$ and have masses $M_{H_i}$,
 \begin{equation}\label{eq:Delta_H}
  \Sigma_H =
  \frac{g_G^2\,\hat{M}}{16\,\pi^2}
  \sum_i
  \left(
  C_2^{H^G_i}(r^\xi_{H^G_i})
  + \frac{s_i^2}{c_i^2\,k_i^2}\,C_2^{H'_i}(R_{H'_i})
  - \frac{s_i^2}{k_i^2}\,C_2^{H_i}(r^\xi_{H_i})
  \right)
  \left[
  f\left(\tfrac{M_{H_i}}{\hat{M}}\right)
  +K(\hat{M},\mu)
  \right]
 \end{equation}
 \item From the massless vector bosons that transform in the adjoint representation of $H$,
 \begin{equation}\label{eq:Delta_0}
  \Sigma_0 =
  \frac{g_G^2\,\hat{M}}{16\,\pi^2}
  \sum_i
  \,
  \frac{s_i^2}{k_i^2}\,C_2^{H_i}(r^\xi_{H_i})
  \,
  K(\hat{M},\mu)
 \end{equation}
 \item From the massive vector bosons that correspond to the $G/H$ coset and have mass $M_{G/H}$,
 \begin{equation}\label{eq:Delta_G/H}
  \Sigma_{G/H} =
  \frac{g_G^2\,\hat{M}}{16\,\pi^2}
  \left(
  C_2^G(R_G)-
  \sum_i
  C_2^{H^G_i}(r^\xi_{H^G_i})
  \right)
  \left[
  f\left(\tfrac{M_{G/H}}{\hat{M}}\right)
  +K(\hat{M},\mu)
  \right]
 \end{equation}
\end{enumerate}
In the above expressions, $K(\hat{M},\mu)$ is a divergent term that depends only on the VL fermion mass and on the renormalization scale $\mu$, while $f(r)$ is a finite loop function given by (cf.~\cite{Cirelli:2005uq})
\begin{equation}\label{eq:f_loop_function}
 f(r) =
 r^4\,\ln r - r^2 +
 \frac{r}{2}\,\sqrt{r^2-4}\,\left(r^2+2\right)\,\ln \left(
 \frac{r^2}{2}-\frac{r}{2}\,\sqrt{r^2-4}-1
 \right)\,.
\end{equation}
The $\xi$ pole mass $M_\xi$ can then be written as
\begin{equation}
 M_\xi = \hat{M} - \Sigma_H - \Sigma_0 - \Sigma_{G/H}\,.
\end{equation}
It is interesting to note that the contributions from the massless vector bosons, eq.~\eqref{eq:Delta_0}, exactly cancel the divergent and scale-dependent terms in eq.~\eqref{eq:Delta_H} that are proportional to $C_2^{H_i}(r^\xi_{H_i})$.
Similarly, the divergent and scale-dependent terms proportional to $C_2^{H^G_i}(r^\xi_{{H_G}_i})$ cancel between eq.~\eqref{eq:Delta_H} and eq.~\eqref{eq:Delta_G/H}.
All remaining divergent and scale-dependent terms are either proportional to $C_2^G(R_G)$ or $C_2^{H'_i}(R_{H'_i})$, i.e.\ these terms are the same for any multiplet component and therefore cancel in the mass differences.
Consequently, the $\xi$ pole mass can be written as
\begin{equation}
 M_\xi =
 \hat{M}
 - I
 - \frac{g_G^2\,\hat{M}}{16\,\pi^2}
 \sum_i
 \left\{
 C_2^{H^G_i}(r^\xi_{H^G_i})
 \left[
 f\left(\tfrac{M_{H_i}}{\hat{M}}\right)
 -
 f\left(\tfrac{M_{G/H}}{\hat{M}}\right)
 \right]
 -
 \frac{s_i^2}{k_i^2}\,C_2^{H_i}(r^\xi_{H_i})\,f\left(\tfrac{M_{H_i}}{\hat{M}}\right)
 \right\}\,,
\end{equation}
where $I$ collects all terms that are the same for each multiplet component and is given by
\begin{equation}
 I =
 \frac{g_G^2\,\hat{M}}{16\,\pi^2}
 \left\{
 C_2^G(R_G)
  \left[
  f\left(\tfrac{M_{G/H}}{\hat{M}}\right)
  +K(\hat{M},\mu)
  \right]
  +
  \sum_i
  \frac{s_i^2}{c_i^2\,k_i^2}\,C_2^{H'_i}(R_{H'_i})
  \left[
  f\left(\tfrac{M_{H_i}}{\hat{M}}\right)
  +K(\hat{M},\mu)
  \right]
  \right\}\,.
\end{equation}
Using the result for the one-loop pole mass, we find a generic expression for the relative mass splitting between the components $\xi$ and $\eta$ of a VL fermion multiplet,
\begin{equation}\label{eq:generic_mass_splitting}
\begin{aligned}
 \Delta_{\xi\eta}
 &=\frac{M_\xi-M_\eta}{\hat{M}}
 =
 \frac{g_G^2}{16\,\pi^2}
 \sum_i
 \bigg\{
 &\!\!\!\!
 \left(
 C_2^{H^G_i}(r^\xi_{H^G_i})
 -
 C_2^{H^G_i}(r^\eta_{H^G_i})
 \right)
 \left[
 f\left(\tfrac{M_{G/H}}{\hat{M}}\right)
 -
 f\left(\tfrac{M_{H_i}}{\hat{M}}\right)
 \right]
 &
 \\
 &&
 +
 \frac{s_i^2}{k_i^2}\,
 \left(
 C_2^{H_i}(r^\xi_{H_i})
 -
 C_2^{H_i}(r^\eta_{H_i})
 \right)
 \,f\left(\tfrac{M_{H_i}}{\hat{M}}\right)
 &\bigg\}\,,
\end{aligned}
\end{equation}
which is finite and scale-independent at one-loop.

Given the generic result, it is straight forward to consider the special cases of the EW gauge group and of the 4321 models.
\begin{itemize}
 \item For the EW gauge group, the unbroken group is abelian. Thus, the quadratic Casimir invariants are simply given in terms of squares of $U(1)$ charges.
 This yields
 \begin{equation}
  C_2^{H^G} = T_3\,T_3 = (Q-Y)^2\,,
  \quad\quad
  C_2^{H} = Q^2\,,
 \end{equation}
 and we find
 \begin{equation}
 \begin{aligned}
 \Delta_{\xi\eta}
 &=
 \frac{g^2}{16\,\pi^2}
 \bigg\{
 &\!\!\!\!
 \left(
 (Q_\xi-Y)^2
 -
 (Q_\eta-Y)^2
 \right)
 \left[
 f\left(\tfrac{M_{W}}{\hat{M}}\right)
 -
 f\left(\tfrac{M_{Z}}{\hat{M}}\right)
 \right]
 &
 \\
 &&
 +
 s_W^2
 \,
 (
 Q_\xi^2
 -
 Q_\eta^2
 )
 \,f\left(\tfrac{M_{Z}}{\hat{M}}\right)
 &\bigg\}\,,
 \end{aligned}
 \end{equation}
 which coincides with the well-known result (cf.\ e.g.~\cite{Cirelli:2005uq}).
 \item For the 43(2)1 gauge group, the unbroken group contains one abelian and one non-abelian factor.
 The quadratic Casimir invariants are
 \begin{equation}
 \begin{aligned}
  C_2^{H^G_1} = T^{15}\,T^{15} = \frac{3}{2}(Y-Y')^2\,,
  \quad\quad
  C_2^{H_1} = Y^2\,,
  \\
  C_2^{H^G_2} = C_2^{SU(3)_4}\,,
  \quad\quad
  C_2^{H_2} = C_2^{SU(3)_c}\,,
 \end{aligned}
 \end{equation}
i.e.\ the Casimir invariants of the abelian groups can be expressed by the $U(1)$ charges $Y$ and $Y'$, while those of the non-abelian factors are $SU(3)$ Casimir invariants.

For our DM candidate $\chi$ and its co-annihilation partner $\psi$, the Casimir invariants are given by
 \begin{equation}
 \begin{aligned}
  C_2^{H^G_1}(\chi) = \frac{3}{8}\,,
  \quad\quad
  C_2^{H_1}(\chi) = 0\,,
  \quad\quad
  C_2^{H^G_2}(\chi) = 0\,,
  \quad\quad
  C_2^{H_2}(\chi) = 0\,,
  \\
  C_2^{H^G_1}(\psi) = \frac{1}{24}\,,
  \quad\quad
  C_2^{H_1}(\psi) = \frac{4}{9}\,,
  \quad\quad
  C_2^{H^G_2}(\psi) = \frac{4}{3}\,,
  \quad\quad
  C_2^{H_2}(\psi) = \frac{4}{3}\,,
 \end{aligned}
 \end{equation}
 such that the $\psi-\chi$ mass splitting is
 \begin{equation}
 \begin{aligned}
 \Delta_{\psi\chi}
 &=
 \frac{g_4^2}{16\,\pi^2}
 \bigg\{
 \!\!\!\!\!
 &
 \left(
 \frac{1}{24}
 -
 \frac{3}{8}
 \right)
 \left[
 f\left(\tfrac{M_{U}}{\hat{M}}\right)
 -
 f\left(\tfrac{M_{Z'}}{\hat{M}}\right)
 \right]
 +
 \frac{3\,s_{41}^2}{2}
 \,
 \frac{4}{9}
 \,f\left(\tfrac{M_{Z'}}{\hat{M}}\right)
 &
 \\
 &&
 +
 \frac{4}{3}
 \left[
 f\left(\tfrac{M_{U}}{\hat{M}}\right)
 -
 f\left(\tfrac{M_{G'}}{\hat{M}}\right)
 \right]
 +
 s_{43}^2
 \,
 \frac{4}{3}
 \,f\left(\tfrac{M_{G'}}{\hat{M}}\right)
 &\bigg\}\,,
 \end{aligned}
 \end{equation}
 which can be simplified to
 \begin{equation}
 \Delta_{\psi\chi}
 =
 \frac{g_4^2}{16\,\pi^2}
 \bigg\{
 f\left(\tfrac{M_{U}}{\hat{M}}\right)
 +
 \frac{1}{3}
 (2\,s_{41}^2+1)\,
 f\left(\tfrac{M_{Z'}}{\hat{M}}\right)
 +
 \frac{4}{3}(s_{43}^2-1)\,
 f\left(\tfrac{M_{G'}}{\hat{M}}\right)
 \bigg\}\,.
 \end{equation}

\end{itemize}
%%% END mass splitting app

%:%% BEGIN cross sections app
\section{Cross sections of processes entering the estimation of the relic density}\label{app:Sigma_eff}

Due to the dependence of the cross-sections on the fourth power of the couplings, contributions to the effective cross section eq.~(\ref{eq:general_sigma_eff}) boils down to those involving new heavy gauge bosons and gluons.
The contribution of the latter being additionally suppressed by the mass of the DM candidate, we found
\begin{equation}
  \begin{aligned}
    \sigma^{\chi\chi} \,\approx\,
    &\frac{g_4^4}{512\,\pi\, c_{41}^4} \frac{s \left(2\,M_\chi^2+s\right)}{\sqrt{s^2-4sM_\chi^2}\,\left(s-M_{Z'}^2\right)^2}\\
    & \hspace{3cm} \times \sum_{i=1}^3 \left(2|\xi_q^i|^2 + |\xi_u^i|^2 + |\xi_d^i|^2 + 3(2|\xi_\ell^i|^2 + |\xi_e^i|^2 + |\xi_\nu^i|^2)\right) \,,
  \end{aligned}
\end{equation}

\begin{equation}
 \begin{aligned}
   \sigma^{\chi\psi} \,\approx\,
   &\frac{g_4^4}{576\,\pi} \frac{\left((M_\chi-M_\psi)^2-s\right) \left((M_\chi+M_\psi)^2+2s\right)}{\sqrt{s^2-2s\,(M_\chi^2+M_\psi^2)+(M_\chi^2-M_\psi^2)^2} \, \left(s-M_{U}^2\right)^2} \\
   & \hspace{5cm} \times \left(2\sum_{i,j=1}^3|\beta_{ij}^{q\ell}|^2 + \sum_{i=1}^3|\beta_i^{de}|^2 + \sum_{i=1}^3|\beta_i^{u\nu}|^2\right) \,,
  \end{aligned}
 \end{equation}

\begin{equation}
 \sigma^{\psi\psi} \approx
 \frac{g_4^4 \, |\kappa_\psi|^2}{324 \,\pi\, c_{43}^4} \frac{s\left(2\,M_\psi^2+s\right)}{\sqrt{s^2-4sM_\psi^2} \,\left(s-M_{G'}^2\right)^2}
 \sum_{i=1}^3 \left(2|\kappa_q^i|^2 + |\kappa_u^i|^2 + |\kappa_d^i|^2\right) \,.
\end{equation}
%%% END cross sections app

%:%% BEGIN fermions app
\section{Fermions in 4321 models}\label{app:fermions}
The fermion sector of the model contains the SM fermions, the $\psi$ and $\chi$, as well as other heavy vector-like (VL) fermions that mix with the SM fermions.
While the mixing of the SM fermions with the heavy VL fermions is important for the couplings of the SM fermions to $U_1$, $Z'$, and $G'$, the VL fermions themselves are not relevant for the DM dynamics as long as their masses are larger than $M_\psi + M_\chi$, which we assume in the following.

\begin{table}[h]
\begin{center}
\def\arraystretch{1.5}
\begin{tabular}{c||c|c|c|c}
&$SU(4)$ & $SU(3)'$ & $SU(2)_L$ & $U(1)'$ \\
\hline
\hline
$\Psi_{4,2}$&${\bf 4}$ & ${\bf 1}$ & ${\bf 2}$ & $0$ \\
$\Psi_{1,2}^q \oplus \Psi_{1,2}^l$&${\bf 1}$ & ${\bf 3}\oplus{\bf 1}$ & ${\bf 2}$ & $+\tfrac{1}{6}\oplus -\tfrac{1}{2}$ \\
[0.1cm]
\hline
$\Psi_{4,1}^\uparrow \oplus \Psi_{4,1}^\downarrow$&${\bf 4}$ & ${\bf 1}$ & ${\bf 1}$ & $+\tfrac{1}{2}\oplus -\tfrac{1}{2}$ \\
$(\Psi_{1,1}^u \oplus \Psi_{1,1}^d)\oplus (\Psi_{1,1}^\nu \oplus \Psi_{1,1}^e)$&${\bf 1}$ & ${\bf 3}\oplus{\bf 1}$ & ${\bf 1}$ & $(+\tfrac{2}{3}\oplus -\tfrac{1}{3})\oplus(0\oplus -1)$ \\
[0.1cm]
\hline
$\Psi_{\rm DM}$&${\bf 4}$ & ${\bf 1}$ & ${\bm N}$ & $+\tfrac{1}{2}$ \\
\end{tabular}
\end{center}
\caption{Quantum numbers of fermions in the model.
Fields with the same quantum numbers as the SM fermions are denoted by $\Psi_{IJ}$, where $I\in\{4,1\}$ and $J\in\{2,1\}$ are the dimensions of the $SU(4)$ and $SU(2)_L$ representations, respectively.
$\Psi_{\rm DM}$ is the multiplet containing $\chi$ and $\psi$.
}
\label{tab:app_fermions}
\end{table}
\subsection{The SM fermions}\label{sec:SMfermions}
Due to the mixing with the heavy VL fermions, the SM $SU(2)_L$ doublets are in general linear combinations of the fields $\Psi_{4,2}$ and $\Psi_{1,2}$ shown in table~\ref{tab:app_fermions}, while the SM $SU(2)_L$ singlets are linear combinations of $\Psi_{4,1}$ and $\Psi_{1,1}$.
To avoid large flavour violating effects, we align the mixings between SM fermions and heavy VL fermions in the basis in which the down-quark mass matrix is diagonal~(cf.\ e.g.~\cite{DiLuzio:2018zxy}) such that the mixings are flavour-diagonal for the fields
\begin{equation}
  q^i = \begin{pmatrix} V_{ji}^*\,u^j_L \\ d^i_L \end{pmatrix} \,, \qquad
  \ell^j =  \begin{pmatrix} \nu^j_L \\ e^j_L \end{pmatrix} \,, \qquad
  u^i_R \,, \qquad d^i_R \,, \qquad e^i_R \,, \qquad \nu^i_R~,
\end{equation}
where $V$ is the CKM matrix.
A misalignment between the quark and lepton components of $\Psi_{4,2}$ is implemented by embedding the components $\Psi_{4,2}^q$ and  $\Psi_{4,2}^\ell$ that have a flavour-diagonal mixing with $\Psi_{1,2}^q$ and  $\Psi_{1,2}^\ell$, respectively, as
\begin{equation}
 \Psi_{4,2}^i
  = \begin{pmatrix} {\phantom{}\tilde{\Psi}_{4,2}^q}^i \\ {\phantom{}\tilde{\Psi}_{4,2}^\ell}^i  \end{pmatrix}
  = \begin{pmatrix} {\Psi_{4,2}^q}^i \\ W_{ij}\, {\Psi_{4,2}^\ell}^j  \end{pmatrix}\,,
\end{equation}
where $W$ is a unitary matrix parameterizing the misalignment.
For simplicity, no misalignment but only a phase difference is introduced for the quark and lepton components ($\Psi_{4,1}^{u}$ and $\Psi_{4,1}^{\nu}$), and ($\Psi_{4,1}^{d}$ and $\Psi_{4,1}^{e}$) of $\Psi_{4,1}^{\uparrow}$, and $\Psi_{4,1}^{\downarrow}$, respectively, i.e.\
\begin{equation}
 {\Psi_{4,1}^{\uparrow}}^i
  = \begin{pmatrix} {\Psi_{4,1}^u}^i \\ e^{i \phi_{\nu_i}}{\Psi_{4,1}^\nu}^i  \end{pmatrix}\,,
  \qquad
 {\Psi_{4,1}^{\downarrow}}^i
  = \begin{pmatrix} {\Psi_{4,1}^d}^i \\ e^{i \phi_{e_i}} {\Psi_{4,1}^e}^i  \end{pmatrix}\,.
\end{equation}
Consequently, the SM fields can be expressed as
\begin{equation}
\begin{aligned}
 q^i &= \cos\theta_{q_i}\,  (\Psi_{1,2}^q)^i + \sin\theta_{q_i}\, (\Psi_{4,2}^q)^i,
 \\
 \ell^i &= \cos\theta_{\ell_i}\,  (\Psi_{1,2}^\ell)^i + \sin\theta_{\ell_i}\, (\Psi_{4,2}^\ell)^i,
 \\
 u_R^i &= \cos\theta_{u_i}\,  (\Psi_{1,1}^u)^i + \sin\theta_{u_i}\, (\Psi_{4,1}^{ u})^i,
 \\
 d_R^i &= \cos\theta_{d_i}\,  (\Psi_{1,1}^d)^i + \sin\theta_{d_i}\, (\Psi_{4,1}^{ d})^i,
 \\
 e_R^i &= \cos\theta_{e_i}\,  (\Psi_{1,1}^e)^i + \sin\theta_{e_i}\, (\Psi_{4,1}^{ e})^i,
 \\
 \nu_R^i &=  \cos\theta_{\nu_i}\,  (\Psi_{1,1}^\nu)^i + \sin\theta_{\nu_i}\, (\Psi_{4,1}^{ \nu})^i.
\end{aligned}
\end{equation}
The couplings of the SM fermions to the new vector bosons are given by
\begin{equation}\label{eq:app_couplings_fermions_vectors}
\begin{aligned}
 \mathcal{L}_{Z'}
 &\supset
 \frac{g_{4}}{2\sqrt{6}\cos\theta_{41}}\,Z_\mu^\prime
 \begin{aligned}[t]
 \Big(&
   \xi_q^i\,\bar{q}_L^i\gamma^\mu q_L^i
   +
   \xi_u^i\,\bar{u}_R^i\gamma^\mu u_R^i
   +
   \xi_d^i\,\bar{d}_R^i\gamma^\mu d_R^i
   \\
  & -3\left(
   \xi_\ell^i\,\bar{\ell}_L^i\gamma^\mu \ell_L^i
   +
   \xi_e^i\,\bar{e}_R^i\gamma^\mu e_R^i
   +
   \xi_\nu\,\bar{\nu}_R^i\gamma^\mu \nu_R^i
   \right)\!\Big),
 \end{aligned}
   \\
 \mathcal{L}_{G'}
 &\supset
 \frac{g_{4}}{\cos\theta_{43}}\,G_\mu^{\prime a}
 \Big(
   \kappa_q^i\,\bar{q}^i\gamma^\mu T^a q^i
   +
   \kappa_u^i\,\bar{u}_R^i\gamma^\mu T^a u_R^i
   +
   \kappa_d^i\,\bar{d}_R^i\gamma^\mu T^a d_R^i
   \Big),
   \\
 \mathcal{L}_{U_1}
 &\supset
 \frac{g_4}{\sqrt{2}}\,U_\mu^+
 \Big(
   \beta_{q\ell}^{ij}\,\bar{q}^i\gamma^\mu \ell^j
   +
   \beta_{de}^i\,\bar{d}_R^i\gamma^\mu e_R^i
   +
   \beta_{u\nu}^i\,\bar{u}_R^i\gamma^\mu \nu_R^i
   \Big)
   +h.c.\,,
\end{aligned}
\end{equation}
where
\begin{equation}
\label{eq:kappa_xi_beta}
 \begin{aligned}
  \kappa_q^i
  &=
   \sin^2\theta_{q_i}
   -\sin^2\theta_{43}\,,
   \qquad&
  \kappa_d^i
  &=
   \sin^2\theta_{d_i}
   -\sin^2\theta_{43}\,,
   \qquad&
  \kappa_u^i
  &=
   \sin^2\theta_{u_i}
   -\sin^2\theta_{43}\,,
   \\
  \xi_q^i
  &=
   \sin^2\theta_{q_i}
   -\sin^2\theta_{41}\,,
   \qquad&
  \xi_d^i
  &=
   \sin^2\theta_{d_i}
   +2\,\sin^2\theta_{41}\,,
   \qquad&
  \xi_u^i
  &=
   \sin^2\theta_{u_i}
   -4\,\sin^2\theta_{41}\,,
   \\
  \xi_\ell^i
  &=
   \sin^2\theta_{\ell_i}
   -\sin^2\theta_{41}\,,
   \qquad&
  \xi_e^i
  &=
   \sin^2\theta_{e_i}
   -2\,\sin^2\theta_{41}\,,
   \qquad&
  \xi_\nu^i
  &=
   \sin^2\theta_{\nu_i}\,,
   \\
 \beta_{q\ell}^{ij}
 &=
 \sin\theta_{q_i}\,
   W_{ij}\,
 \sin\theta_{\ell_j}\,,
   \qquad&
 \beta_{de}^{i}
 &=
 \sin\theta_{d_i}\,
 \sin\theta_{e_i}\,
 e^{i \phi_{e_i}}\,,
   \qquad&
 \beta_{u\nu}^{i}
 &=
 \sin\theta_{u_i}\,
 \sin\theta_{\nu_i}\,
 e^{i \phi_{\nu_i}}\,.
 \end{aligned}
\end{equation}
The above parameterization is general enough to recover the couplings between SM fermions and the heavy vector bosons in several 4321 models in the literature.
In particular, the following special cases can be considered.
\begin{itemize}
 \item \textbf{Traditional 4321 models:}
 In ``traditional'' 4321 models~\cite{DiLuzio:2017vat,DiLuzio:2018zxy}, all three generations of left-handed SM fermions are each a mixture of a $\mathbf{4}$ and a $\mathbf{1}$ of $SU(4)$, while all right-handed SM fermions are purely singlets of $SU(4)$.
 This corresponds to the choice
\begin{equation}
\begin{aligned}
 \sin\theta_{u_i}&=\sin\theta_{d_i}=\sin\theta_{e_i}=\sin\theta_{\nu_i}=0,
\end{aligned}
\end{equation}
The misalignment matrix $W$ is usually chosen to be $C\!P$-conserving and to mix only the second and third generation, i.e.\
\begin{equation}\label{eq:W_matrix}
 W=
 \begin{pmatrix}
  1 & 0 & 0 \\
  0 & \cos \theta_{LQ} & \sin \theta_{LQ} \\
  0 & -\sin \theta_{LQ} & \cos \theta_{LQ} \\
 \end{pmatrix}.
\end{equation}
Consequently, the only free parameters in the fermion sector are
\begin{equation}
 \theta_{q_1},
 \quad
 \theta_{q_2},
 \quad
 \theta_{q_3},
 \quad
 \theta_{\ell_1},
 \quad
 \theta_{\ell_2},
 \quad
 \theta_{\ell_3},
 \quad
 \theta_{LQ}.
\end{equation}
The number of parameters can be further reduced by the following phenomenologically motivated assumptions~\cite{DiLuzio:2018zxy}:
\begin{itemize}
 \item A $U(2)$ symmetry in the quark sector, i.e.\ $\theta_{q_1}=\theta_{q_2}$, can be employed to suppress tree-level FCNC in the up-quark sector that are mediated by the $Z'$ and $G'$.
 Without such a $U(2)$ protection, excessive contributions to $\Delta C=2$ observables would be possible.
 \item The first-generation lepton doublet can be taken to be purely a singlet of $SU(4)$, i.e.\ $\theta_{\ell_1}=0$, to be safe from LFV due to $U_1$ couplings involving the electron.
\end{itemize}
Making both of the two above assumptions and defining $\theta_{q_{12}}=\theta_{q_1}=\theta_{q_2}$, the only free parameters in the fermion sector are
\begin{equation}
 \theta_{q_{12}},
 \quad
 \theta_{q_3},
 \quad
 \theta_{\ell_2},
 \quad
 \theta_{\ell_3},
 \quad
 \theta_{LQ}.
\end{equation}
If one further maximizes third generation couplings\footnote{%
Maximizing only the left-handed third-generation couplings, i.e.\ unifying the third-generation quark and lepton doublets in a pure $\mathbf{4}$ of $SU(4)$ while keeping the right-handed third-generation fermions pure singlets of $SU(4)$ might be problematic for generating the large Higgs Yukawa coupling in the third generation.
In such a case, a ``flavoured 4321'' as described below might be preferable.}
by taking $\theta_{q_3}=\theta_{\ell_3}=\tfrac{\pi}{2}$, the set of parameters further reduces to
\begin{equation}
\label{eq:flavour_pars_4321T}
 \theta_{q_{12}},
 \quad
 \theta_{\ell_2},
 \quad
 \theta_{LQ}.
\end{equation}

 \item \textbf{Flavoured 4321 models:}
In ``flavoured'' 4321 models~\cite{Bordone:2017bld,Bordone:2018nbg,Greljo:2018tuh,Cornella:2019hct}, all third generation SM fermions are fully unified into $\mathbf{4}$ representations of $SU(4)$ and only the left-handed first- and second-generation fermions are each a mixture of a $\mathbf{4}$ and a $\mathbf{1}$ of $SU(4)$.
This corresponds to the choice
\begin{equation}
  \label{eq:FlavouredModels}
\begin{aligned}
 \sin\theta_{q_3}&=\sin\theta_{\ell_3}=\sin\theta_{u_3}=\sin\theta_{d_3}=\sin\theta_{e_3}=\sin\theta_{\nu_3}=1,
 \\
 \sin\theta_{u_1}&=\sin\theta_{u_2}=\sin\theta_{d_1}=\sin\theta_{d_2}=\sin\theta_{e_1}=\sin\theta_{e_2}=\sin\theta_{\nu_1}=\sin\theta_{\nu_2}=0,
 \\
 e^{i \phi_{e_3}}&=-1.
\end{aligned}
\end{equation}
The misalignment matrix $W$ is usually chosen as in eq.~(\ref{eq:W_matrix}).
Consequently, the only free parameters in the fermion sector are
\begin{equation}
 \theta_{q_1},
 \quad
 \theta_{q_2},
 \quad
 \theta_{\ell_1},
 \quad
 \theta_{\ell_2},
 \quad
 \theta_{LQ}.
\end{equation}
Making the above described assumptions to reduce contributions to $\Delta C=2$ observables and LFV electron couplings, the set of free parameters in the fermion sector is reduced to
\begin{equation}
\label{eq:flavour_pars_4321F}
 \theta_{q_{12}},
 \quad
 \theta_{\ell_2},
 \quad
 \theta_{LQ}.
\end{equation}
\end{itemize}

\subsection{The fermions in the DM sector}
We consider a DM candidate $\chi$ that, together with its coannihilation partner $\psi$, is part of a vector-like $\mathbf{4}$ of $SU(4)$ denoted by $\Psi_{\rm DM}$ (cf.\ table~\ref{tab:app_fermions}).
The couplings of $\chi$ and $\psi$ to the new vector bosons and the gluons are thus given by
\begin{equation}
\begin{aligned}
 \mathcal{L}_{Z'}
 &\supset
 \frac{g_{4}}{2\sqrt{6}\cos\theta_{41}}\,Z_\mu^\prime
 \left(
   \xi_\psi\,\bar{\psi} \gamma^\mu \psi
   -3\,
   \xi_\chi\,\bar{\chi} \gamma^\mu \chi
   \right),
   \\
 \mathcal{L}_{G'}
 &\supset
 \frac{g_{4}}{\cos\theta_{43}}\,\kappa_\psi\,G_\mu^{\prime a}
   \,
   \bar{\psi} \gamma^\mu T^a \psi\,,
   \\
 \mathcal{L}_{G\phantom{'}}
 &\supset
 g_{s}\,G_\mu^{a}
   \,
   \bar{\psi} \gamma^\mu T^a \psi\,,
   \\
 \mathcal{L}_{U_1}
 &\supset
 \tfrac{g_4}{\sqrt{2}}\,U_\mu^+
 \,
   \bar{\psi} \gamma^\mu \chi
   +h.c.\,,
\end{aligned}
\end{equation}
where
\begin{equation}
 \xi_\psi
  =
   1
   -4\,\sin^2\theta_{41}\,,
   \qquad
  \xi_\chi
  =
   1\,,
   \qquad
  \kappa_\psi
  =
   \cos^2\theta_{43}\,.
\end{equation}

While the above couplings are independent of the representation ${\bm N}$ of $SU(2)_L$ under which $\Psi_{\rm DM}$ transforms, the couplings to the $W$ and $Z$ bosons are clearly different for different representations.
The coupling of $Z$ to a field $\Psi_{\bm N}$ transforming as a ${\bm N}$ of $SU(2)_L$ and having hypercharge $Y$ is given by
\begin{equation}
 \mathcal{L}_Z
 \supset
 \frac{g}{\cos\theta_W}\,\bar\Psi_{\bm N}\,(T^3_{\bm N} - \sin^2\theta_W\,Q)\,\gamma^\mu\, \Psi_{\bm N}\,
    Z_\mu,
\end{equation}
where $T^3_{\bm N}$ is the diagonal generator of $SU(2)$ in the ${\bm N}$ representation and the electric charge is defined by $Q=T^3_{\bm N}+Y$.
 The coupling of $W^\pm$ to a field $\Psi_{\bm N}$ transforming as a ${\bm N}$ of $SU(2)_L$ is derived from the covariant derivative
\begin{equation}
 \begin{aligned}
   \bar\Psi_{\bm N}\,i\,D_\mu\,\gamma^\mu\, \Psi_{\bm N}
   &\supset
    \bar\Psi_{\bm N}\left(i\,\partial_\mu + g\, W_\mu^a\, T^a_{\bm N}\right)\gamma^\mu\, \Psi_{\bm N}
   \\
   &\supset
   \frac{g}{\sqrt{2}}\, \bar\Psi_{\bm N}\, (W_\mu^+\,T^+_{\bm N} + W_\mu^-\,T^-_{\bm N})\, \gamma^\mu\, \Psi_{\bm N}
   \\
   &=
   \frac{g}{\sqrt{2}}\,W_\mu^+\, \bar\Psi_{\bm N}\, \gamma^\mu\, T^+_{\bm N}\, \Psi_{\bm N}
   +h.c.
 \end{aligned}
\end{equation}
where $T^{\pm}_{\bm N}=T^1_{\bm N}\pm i\, T^2_{\bm N}$.
%%% END fermions app

\bibliographystyle{JHEP}
\bibliography{bibliography}

\providecommand{\href}[2]{#2}\begingroup\raggedright\begin{thebibliography}{100}

\bibitem{Aaij:2013qta}
{\scshape LHCb} collaboration, R.~Aaij et~al., \emph{{Measurement of
  Form-Factor-Independent Observables in the Decay $B^{0} \to K^{*0} \mu^+
  \mu^-$}}, \href{https://doi.org/10.1103/PhysRevLett.111.191801}{\emph{Phys.
  Rev. Lett.} {\bfseries 111} (2013) 191801},
  [\href{https://arxiv.org/abs/1308.1707}{{\ttfamily 1308.1707}}].

\bibitem{Aaij:2014pli}
{\scshape LHCb} collaboration, R.~Aaij et~al., \emph{{Differential branching
  fractions and isospin asymmetries of $B \to K^{(*)} \mu^+ \mu^-$ decays}},
  \href{https://doi.org/10.1007/JHEP06(2014)133}{\emph{JHEP} {\bfseries 06}
  (2014) 133}, [\href{https://arxiv.org/abs/1403.8044}{{\ttfamily 1403.8044}}].

\bibitem{Aaij:2015esa}
{\scshape LHCb} collaboration, R.~Aaij et~al., \emph{{Angular analysis and
  differential branching fraction of the decay $B^0_s\to\phi\mu^+\mu^-$}},
  \href{https://doi.org/10.1007/JHEP09(2015)179}{\emph{JHEP} {\bfseries 09}
  (2015) 179}, [\href{https://arxiv.org/abs/1506.08777}{{\ttfamily
  1506.08777}}].

\bibitem{Aaij:2014ora}
{\scshape LHCb} collaboration, R.~Aaij et~al., \emph{{Test of lepton
  universality using $B^{+}\rightarrow K^{+}\ell^{+}\ell^{-}$ decays}},
  \href{https://doi.org/10.1103/PhysRevLett.113.151601}{\emph{Phys. Rev. Lett.}
  {\bfseries 113} (2014) 151601},
  [\href{https://arxiv.org/abs/1406.6482}{{\ttfamily 1406.6482}}].

\bibitem{Aaij:2015oid}
{\scshape LHCb} collaboration, R.~Aaij et~al., \emph{{Angular analysis of the
  $B^{0} \to K^{*0} \mu^{+} \mu^{-}$ decay using 3 fb$^{-1}$ of integrated
  luminosity}}, \href{https://doi.org/10.1007/JHEP02(2016)104}{\emph{JHEP}
  {\bfseries 02} (2016) 104},
  [\href{https://arxiv.org/abs/1512.04442}{{\ttfamily 1512.04442}}].

\bibitem{ATLAS:2017dlm}
{\scshape ATLAS} collaboration, T.~A. collaboration, \emph{{Angular analysis of
  $B^0_d \to K^{*}\mu^+\mu^-$ decays in $pp$ collisions at $\sqrt{s}= 8$ TeV
  with the ATLAS detector}}, .

\bibitem{CMS:2017ivg}
{\scshape CMS} collaboration, C.~Collaboration, \emph{{Measurement of the $P_1$
  and $P_5'$ angular parameters of the decay $\mathrm{B}^0 \to \mathrm{K}^{*0}
  \mu^+ \mu^-$ in proton-proton collisions at $\sqrt{s}=8~\mathrm{TeV}$}}, .

\bibitem{Khachatryan:2015isa}
{\scshape CMS} collaboration, V.~Khachatryan et~al., \emph{{Angular analysis of
  the decay $B^0 \to K^{*0} \mu^+ \mu^-$ from pp collisions at $\sqrt s = 8$
  TeV}}, \href{https://doi.org/10.1016/j.physletb.2015.12.020}{\emph{Phys.
  Lett.} {\bfseries B753} (2016) 424--448},
  [\href{https://arxiv.org/abs/1507.08126}{{\ttfamily 1507.08126}}].

\bibitem{Aaij:2017vbb}
{\scshape LHCb} collaboration, R.~Aaij et~al., \emph{{Test of lepton
  universality with $B^{0} \rightarrow K^{*0}\ell^{+}\ell^{-}$ decays}},
  \href{https://doi.org/10.1007/JHEP08(2017)055}{\emph{JHEP} {\bfseries 08}
  (2017) 055}, [\href{https://arxiv.org/abs/1705.05802}{{\ttfamily
  1705.05802}}].

\bibitem{Aaij:2019wad}
{\scshape LHCb} collaboration, R.~Aaij et~al., \emph{{Search for
  lepton-universality violation in $B^+\to K^+\ell^+\ell^-$ decays}},
  \href{https://doi.org/10.1103/PhysRevLett.122.191801}{\emph{Phys. Rev. Lett.}
  {\bfseries 122} (2019) 191801},
  [\href{https://arxiv.org/abs/1903.09252}{{\ttfamily 1903.09252}}].

\bibitem{Abdesselam:2019wac}
{\scshape Belle} collaboration, A.~Abdesselam et~al., \emph{{Test of lepton
  flavor universality in ${B\to K^\ast\ell^+\ell^-}$ decays at Belle}},
  \href{https://arxiv.org/abs/1904.02440}{{\ttfamily 1904.02440}}.

\bibitem{Lees:2012xj}
{\scshape BaBar} collaboration, J.~P. Lees et~al., \emph{{Evidence for an
  excess of $\bar{B} \to D^{(*)} \tau^-\bar{\nu}_\tau$ decays}},
  \href{https://doi.org/10.1103/PhysRevLett.109.101802}{\emph{Phys. Rev. Lett.}
  {\bfseries 109} (2012) 101802},
  [\href{https://arxiv.org/abs/1205.5442}{{\ttfamily 1205.5442}}].

\bibitem{Lees:2013uzd}
{\scshape BaBar} collaboration, J.~P. Lees et~al., \emph{{Measurement of an
  Excess of $\bar{B} \to D^{(*)}\tau^- \bar{\nu}_\tau$ Decays and Implications
  for Charged Higgs Bosons}},
  \href{https://doi.org/10.1103/PhysRevD.88.072012}{\emph{Phys. Rev.}
  {\bfseries D88} (2013) 072012},
  [\href{https://arxiv.org/abs/1303.0571}{{\ttfamily 1303.0571}}].

\bibitem{Huschle:2015rga}
{\scshape Belle} collaboration, M.~Huschle et~al., \emph{{Measurement of the
  branching ratio of $\bar{B} \to D^{(\ast)} \tau^- \bar{\nu}_\tau$ relative to
  $\bar{B} \to D^{(\ast)} \ell^- \bar{\nu}_\ell$ decays with hadronic tagging
  at Belle}}, \href{https://doi.org/10.1103/PhysRevD.92.072014}{\emph{Phys.
  Rev.} {\bfseries D92} (2015) 072014},
  [\href{https://arxiv.org/abs/1507.03233}{{\ttfamily 1507.03233}}].

\bibitem{Aaij:2015yra}
{\scshape LHCb} collaboration, R.~Aaij et~al., \emph{{Measurement of the ratio
  of branching fractions $\mathcal{B}(\bar{B}^0 \to
  D^{*+}\tau^{-}\bar{\nu}_{\tau})/\mathcal{B}(\bar{B}^0 \to
  D^{*+}\mu^{-}\bar{\nu}_{\mu})$}},
  \href{https://doi.org/10.1103/PhysRevLett.115.159901,
  10.1103/PhysRevLett.115.111803}{\emph{Phys. Rev. Lett.} {\bfseries 115}
  (2015) 111803}, [\href{https://arxiv.org/abs/1506.08614}{{\ttfamily
  1506.08614}}].

\bibitem{Sato:2016svk}
{\scshape Belle} collaboration, Y.~Sato et~al., \emph{{Measurement of the
  branching ratio of $\bar{B}^0 \rightarrow D^{*+} \tau^- \bar{\nu}_{\tau}$
  relative to $\bar{B}^0 \rightarrow D^{*+} \ell^- \bar{\nu}_{\ell}$ decays
  with a semileptonic tagging method}},
  \href{https://doi.org/10.1103/PhysRevD.94.072007}{\emph{Phys. Rev.}
  {\bfseries D94} (2016) 072007},
  [\href{https://arxiv.org/abs/1607.07923}{{\ttfamily 1607.07923}}].

\bibitem{Hirose:2016wfn}
{\scshape Belle} collaboration, S.~Hirose et~al., \emph{{Measurement of the
  $\tau$ lepton polarization and $R(D^*)$ in the decay $\bar{B} \to D^* \tau^-
  \bar{\nu}_\tau$}},
  \href{https://doi.org/10.1103/PhysRevLett.118.211801}{\emph{Phys. Rev. Lett.}
  {\bfseries 118} (2017) 211801},
  [\href{https://arxiv.org/abs/1612.00529}{{\ttfamily 1612.00529}}].

\bibitem{Aaij:2017tyk}
{\scshape LHCb} collaboration, R.~Aaij et~al., \emph{{Measurement of the ratio
  of branching fractions
  $\mathcal{B}(B_c^+\,\to\,J/\psi\tau^+\nu_\tau)$/$\mathcal{B}(B_c^+\,\to\,J/\psi\mu^+\nu_\mu)$}},
  \href{https://doi.org/10.1103/PhysRevLett.120.121801}{\emph{Phys. Rev. Lett.}
  {\bfseries 120} (2018) 121801},
  [\href{https://arxiv.org/abs/1711.05623}{{\ttfamily 1711.05623}}].

\bibitem{Aaij:2017uff}
{\scshape LHCb} collaboration, R.~Aaij et~al., \emph{{Measurement of the ratio
  of the $B^0 \to D^{*-} \tau^+ \nu_{\tau}$ and $B^0 \to D^{*-} \mu^+
  \nu_{\mu}$ branching fractions using three-prong $\tau$-lepton decays}},
  \href{https://doi.org/10.1103/PhysRevLett.120.171802}{\emph{Phys. Rev. Lett.}
  {\bfseries 120} (2018) 171802},
  [\href{https://arxiv.org/abs/1708.08856}{{\ttfamily 1708.08856}}].

\bibitem{Abdesselam:2019dgh}
{\scshape Belle} collaboration, A.~Abdesselam et~al., \emph{{Measurement of
  $\mathcal{R}(D)$ and $\mathcal{R}(D^{\ast})$ with a semileptonic tagging
  method}},  \href{https://arxiv.org/abs/1904.08794}{{\ttfamily 1904.08794}}.

\bibitem{Belle:2019rba}
{\scshape Belle} collaboration, G.~Caria et~al., \emph{{Measurement of
  $\mathcal{R}(D)$ and $\mathcal{R}(D^*)$ with a semileptonic tagging method}},
  \href{https://doi.org/10.1103/PhysRevLett.124.161803}{\emph{Phys. Rev. Lett.}
  {\bfseries 124} (2020) 161803},
  [\href{https://arxiv.org/abs/1910.05864}{{\ttfamily 1910.05864}}].

\bibitem{Alonso:2015sja}
R.~Alonso, B.~Grinstein and J.~Martin~Camalich, \emph{{Lepton universality
  violation and lepton flavor conservation in $B$-meson decays}},
  \href{https://doi.org/10.1007/JHEP10(2015)184}{\emph{JHEP} {\bfseries 10}
  (2015) 184}, [\href{https://arxiv.org/abs/1505.05164}{{\ttfamily
  1505.05164}}].

\bibitem{Calibbi:2015kma}
L.~Calibbi, A.~Crivellin and T.~Ota, \emph{{Effective Field Theory Approach to
  $b \rightarrow s \ell \ell(^\prime)$, $B \rightarrow K^{(*)}\nu
  \overline{\nu}$ and $B \rightarrow D^{(*)} \tau \nu$ with Third Generation
  Couplings}},
  \href{https://doi.org/10.1103/PhysRevLett.115.181801}{\emph{Phys. Rev. Lett.}
  {\bfseries 115} (2015) 181801},
  [\href{https://arxiv.org/abs/1506.02661}{{\ttfamily 1506.02661}}].

\bibitem{Barbieri:2015yvd}
R.~Barbieri, G.~Isidori, A.~Pattori and F.~Senia, \emph{{Anomalies in
  $B$-decays and $U(2)$ flavour symmetry}},
  \href{https://doi.org/10.1140/epjc/s10052-016-3905-3}{\emph{Eur. Phys. J.}
  {\bfseries C76} (2016) 67},
  [\href{https://arxiv.org/abs/1512.01560}{{\ttfamily 1512.01560}}].

\bibitem{Hiller:2016kry}
G.~Hiller, D.~Loose and K.~Schönwald, \emph{{Leptoquark Flavor Patterns \& B
  Decay Anomalies}}, \href{https://doi.org/10.1007/JHEP12(2016)027}{\emph{JHEP}
  {\bfseries 12} (2016) 027},
  [\href{https://arxiv.org/abs/1609.08895}{{\ttfamily 1609.08895}}].

\bibitem{Bhattacharya:2016mcc}
B.~Bhattacharya, A.~Datta, J.-P. Guévin, D.~London and R.~Watanabe,
  \emph{{Simultaneous Explanation of the $R_K$ and $R_{D^{(*)}}$ Puzzles: a
  Model Analysis}}, \href{https://doi.org/10.1007/JHEP01(2017)015}{\emph{JHEP}
  {\bfseries 01} (2017) 015},
  [\href{https://arxiv.org/abs/1609.09078}{{\ttfamily 1609.09078}}].

\bibitem{Buttazzo:2017ixm}
D.~Buttazzo, A.~Greljo, G.~Isidori and D.~Marzocca, \emph{{B-physics anomalies:
  a guide to combined explanations}},
  \href{https://doi.org/10.1007/JHEP11(2017)044}{\emph{JHEP} {\bfseries 11}
  (2017) 044}, [\href{https://arxiv.org/abs/1706.07808}{{\ttfamily
  1706.07808}}].

\bibitem{Calibbi:2017qbu}
L.~Calibbi, A.~Crivellin and T.~Li, \emph{{Model of vector leptoquarks in view
  of the $B$-physics anomalies}},
  \href{https://doi.org/10.1103/PhysRevD.98.115002}{\emph{Phys. Rev.}
  {\bfseries D98} (2018) 115002},
  [\href{https://arxiv.org/abs/1709.00692}{{\ttfamily 1709.00692}}].

\bibitem{Angelescu:2018tyl}
A.~Angelescu, D.~Bečirević, D.~A. Faroughy and O.~Sumensari, \emph{{Closing
  the window on single leptoquark solutions to the $B$-physics anomalies}},
  \href{https://doi.org/10.1007/JHEP10(2018)183}{\emph{JHEP} {\bfseries 10}
  (2018) 183}, [\href{https://arxiv.org/abs/1808.08179}{{\ttfamily
  1808.08179}}].

\bibitem{Kumar:2018kmr}
J.~Kumar, D.~London and R.~Watanabe, \emph{{Combined Explanations of the $b \to
  s \mu^+ \mu^-$ and $b \to c \tau^- {\bar\nu}$ Anomalies: a General Model
  Analysis}}, \href{https://doi.org/10.1103/PhysRevD.99.015007}{\emph{Phys.
  Rev.} {\bfseries D99} (2019) 015007},
  [\href{https://arxiv.org/abs/1806.07403}{{\ttfamily 1806.07403}}].

\bibitem{Das:2016vkr}
D.~Das, C.~Hati, G.~Kumar and N.~Mahajan, \emph{{Towards a unified explanation
  of $R_{D^{(\ast)}}$, $R_{K}$ and $(g-2)_{\mu}$ anomalies in a left-right
  model with leptoquarks}},
  \href{https://doi.org/10.1103/PhysRevD.94.055034}{\emph{Phys. Rev.}
  {\bfseries D94} (2016) 055034},
  [\href{https://arxiv.org/abs/1605.06313}{{\ttfamily 1605.06313}}].

\bibitem{Crivellin:2017zlb}
A.~Crivellin, D.~Müller and T.~Ota, \emph{{Simultaneous explanation of
  R(D$^{(*)}$) and $b \to s \mu^{+} \mu^{-}$: the last scalar leptoquarks
  standing}}, \href{https://doi.org/10.1007/JHEP09(2017)040}{\emph{JHEP}
  {\bfseries 09} (2017) 040},
  [\href{https://arxiv.org/abs/1703.09226}{{\ttfamily 1703.09226}}].

\bibitem{Marzocca:2018wcf}
D.~Marzocca, \emph{{Addressing the B-physics anomalies in a fundamental
  Composite Higgs Model}},
  \href{https://doi.org/10.1007/JHEP07(2018)121}{\emph{JHEP} {\bfseries 07}
  (2018) 121}, [\href{https://arxiv.org/abs/1803.10972}{{\ttfamily
  1803.10972}}].

\bibitem{Becirevic:2018afm}
D.~Bečirević, I.~Doršner, S.~Fajfer, N.~Košnik, D.~A. Faroughy and
  O.~Sumensari, \emph{{Scalar leptoquarks from grand unified theories to
  accommodate the $B$-physics anomalies}},
  \href{https://doi.org/10.1103/PhysRevD.98.055003}{\emph{Phys. Rev.}
  {\bfseries D98} (2018) 055003},
  [\href{https://arxiv.org/abs/1806.05689}{{\ttfamily 1806.05689}}].

\bibitem{Bigaran:2019bqv}
I.~Bigaran, J.~Gargalionis and R.~R. Volkas, \emph{{A near-minimal leptoquark
  model for reconciling flavour anomalies and generating radiative neutrino
  masses}}, \href{https://doi.org/10.1007/JHEP10(2019)106}{\emph{JHEP}
  {\bfseries 10} (2019) 106},
  [\href{https://arxiv.org/abs/1906.01870}{{\ttfamily 1906.01870}}].

\bibitem{Datta:2019bzu}
A.~Datta, J.~L. Feng, S.~Kamali and J.~Kumar, \emph{{Resolving the
  $(g-2)_{\mu}$ and $B$ Anomalies with Leptoquarks and a Dark Higgs Boson}},
  \href{https://doi.org/10.1103/PhysRevD.101.035010}{\emph{Phys. Rev.}
  {\bfseries D101} (2020) 035010},
  [\href{https://arxiv.org/abs/1908.08625}{{\ttfamily 1908.08625}}].

\bibitem{Altmannshofer:2020axr}
W.~Altmannshofer, P.~S.~B. Dev, A.~Soni and Y.~Sui, \emph{{Addressing
  $R_{D^{(*)}}$, $R_{K^{(*)}}$, muon $g-2$ and ANITA anomalies in a minimal
  $R$-parity violating supersymmetric framework}},
  \href{https://arxiv.org/abs/2002.12910}{{\ttfamily 2002.12910}}.

\bibitem{Aebischer:2019mlg}
J.~Aebischer, W.~Altmannshofer, D.~Guadagnoli, M.~Reboud, P.~Stangl and D.~M.
  Straub, \emph{{$B$-decay discrepancies after Moriond 2019}},
  \href{https://arxiv.org/abs/1903.10434}{{\ttfamily 1903.10434}}.

\bibitem{Alguero:2019ptt}
M.~Algueró, B.~Capdevila, A.~Crivellin, S.~Descotes-Genon, P.~Masjuan,
  J.~Matias et~al., \emph{{Emerging patterns of New Physics with and without
  Lepton Flavour Universal contributions}},
  \href{https://doi.org/10.1140/epjc/s10052-019-7216-3}{\emph{Eur. Phys. J.}
  {\bfseries C79} (2019) 714},
  [\href{https://arxiv.org/abs/1903.09578}{{\ttfamily 1903.09578}}].

\bibitem{Crivellin:2018yvo}
A.~Crivellin, C.~Greub, D.~Müller and F.~Saturnino, \emph{{Importance of Loop
  Effects in Explaining the Accumulated Evidence for New Physics in B Decays
  with a Vector Leptoquark}},
  \href{https://doi.org/10.1103/PhysRevLett.122.011805}{\emph{Phys. Rev. Lett.}
  {\bfseries 122} (2019) 011805},
  [\href{https://arxiv.org/abs/1807.02068}{{\ttfamily 1807.02068}}].

\bibitem{Barbieri:2016las}
R.~Barbieri, C.~W. Murphy and F.~Senia, \emph{{B-decay Anomalies in a Composite
  Leptoquark Model}},
  \href{https://doi.org/10.1140/epjc/s10052-016-4578-7}{\emph{Eur. Phys. J.}
  {\bfseries C77} (2017) 8},
  [\href{https://arxiv.org/abs/1611.04930}{{\ttfamily 1611.04930}}].

\bibitem{Pati:1974yy}
J.~C. Pati and A.~Salam, \emph{{Lepton Number as the Fourth Color}},
  \href{https://doi.org/10.1103/PhysRevD.10.275,
  10.1103/PhysRevD.11.703.2}{\emph{Phys. Rev.} {\bfseries D10} (1974)
  275--289}.

\bibitem{Georgi:2016xhm}
H.~Georgi and Y.~Nakai, \emph{{Diphoton resonance from a new strong force}},
  \href{https://doi.org/10.1103/PhysRevD.94.075005}{\emph{Phys. Rev.}
  {\bfseries D94} (2016) 075005},
  [\href{https://arxiv.org/abs/1606.05865}{{\ttfamily 1606.05865}}].

\bibitem{Diaz:2017lit}
B.~Diaz, M.~Schmaltz and Y.-M. Zhong, \emph{{The leptoquark Hunter’s guide:
  Pair production}}, \href{https://doi.org/10.1007/JHEP10(2017)097}{\emph{JHEP}
  {\bfseries 10} (2017) 097},
  [\href{https://arxiv.org/abs/1706.05033}{{\ttfamily 1706.05033}}].

\bibitem{DiLuzio:2017vat}
L.~Di~Luzio, A.~Greljo and M.~Nardecchia, \emph{{Gauge leptoquark as the origin
  of B-physics anomalies}},
  \href{https://doi.org/10.1103/PhysRevD.96.115011}{\emph{Phys. Rev.}
  {\bfseries D96} (2017) 115011},
  [\href{https://arxiv.org/abs/1708.08450}{{\ttfamily 1708.08450}}].

\bibitem{Blanke:2018sro}
M.~Blanke and A.~Crivellin, \emph{{$B$ Meson Anomalies in a Pati-Salam Model
  within the Randall-Sundrum Background}},
  \href{https://doi.org/10.1103/PhysRevLett.121.011801}{\emph{Phys. Rev. Lett.}
  {\bfseries 121} (2018) 011801},
  [\href{https://arxiv.org/abs/1801.07256}{{\ttfamily 1801.07256}}].

\bibitem{DiLuzio:2018zxy}
L.~Di~Luzio, J.~Fuentes-Martin, A.~Greljo, M.~Nardecchia and S.~Renner,
  \emph{{Maximal Flavour Violation: a Cabibbo mechanism for leptoquarks}},
  \href{https://doi.org/10.1007/JHEP11(2018)081}{\emph{JHEP} {\bfseries 11}
  (2018) 081}, [\href{https://arxiv.org/abs/1808.00942}{{\ttfamily
  1808.00942}}].

\bibitem{Bordone:2017bld}
M.~Bordone, C.~Cornella, J.~Fuentes-Martin and G.~Isidori, \emph{{A three-site
  gauge model for flavor hierarchies and flavor anomalies}},
  \href{https://doi.org/10.1016/j.physletb.2018.02.011}{\emph{Phys. Lett.}
  {\bfseries B779} (2018) 317--323},
  [\href{https://arxiv.org/abs/1712.01368}{{\ttfamily 1712.01368}}].

\bibitem{Greljo:2018tuh}
A.~Greljo and B.~A. Stefanek, \emph{{Third family quark–lepton unification at
  the TeV scale}},
  \href{https://doi.org/10.1016/j.physletb.2018.05.033}{\emph{Phys. Lett.}
  {\bfseries B782} (2018) 131--138},
  [\href{https://arxiv.org/abs/1802.04274}{{\ttfamily 1802.04274}}].

\bibitem{Bordone:2018nbg}
M.~Bordone, C.~Cornella, J.~Fuentes-Martín and G.~Isidori, \emph{{Low-energy
  signatures of the $\mathrm{PS}^3$ model: from $B$-physics anomalies to LFV}},
  \href{https://doi.org/10.1007/JHEP10(2018)148}{\emph{JHEP} {\bfseries 10}
  (2018) 148}, [\href{https://arxiv.org/abs/1805.09328}{{\ttfamily
  1805.09328}}].

\bibitem{Cornella:2019hct}
C.~Cornella, J.~Fuentes-Martin and G.~Isidori, \emph{{Revisiting the vector
  leptoquark explanation of the B-physics anomalies}},
  \href{https://arxiv.org/abs/1903.11517}{{\ttfamily 1903.11517}}.

\bibitem{Fuentes-Martin:2020bnh}
J.~Fuentes-Martín and P.~Stangl, \emph{{Third-family quark-lepton unification
  with a fundamental composite Higgs}},
  \href{https://arxiv.org/abs/2004.11376}{{\ttfamily 2004.11376}}.

\bibitem{Barbieri:2011ci}
R.~Barbieri, G.~Isidori, J.~Jones-Perez, P.~Lodone and D.~M. Straub,
  \emph{{$U(2)$ and Minimal Flavour Violation in Supersymmetry}},
  \href{https://doi.org/10.1140/epjc/s10052-011-1725-z}{\emph{Eur. Phys. J.}
  {\bfseries C71} (2011) 1725},
  [\href{https://arxiv.org/abs/1105.2296}{{\ttfamily 1105.2296}}].

\bibitem{Blankenburg:2012nx}
G.~Blankenburg, G.~Isidori and J.~Jones-Perez, \emph{{Neutrino Masses and LFV
  from Minimal Breaking of $U(3)^5$ and $U(2)^5$ flavor Symmetries}},
  \href{https://doi.org/10.1140/epjc/s10052-012-2126-7}{\emph{Eur. Phys. J.}
  {\bfseries C72} (2012) 2126},
  [\href{https://arxiv.org/abs/1204.0688}{{\ttfamily 1204.0688}}].

\bibitem{Barbieri:2012uh}
R.~Barbieri, D.~Buttazzo, F.~Sala and D.~M. Straub, \emph{{Flavour physics from
  an approximate $U(2)^3$ symmetry}},
  \href{https://doi.org/10.1007/JHEP07(2012)181}{\emph{JHEP} {\bfseries 07}
  (2012) 181}, [\href{https://arxiv.org/abs/1203.4218}{{\ttfamily 1203.4218}}].

\bibitem{Bertone:2004pz}
G.~Bertone, D.~Hooper and J.~Silk, \emph{{Particle dark matter: Evidence,
  candidates and constraints}},
  \href{https://doi.org/10.1016/j.physrep.2004.08.031}{\emph{Phys. Rept.}
  {\bfseries 405} (2005) 279--390},
  [\href{https://arxiv.org/abs/hep-ph/0404175}{{\ttfamily hep-ph/0404175}}].

\bibitem{Roszkowski:2017nbc}
L.~Roszkowski, E.~M. Sessolo and S.~Trojanowski, \emph{{WIMP dark matter
  candidates and searches—current status and future prospects}},
  \href{https://doi.org/10.1088/1361-6633/aab913}{\emph{Rept. Prog. Phys.}
  {\bfseries 81} (2018) 066201},
  [\href{https://arxiv.org/abs/1707.06277}{{\ttfamily 1707.06277}}].

\bibitem{Cirelli:2005uq}
M.~Cirelli, N.~Fornengo and A.~Strumia, \emph{{Minimal dark matter}},
  \href{https://doi.org/10.1016/j.nuclphysb.2006.07.012}{\emph{Nucl. Phys.}
  {\bfseries B753} (2006) 178--194},
  [\href{https://arxiv.org/abs/hep-ph/0512090}{{\ttfamily hep-ph/0512090}}].

\bibitem{Baker:2015qna}
M.~J. Baker et~al., \emph{{The Coannihilation Codex}},
  \href{https://doi.org/10.1007/JHEP12(2015)120}{\emph{JHEP} {\bfseries 12}
  (2015) 120}, [\href{https://arxiv.org/abs/1510.03434}{{\ttfamily
  1510.03434}}].

\bibitem{Faroughy:2016osc}
D.~A. Faroughy, A.~Greljo and J.~F. Kamenik, \emph{{Confronting lepton flavor
  universality violation in B decays with high-$p_T$ tau lepton searches at
  LHC}}, \href{https://doi.org/10.1016/j.physletb.2016.11.011}{\emph{Phys.
  Lett.} {\bfseries B764} (2017) 126--134},
  [\href{https://arxiv.org/abs/1609.07138}{{\ttfamily 1609.07138}}].

\bibitem{Schmaltz:2018nls}
M.~Schmaltz and Y.-M. Zhong, \emph{{The leptoquark Hunter’s guide: large
  coupling}}, \href{https://doi.org/10.1007/JHEP01(2019)132}{\emph{JHEP}
  {\bfseries 01} (2019) 132},
  [\href{https://arxiv.org/abs/1810.10017}{{\ttfamily 1810.10017}}].

\bibitem{Greljo:2018tzh}
A.~Greljo, J.~Martin~Camalich and J.~D. Ruiz-Álvarez, \emph{{Mono-$\tau$
  Signatures at the LHC Constrain Explanations of $B$-decay Anomalies}},
  \href{https://doi.org/10.1103/PhysRevLett.122.131803}{\emph{Phys. Rev. Lett.}
  {\bfseries 122} (2019) 131803},
  [\href{https://arxiv.org/abs/1811.07920}{{\ttfamily 1811.07920}}].

\bibitem{Baker:2019sli}
M.~J. Baker, J.~Fuentes-Martín, G.~Isidori and M.~König, \emph{{High- $p_T$
  signatures in vector–leptoquark models}},
  \href{https://doi.org/10.1140/epjc/s10052-019-6853-x}{\emph{Eur. Phys. J.}
  {\bfseries C79} (2019) 334},
  [\href{https://arxiv.org/abs/1901.10480}{{\ttfamily 1901.10480}}].

\bibitem{Fuentes-Martin:2019ign}
J.~Fuentes-Martín, G.~Isidori, M.~König and N.~Selimović, \emph{{Vector
  Leptoquarks Beyond Tree Level}},
  \href{https://arxiv.org/abs/1910.13474}{{\ttfamily 1910.13474}}.

\bibitem{Griest:1990kh}
K.~Griest and D.~Seckel, \emph{{Three exceptions in the calculation of relic
  abundances}}, \href{https://doi.org/10.1103/PhysRevD.43.3191}{\emph{Phys.
  Rev.} {\bfseries D43} (1991) 3191--3203}.

\bibitem{Kolb:1990vq}
E.~W. Kolb and M.~S. Turner, \emph{{The Early Universe}}, {\emph{Front. Phys.}
  {\bfseries 69} (1990) 1--547}.

\bibitem{Jungman:1995df}
G.~Jungman, M.~Kamionkowski and K.~Griest, \emph{{Supersymmetric dark matter}},
  \href{https://doi.org/10.1016/0370-1573(95)00058-5}{\emph{Phys. Rept.}
  {\bfseries 267} (1996) 195--373},
  [\href{https://arxiv.org/abs/hep-ph/9506380}{{\ttfamily hep-ph/9506380}}].

\bibitem{Srednicki:1988ce}
M.~Srednicki, R.~Watkins and K.~A. Olive, \emph{{Calculations of Relic
  Densities in the Early Universe}},
  \href{https://doi.org/10.1016/0550-3213(88)90099-5}{\emph{Nucl. Phys.}
  {\bfseries B310} (1988) 693}.

\bibitem{Cannoni:2013bza}
M.~Cannoni, \emph{{Relativistic $<\sigma v_\text{rel}>$ in the calculation of
  relics abundances: a closer look}},
  \href{https://doi.org/10.1103/PhysRevD.89.103533}{\emph{Phys. Rev.}
  {\bfseries D89} (2014) 103533},
  [\href{https://arxiv.org/abs/1311.4508}{{\ttfamily 1311.4508}}].

\bibitem{Akerib:2016vxi}
{\scshape LUX} collaboration, D.~S. Akerib et~al., \emph{{Results from a search
  for dark matter in the complete LUX exposure}},
  \href{https://doi.org/10.1103/PhysRevLett.118.021303}{\emph{Phys. Rev. Lett.}
  {\bfseries 118} (2017) 021303},
  [\href{https://arxiv.org/abs/1608.07648}{{\ttfamily 1608.07648}}].

\bibitem{Cui:2017nnn}
{\scshape PandaX-II} collaboration, X.~Cui et~al., \emph{{Dark Matter Results
  From 54-Ton-Day Exposure of PandaX-II Experiment}},
  \href{https://doi.org/10.1103/PhysRevLett.119.181302}{\emph{Phys. Rev. Lett.}
  {\bfseries 119} (2017) 181302},
  [\href{https://arxiv.org/abs/1708.06917}{{\ttfamily 1708.06917}}].

\bibitem{Aprile:2018dbl}
{\scshape XENON} collaboration, E.~Aprile et~al., \emph{{Dark Matter Search
  Results from a One Ton-Year Exposure of XENON1T}},
  \href{https://doi.org/10.1103/PhysRevLett.121.111302}{\emph{Phys. Rev. Lett.}
  {\bfseries 121} (2018) 111302},
  [\href{https://arxiv.org/abs/1805.12562}{{\ttfamily 1805.12562}}].

\bibitem{Salati:2007zz}
P.~Salati, \emph{{Indirect and direct dark matter detection}},
  \href{https://doi.org/10.22323/1.049.0009}{\emph{PoS} {\bfseries CARGESE2007}
  (2007) 009}.

\bibitem{Anand:2013yka}
N.~Anand, A.~L. Fitzpatrick and W.~C. Haxton, \emph{{Weakly interacting massive
  particle-nucleus elastic scattering response}},
  \href{https://doi.org/10.1103/PhysRevC.89.065501}{\emph{Phys. Rev.}
  {\bfseries C89} (2014) 065501},
  [\href{https://arxiv.org/abs/1308.6288}{{\ttfamily 1308.6288}}].

\bibitem{Fan:2010gt}
J.~Fan, M.~Reece and L.-T. Wang, \emph{{Non-relativistic effective theory of
  dark matter direct detection}},
  \href{https://doi.org/10.1088/1475-7516/2010/11/042}{\emph{JCAP} {\bfseries
  1011} (2010) 042}, [\href{https://arxiv.org/abs/1008.1591}{{\ttfamily
  1008.1591}}].

\bibitem{Fitzpatrick:2012ix}
A.~L. Fitzpatrick, W.~Haxton, E.~Katz, N.~Lubbers and Y.~Xu, \emph{{The
  Effective Field Theory of Dark Matter Direct Detection}},
  \href{https://doi.org/10.1088/1475-7516/2013/02/004}{\emph{JCAP} {\bfseries
  1302} (2013) 004}, [\href{https://arxiv.org/abs/1203.3542}{{\ttfamily
  1203.3542}}].

\bibitem{Fitzpatrick:2012ib}
A.~L. Fitzpatrick, W.~Haxton, E.~Katz, N.~Lubbers and Y.~Xu, \emph{{Model
  Independent Direct Detection Analyses}},
  \href{https://arxiv.org/abs/1211.2818}{{\ttfamily 1211.2818}}.

\bibitem{Cirigliano:2012pq}
V.~Cirigliano, M.~L. Graesser and G.~Ovanesyan, \emph{{WIMP-nucleus scattering
  in chiral effective theory}},
  \href{https://doi.org/10.1007/JHEP10(2012)025}{\emph{JHEP} {\bfseries 10}
  (2012) 025}, [\href{https://arxiv.org/abs/1205.2695}{{\ttfamily 1205.2695}}].

\bibitem{DelNobile:2013sia}
M.~Cirelli, E.~Del~Nobile and P.~Panci, \emph{{Tools for model-independent
  bounds in direct dark matter searches}},
  \href{https://doi.org/10.1088/1475-7516/2013/10/019}{\emph{JCAP} {\bfseries
  1310} (2013) 019}, [\href{https://arxiv.org/abs/1307.5955}{{\ttfamily
  1307.5955}}].

\bibitem{Barello:2014uda}
G.~Barello, S.~Chang and C.~A. Newby, \emph{{A Model Independent Approach to
  Inelastic Dark Matter Scattering}},
  \href{https://doi.org/10.1103/PhysRevD.90.094027}{\emph{Phys. Rev.}
  {\bfseries D90} (2014) 094027},
  [\href{https://arxiv.org/abs/1409.0536}{{\ttfamily 1409.0536}}].

\bibitem{Hill:2014yxa}
R.~J. Hill and M.~P. Solon, \emph{{Standard Model anatomy of WIMP dark matter
  direct detection II: QCD analysis and hadronic matrix elements}},
  \href{https://doi.org/10.1103/PhysRevD.91.043505}{\emph{Phys. Rev.}
  {\bfseries D91} (2015) 043505},
  [\href{https://arxiv.org/abs/1409.8290}{{\ttfamily 1409.8290}}].

\bibitem{Hoferichter:2015ipa}
M.~Hoferichter, P.~Klos and A.~Schwenk, \emph{{Chiral power counting of one-
  and two-body currents in direct detection of dark matter}},
  \href{https://doi.org/10.1016/j.physletb.2015.05.041}{\emph{Phys. Lett.}
  {\bfseries B746} (2015) 410--416},
  [\href{https://arxiv.org/abs/1503.04811}{{\ttfamily 1503.04811}}].

\bibitem{Catena:2014uqa}
R.~Catena and P.~Gondolo, \emph{{Global fits of the dark matter-nucleon
  effective interactions}},
  \href{https://doi.org/10.1088/1475-7516/2014/09/045}{\emph{JCAP} {\bfseries
  1409} (2014) 045}, [\href{https://arxiv.org/abs/1405.2637}{{\ttfamily
  1405.2637}}].

\bibitem{Hill:2013hoa}
R.~J. Hill and M.~P. Solon, \emph{{WIMP-nucleon scattering with heavy WIMP
  effective theory}},
  \href{https://doi.org/10.1103/PhysRevLett.112.211602}{\emph{Phys. Rev. Lett.}
  {\bfseries 112} (2014) 211602},
  [\href{https://arxiv.org/abs/1309.4092}{{\ttfamily 1309.4092}}].

\bibitem{Hill:2011be}
R.~J. Hill and M.~P. Solon, \emph{{Universal behavior in the scattering of
  heavy, weakly interacting dark matter on nuclear targets}},
  \href{https://doi.org/10.1016/j.physletb.2012.01.013}{\emph{Phys. Lett.}
  {\bfseries B707} (2012) 539--545},
  [\href{https://arxiv.org/abs/1111.0016}{{\ttfamily 1111.0016}}].

\bibitem{Hoferichter:2016nvd}
M.~Hoferichter, P.~Klos, J.~Menéndez and A.~Schwenk, \emph{{Analysis
  strategies for general spin-independent WIMP-nucleus scattering}},
  \href{https://doi.org/10.1103/PhysRevD.94.063505}{\emph{Phys. Rev.}
  {\bfseries D94} (2016) 063505},
  [\href{https://arxiv.org/abs/1605.08043}{{\ttfamily 1605.08043}}].

\bibitem{Kurylov:2003ra}
A.~Kurylov and M.~Kamionkowski, \emph{{Generalized analysis of weakly
  interacting massive particle searches}},
  \href{https://doi.org/10.1103/PhysRevD.69.063503}{\emph{Phys. Rev.}
  {\bfseries D69} (2004) 063503},
  [\href{https://arxiv.org/abs/hep-ph/0307185}{{\ttfamily hep-ph/0307185}}].

\bibitem{Pospelov:2000bq}
M.~Pospelov and T.~ter Veldhuis, \emph{{Direct and indirect limits on the
  electromagnetic form-factors of WIMPs}},
  \href{https://doi.org/10.1016/S0370-2693(00)00358-0}{\emph{Phys. Lett.}
  {\bfseries B480} (2000) 181--186},
  [\href{https://arxiv.org/abs/hep-ph/0003010}{{\ttfamily hep-ph/0003010}}].

\bibitem{Bagnasco:1993st}
J.~Bagnasco, M.~Dine and S.~D. Thomas, \emph{{Detecting technibaryon dark
  matter}}, \href{https://doi.org/10.1016/0370-2693(94)90830-3}{\emph{Phys.
  Lett.} {\bfseries B320} (1994) 99--104},
  [\href{https://arxiv.org/abs/hep-ph/9310290}{{\ttfamily hep-ph/9310290}}].

\bibitem{Bishara:2016hek}
F.~Bishara, J.~Brod, B.~Grinstein and J.~Zupan, \emph{{Chiral Effective Theory
  of Dark Matter Direct Detection}},
  \href{https://doi.org/10.1088/1475-7516/2017/02/009}{\emph{JCAP} {\bfseries
  1702} (2017) 009}, [\href{https://arxiv.org/abs/1611.00368}{{\ttfamily
  1611.00368}}].

\bibitem{Bishara:2017pfq}
F.~Bishara, J.~Brod, B.~Grinstein and J.~Zupan, \emph{{From quarks to nucleons
  in dark matter direct detection}},
  \href{https://doi.org/10.1007/JHEP11(2017)059}{\emph{JHEP} {\bfseries 11}
  (2017) 059}, [\href{https://arxiv.org/abs/1707.06998}{{\ttfamily
  1707.06998}}].

\bibitem{Bishara:2017nnn}
F.~Bishara, J.~Brod, B.~Grinstein and J.~Zupan, \emph{{DirectDM: a tool for
  dark matter direct detection}},
  \href{https://arxiv.org/abs/1708.02678}{{\ttfamily 1708.02678}}.

\bibitem{Sierra:2015fma}
D.~Aristizabal~Sierra, F.~Staub and A.~Vicente, \emph{{Shedding light on the
  $b\to s$ anomalies with a dark sector}},
  \href{https://doi.org/10.1103/PhysRevD.92.015001}{\emph{Phys. Rev.}
  {\bfseries D92} (2015) 015001},
  [\href{https://arxiv.org/abs/1503.06077}{{\ttfamily 1503.06077}}].

\bibitem{Belanger:2015nma}
G.~Bélanger, C.~Delaunay and S.~Westhoff, \emph{{A Dark Matter Relic From Muon
  Anomalies}}, \href{https://doi.org/10.1103/PhysRevD.92.055021}{\emph{Phys.
  Rev.} {\bfseries D92} (2015) 055021},
  [\href{https://arxiv.org/abs/1507.06660}{{\ttfamily 1507.06660}}].

\bibitem{Allanach:2015gkd}
B.~Allanach, F.~S. Queiroz, A.~Strumia and S.~Sun, \emph{{$Z'$ models for the
  LHCb and $g-2$ muon anomalies}},
  \href{https://doi.org/10.1103/PhysRevD.93.055045,
  10.1103/PhysRevD.95.119902}{\emph{Phys. Rev.} {\bfseries D93} (2016) 055045},
  [\href{https://arxiv.org/abs/1511.07447}{{\ttfamily 1511.07447}}].

\bibitem{Bauer:2015boy}
M.~Bauer and M.~Neubert, \emph{{Flavor anomalies, the 750 GeV diphoton excess,
  and a dark matter candidate}},
  \href{https://doi.org/10.1103/PhysRevD.93.115030}{\emph{Phys. Rev.}
  {\bfseries D93} (2016) 115030},
  [\href{https://arxiv.org/abs/1512.06828}{{\ttfamily 1512.06828}}].

\bibitem{Celis:2016ayl}
A.~Celis, W.-Z. Feng and M.~Vollmann, \emph{{Dirac dark matter and $b \to s
  \ell^+ \ell^-$ with $\mathrm{U(1)}$ gauge symmetry}},
  \href{https://doi.org/10.1103/PhysRevD.95.035018}{\emph{Phys. Rev.}
  {\bfseries D95} (2017) 035018},
  [\href{https://arxiv.org/abs/1608.03894}{{\ttfamily 1608.03894}}].

\bibitem{Altmannshofer:2016jzy}
W.~Altmannshofer, S.~Gori, S.~Profumo and F.~S. Queiroz, \emph{{Explaining dark
  matter and B decay anomalies with an $L_\mu - L_\tau$ model}},
  \href{https://doi.org/10.1007/JHEP12(2016)106}{\emph{JHEP} {\bfseries 12}
  (2016) 106}, [\href{https://arxiv.org/abs/1609.04026}{{\ttfamily
  1609.04026}}].

\bibitem{Ko:2017quv}
P.~Ko, T.~Nomura and H.~Okada, \emph{{A flavor dependent gauge symmetry,
  Predictive radiative seesaw and LHCb anomalies}},
  \href{https://doi.org/10.1016/j.physletb.2017.07.021}{\emph{Phys. Lett.}
  {\bfseries B772} (2017) 547--552},
  [\href{https://arxiv.org/abs/1701.05788}{{\ttfamily 1701.05788}}].

\bibitem{Ko:2017yrd}
P.~Ko, T.~Nomura and H.~Okada, \emph{{Explaining $B\to K^{(*)}\ell^+ \ell^-$
  anomaly by radiatively induced coupling in $U(1)_{\mu-\tau}$ gauge
  symmetry}}, \href{https://doi.org/10.1103/PhysRevD.95.111701}{\emph{Phys.
  Rev.} {\bfseries D95} (2017) 111701},
  [\href{https://arxiv.org/abs/1702.02699}{{\ttfamily 1702.02699}}].

\bibitem{Cline:2017lvv}
J.~M. Cline, J.~M. Cornell, D.~London and R.~Watanabe, \emph{{Hidden sector
  explanation of $B$-decay and cosmic ray anomalies}},
  \href{https://doi.org/10.1103/PhysRevD.95.095015}{\emph{Phys. Rev.}
  {\bfseries D95} (2017) 095015},
  [\href{https://arxiv.org/abs/1702.00395}{{\ttfamily 1702.00395}}].

\bibitem{Sala:2017ihs}
F.~Sala and D.~M. Straub, \emph{{A New Light Particle in B Decays?}},
  \href{https://doi.org/10.1016/j.physletb.2017.09.072}{\emph{Phys. Lett.}
  {\bfseries B774} (2017) 205--209},
  [\href{https://arxiv.org/abs/1704.06188}{{\ttfamily 1704.06188}}].

\bibitem{Ellis:2017nrp}
J.~Ellis, M.~Fairbairn and P.~Tunney, \emph{{Anomaly-Free Models for Flavour
  Anomalies}}, \href{https://doi.org/10.1140/epjc/s10052-018-5725-0}{\emph{Eur.
  Phys. J.} {\bfseries C78} (2018) 238},
  [\href{https://arxiv.org/abs/1705.03447}{{\ttfamily 1705.03447}}].

\bibitem{Kawamura:2017ecz}
J.~Kawamura, S.~Okawa and Y.~Omura, \emph{{Interplay between the b$\to
  s\ell\ell$ anomalies and dark matter physics}},
  \href{https://doi.org/10.1103/PhysRevD.96.075041}{\emph{Phys. Rev.}
  {\bfseries D96} (2017) 075041},
  [\href{https://arxiv.org/abs/1706.04344}{{\ttfamily 1706.04344}}].

\bibitem{Baek:2017sew}
S.~Baek, \emph{{Dark matter contribution to $b\to s \mu^+ \mu^-$ anomaly in
  local $U(1)_{L_\mu-L_\tau}$ model}},
  \href{https://doi.org/10.1016/j.physletb.2018.04.012}{\emph{Phys. Lett.}
  {\bfseries B781} (2018) 376--382},
  [\href{https://arxiv.org/abs/1707.04573}{{\ttfamily 1707.04573}}].

\bibitem{Cline:2017aed}
J.~M. Cline, \emph{{$B$ decay anomalies and dark matter from vectorlike
  confinement}}, \href{https://doi.org/10.1103/PhysRevD.97.015013}{\emph{Phys.
  Rev.} {\bfseries D97} (2018) 015013},
  [\href{https://arxiv.org/abs/1710.02140}{{\ttfamily 1710.02140}}].

\bibitem{Cline:2017qqu}
J.~M. Cline and J.~M. Cornell, \emph{{$R({K^{(*)}})$ from dark matter
  exchange}}, \href{https://doi.org/10.1016/j.physletb.2018.05.034}{\emph{Phys.
  Lett.} {\bfseries B782} (2018) 232--237},
  [\href{https://arxiv.org/abs/1711.10770}{{\ttfamily 1711.10770}}].

\bibitem{Dhargyal:2017vcu}
L.~Dhargyal, \emph{{A simple model to explain observed muon sector anomalies
  and small neutrino masses}},
  \href{https://doi.org/10.1088/1361-6471/ab4120}{\emph{J. Phys.} {\bfseries
  G46} (2019) 125002}, [\href{https://arxiv.org/abs/1711.09772}{{\ttfamily
  1711.09772}}].

\bibitem{Chiang:2017zkh}
C.-W. Chiang and H.~Okada, \emph{{A simple model for explaining muon-related
  anomalies and dark matter}},
  \href{https://doi.org/10.1142/S0217751X19501069}{\emph{Int. J. Mod. Phys.}
  {\bfseries A34} (2019) 1950106},
  [\href{https://arxiv.org/abs/1711.07365}{{\ttfamily 1711.07365}}].

\bibitem{Vicente:2018xbv}
A.~Vicente, \emph{{Anomalies in $b \to s$ transitions and dark matter}},
  \href{https://doi.org/10.1155/2018/3905848}{\emph{Adv. High Energy Phys.}
  {\bfseries 2018} (2018) 3905848},
  [\href{https://arxiv.org/abs/1803.04703}{{\ttfamily 1803.04703}}].

\bibitem{Falkowski:2018dsl}
A.~Falkowski, S.~F. King, E.~Perdomo and M.~Pierre, \emph{{Flavourful $Z'$
  portal for vector-like neutrino Dark Matter and $R_{K^{(*)}}$}},
  \href{https://doi.org/10.1007/JHEP08(2018)061}{\emph{JHEP} {\bfseries 08}
  (2018) 061}, [\href{https://arxiv.org/abs/1803.04430}{{\ttfamily
  1803.04430}}].

\bibitem{Arcadi:2018tly}
G.~Arcadi, T.~Hugle and F.~S. Queiroz, \emph{{The Dark $L_\mu - L_\tau$ Rises
  via Kinetic Mixing}},
  \href{https://doi.org/10.1016/j.physletb.2018.07.028}{\emph{Phys. Lett.}
  {\bfseries B784} (2018) 151--158},
  [\href{https://arxiv.org/abs/1803.05723}{{\ttfamily 1803.05723}}].

\bibitem{Baek:2018aru}
S.~Baek and C.~Yu, \emph{{Dark matter for $b\to s \mu^+ \mu^-$ anomaly in a
  gauged $U(1)_X$ model}},
  \href{https://doi.org/10.1007/JHEP11(2018)054}{\emph{JHEP} {\bfseries 11}
  (2018) 054}, [\href{https://arxiv.org/abs/1806.05967}{{\ttfamily
  1806.05967}}].

\bibitem{Azatov:2018kzb}
A.~Azatov, D.~Barducci, D.~Ghosh, D.~Marzocca and L.~Ubaldi, \emph{{Combined
  explanations of B-physics anomalies: the sterile neutrino solution}},
  \href{https://doi.org/10.1007/JHEP10(2018)092}{\emph{JHEP} {\bfseries 10}
  (2018) 092}, [\href{https://arxiv.org/abs/1807.10745}{{\ttfamily
  1807.10745}}].

\bibitem{Barman:2018jhz}
B.~Barman, D.~Borah, L.~Mukherjee and S.~Nandi, \emph{{Correlating the
  anomalous results in $b \to s$ decays with inert Higgs doublet dark matter
  and muon $(g-2)$}},
  \href{https://doi.org/10.1103/PhysRevD.100.115010}{\emph{Phys. Rev.}
  {\bfseries D100} (2019) 115010},
  [\href{https://arxiv.org/abs/1808.06639}{{\ttfamily 1808.06639}}].

\bibitem{Cerdeno:2019vpd}
D.~G. Cerdeño, A.~Cheek, P.~Martín-Ramiro and J.~M. Moreno, \emph{{B
  anomalies and dark matter: a complex connection}},
  \href{https://doi.org/10.1140/epjc/s10052-019-6979-x}{\emph{Eur. Phys. J.}
  {\bfseries C79} (2019) 517},
  [\href{https://arxiv.org/abs/1902.01789}{{\ttfamily 1902.01789}}].

\bibitem{Trifinopoulos:2019lyo}
S.~Trifinopoulos, \emph{{B -physics anomalies: The bridge between R -parity
  violating supersymmetry and flavored dark matter}},
  \href{https://doi.org/10.1103/PhysRevD.100.115022}{\emph{Phys. Rev.}
  {\bfseries D100} (2019) 115022},
  [\href{https://arxiv.org/abs/1904.12940}{{\ttfamily 1904.12940}}].

\bibitem{DaRold:2019fiw}
L.~Da~Rold and F.~Lamagna, \emph{{A vector leptoquark for the B-physics
  anomalies from a composite GUT}},
  \href{https://doi.org/10.1007/JHEP12(2019)112}{\emph{JHEP} {\bfseries 12}
  (2019) 112}, [\href{https://arxiv.org/abs/1906.11666}{{\ttfamily
  1906.11666}}].

\bibitem{Han:2019diw}
Z.-L. Han, R.~Ding, S.-J. Lin and B.~Zhu, \emph{{Gauged $U(1)_{L_\mu -L_\tau }$
  scotogenic model in light of $R_{K^{(*)}}$ anomaly and AMS-02 positron
  excess}}, \href{https://doi.org/10.1140/epjc/s10052-019-7526-5}{\emph{Eur.
  Phys. J.} {\bfseries C79} (2019) 1007},
  [\href{https://arxiv.org/abs/1908.07192}{{\ttfamily 1908.07192}}].

\end{thebibliography}\endgroup

\end{document}